\PassOptionsToPackage{unicode}{hyperref}
\PassOptionsToPackage{hyphens}{url}
\PassOptionsToPackage{dvipsnames,svgnames,x11names}{xcolor}

\documentclass[12pt]{article}

\usepackage{amsmath,amssymb}
\usepackage{iftex}
\ifPDFTeX
  \usepackage[T1]{fontenc}
  \usepackage[utf8]{inputenc}
  \usepackage{textcomp}
\else
  \usepackage{unicode-math}
  \defaultfontfeatures{Scale=MatchLowercase}
  \defaultfontfeatures[\rmfamily]{Ligatures=TeX,Scale=1}
\fi
\usepackage{lmodern}
\ifPDFTeX\else  
\fi

\IfFileExists{upquote.sty}{\usepackage{upquote}}{}
\IfFileExists{microtype.sty}{
  \usepackage[]{microtype}
  \UseMicrotypeSet[protrusion]{basicmath}
}{}
\makeatletter
\@ifundefined{KOMAClassName}{
  \IfFileExists{parskip.sty}{
    \usepackage{parskip}
  }{
    \setlength{\parindent}{0pt}
    \setlength{\parskip}{6pt plus 2pt minus 1pt}}
}{
  \KOMAoptions{parskip=half}}
\makeatother
\usepackage{xcolor}
\makeatletter
\ifx\paragraph\undefined\else
  \let\oldparagraph\paragraph
  \renewcommand{\paragraph}{
    \@ifstar
      \xxxParagraphStar
      \xxxParagraphNoStar
  }
  \newcommand{\xxxParagraphStar}[1]{\oldparagraph*{#1}\mbox{}}
  \newcommand{\xxxParagraphNoStar}[1]{\oldparagraph{#1}\mbox{}}
\fi
\ifx\subparagraph\undefined\else
  \let\oldsubparagraph\subparagraph
  \renewcommand{\subparagraph}{
    \@ifstar
      \xxxSubParagraphStar
      \xxxSubParagraphNoStar
  }
  \newcommand{\xxxSubParagraphStar}[1]{\oldsubparagraph*{#1}\mbox{}}
  \newcommand{\xxxSubParagraphNoStar}[1]{\oldsubparagraph{#1}\mbox{}}
\fi
\makeatother

\usepackage{longtable,booktabs,array}
\usepackage{calc}

\usepackage{etoolbox}
\makeatletter
\patchcmd\longtable{\par}{\if@noskipsec\mbox{}\fi\par}{}{}
\makeatother

\IfFileExists{footnotehyper.sty}{\usepackage{footnotehyper}}{\usepackage{footnote}}
\makesavenoteenv{longtable}
\usepackage{graphicx}
\makeatletter
\def\maxwidth{\ifdim\Gin@nat@width>\linewidth\linewidth\else\Gin@nat@width\fi}
\def\maxheight{\ifdim\Gin@nat@height>\textheight\textheight\else\Gin@nat@height\fi}
\makeatother

\setkeys{Gin}{width=\maxwidth,height=\maxheight,keepaspectratio}

\makeatletter
\def\fps@figure{htbp}
\makeatother

\makeatletter
\@ifpackageloaded{caption}{}{\usepackage{caption}}
\AtBeginDocument{
\ifdefined\contentsname
  \renewcommand*\contentsname{Table of contents}
\else
  \newcommand\contentsname{Table of contents}
\fi
\ifdefined\listfigurename
  \renewcommand*\listfigurename{List of Figures}
\else
  \newcommand\listfigurename{List of Figures}
\fi
\ifdefined\listtablename
  \renewcommand*\listtablename{List of Tables}
\else
  \newcommand\listtablename{List of Tables}
\fi
\ifdefined\figurename
  \renewcommand*\figurename{Figure}
\else
  \newcommand\figurename{Figure}
\fi
\ifdefined\tablename
  \renewcommand*\tablename{Table}
\else
  \newcommand\tablename{Table}
\fi
}
\@ifpackageloaded{float}{}{\usepackage{float}}
\floatstyle{ruled}
\@ifundefined{c@chapter}{\newfloat{codelisting}{h}{lop}}{\newfloat{codelisting}{h}{lop}[chapter]}
\floatname{codelisting}{Listing}

\makeatother
\makeatletter
\@ifpackageloaded{caption}{}{\usepackage{caption}}
\@ifpackageloaded{subcaption}{}{\usepackage{subcaption}}
\makeatother

\ifLuaTeX
  \usepackage{selnolig}
\fi
\usepackage[]{natbib}
\usepackage{bookmark}

\IfFileExists{xurl.sty}{\usepackage{xurl}}{}
\hypersetup{
  pdftitle={Recursive Adaptive Importance Sampling with Optimal Replenishment},
  pdfauthor={Daniel Würzler Barreto; Mevin B. Hooten},
  pdfkeywords={Bayesian Inference, Parallelization, Asymptotic Distribution, Gaussian Process, Sea Surface Temperature},
  colorlinks=true,
  linkcolor={blue},
  filecolor={Maroon},
  citecolor={Blue},
  urlcolor={Blue},
  pdfcreator={LaTeX via pandoc}}

\newcommand{\anon}{1}

\usepackage[noend]{algpseudocode}
\usepackage{algorithm,amsthm}
\usepackage{bm}

\newtheorem{theorem}{Theorem}[section]
\newtheorem{lemma}[theorem]{Lemma}
\newtheorem{example}{Example}[section]

\newcommand{\bs}[1]{\bm{#1}}

\usepackage{dsfont}

\begin{document}

\def\spacingset#1{\renewcommand{\baselinestretch}
{#1}\small\normalsize} \spacingset{1}

\if1\anon
{
  \title{\bf Recursive Adaptive Importance Sampling with Optimal Replenishment}
  \author{Daniel Würzler Barreto\\
    Department of Statistics and Data Science, \\ The University of Texas at Austin\\
    and \\
    Mevin B. Hooten \\
    Department of Statistics and Data Science, \\ The University of Texas at Austin}
  \maketitle
} \fi

\if0\anon
{
  \bigskip
  \bigskip
  \bigskip
  \begin{center}
    {\LARGE\bf Recursive Adaptive Importance Sampling with Optimal Replenishment}
\end{center}
  \medskip
} \fi

\bigskip
\begin{abstract}
Increased access to computing resources has led to the development of algorithms that can run efficiently on multi-core processing units or in distributed computing environments. In the context of Bayesian inference, many parallel computing approaches to fit statistical models have been proposed using Markov chain Monte Carlo methods, but they either have limited gains due to high latency cost or rely on model-specific decompositions. Alternatively, adaptive importance sampling, sequential Monte Carlo, and recursive Bayesian methods provide a parallel-friendly and asymptotically exact framework with well-developed theory for error estimation. We consider a recursive adaptive importance sampling approach that alternates  between fast recursive weight updates and sample replenishment steps to balance computational effort while ensuring sample quality. A naive implementation can be unstable and inefficient, so we derive theoretical results to determine the optimal replenishment frequency based on the posterior concentration rate, thus attaining an asymptotically optimal trade-off. We demonstrate the efficacy of our method in simulated experiments and an application of sea surface temperature prediction in the Gulf of Mexico using Gaussian processes.
\end{abstract}

\noindent
{\it Keywords:} Bayesian, Gaussian Process, Parallelization, Sea Surface Temperature, Spatial Statistics
\vfill

\newpage
\spacingset{1} 

\section{Introduction}\label{sec:introduction}

The development of methods to fit Bayesian models has been an active area of research with impact in essentially all fields of applied science. However, with increasing availability of large data sets, it has become necessary to make use of distributed computing environments and widely available multi-core processors to solve large statistical problems quickly while facilitating accurate inference.

To achieve this, within the well-established framework of Markov chain Monte Carlo (MCMC) \citep{metropolis1953equation, hastings1970monte, geman1984stochastic}, many methods have been proposed. Other approaches immediately benefited from the availability of parallel computing resources due to their intrinsic structure, such as sequential Monte Carlo (SMC) \citep{gordon1993novel, kitagawa1996monte}, adaptive importance sampling (AIS) \citep{kloek1978bayesian, naylor1988econometric, oh1992adaptive}, and, more recently, recursive Bayesian (RB) methods \citep{hooten2021making, taylor2024fast}. We propose a general approach to fit Bayesian models that combines AIS and RB strategies. Balancing computational efficiency and estimation accuracy, our method alternates between fast deterministic weight updates and replenishment steps to sample from a sequence of posterior distributions with increasing conditioning sets, similar to \cite{chopin2002sequential} and \cite{del2006sequential}. We derive theoretical results that leverage the concentration rate of the posterior distribution to establish an optimal sequence of target distributions, consequently determining the replenishment frequency that yields an asymptotically optimal trade-off between computational effort and sample quality, even under model misspecification. Our approach uses parallelization to accelerate sampling from general low-dimensional continuous or discrete distributions, and we show it yields appreciable computational gains even when used to fit well-optimized and widely applied models, such as the geostatistical model with Vecchia approximation.

In the context of MCMC, there are three main approaches to leverage parallel computing. The first two aim to either parallelize computation within each step of a single chain (e.g., \citealp{liu2000multiple, brockwell2006parallel, yan2007parallelizing, byrd2008reducing, grazzi2025parallel}) or run chains in parallel while sharing information intermittently to accelerate mixing (e.g., \citealp{geyer1991markov, nishihara2014parallel, biron2025automatic}), both of which result in a high latency cost due to frequent information transmission between parallel processors. This last approach relies on independent parallel chains that are combined afterward. A naive implementation is known to yield diminishing returns due to burn-in \citep{rosenthal2000parallel, wilkinson2006parallel}, and thus modified approaches that decomposed the posterior distribution to reduce the cost of each chain were introduced (e.g., \citealp{neiswanger2014asymptotically, wang2015parallelizing, scott2016bayes}). These are known as embarrassingly parallel MCMC methods. However, thus far efficient decompositions of the posterior distribution rely on strong model assumptions, and hence are model-specific.

As alternatives, AIS and SMC can naturally leverage parallel resources to accelerate computation, similar to importance sampling (IS), while having well-developed error estimation and convergence theory (see \citealp{geweke1989bayesian, delyon2018asymptotic}). Despite their promising features, both approaches also have limited gains with parallelization due to the communication imposed by the operations of normalizing IS weights and updating proposal distributions \citep{rosen2011efficient}. Approaches for mitigating these issues have been considered (e.g., \citealp{chao2010efficient, murray2016parallel}), but latency remains a potential bottleneck during parallelization.

There has been renewed interest in recursive Bayesian methods (e.g., \citealp{taylor2025generative, scharf2025strategy, ren2026multi}), which use the natural prior-to-posterior update from Bayes' theorem to incorporate sample information in a multi-stage approach. RB methods fall between AIS and SMC, and therefore share many of their features and advantages such as ease of parallelization, as well as disadvantages (e.g., latency and sample degeneracy).

We consider a RB approach to fit Bayesian models that uses AIS to minimize issues associated with latency and degeneracy. Following \cite{gelfand1990sampling}, we use $[\cdot]$ to generically represent probability density or mass functions, and we denote the data, prior, and posterior distributions as $[\bs{y}_{1:n} | \bs{\theta}]$, $[\bs{\theta}]$, and $[\bs{\theta} | \bs{y}_{1:n}]$, respectively, where $\bs{y}_{1:n} = (y_{1}, \dots, y_{n})'$ represents the vector of observations and $\bs{\theta}$ is the parameter of interest. Following conventional recursive methods, we decompose the full joint posterior distribution as
\begin{equation}\label{eq:recursion identity}
    [\bs{\theta} | \bs{y}_{1:n}] = \dfrac{[\bs{y}_{1:n} | \bs{\theta}]  [\bs{\theta}]}{[\bs{y}_{1:n}]} = \dfrac{[\bs{y}_{n} | \bs{\theta}, \bs{y}_{1:(n-1)}] [\bs{\theta} | \bs{y}_{1:(n-1)}]}{[\bs{y}_{n} | \bs{y}_{1:(n-1)}]} \propto [\bs{y}_{n} | \bs{\theta}, \bs{y}_{1:(n-1)}] [\bs{\theta} | \bs{y}_{1:(n-1)}]
\end{equation}
to update the IS weights in an ``online'' manner. To counteract degeneracy, we use the weights to build approximations of partially updated posterior distributions that are then used as proposal distributions for sample replenishment. A naive introduction of replenishment (i.e., sampling from a proposal after every weight update) leads to high computational cost due to the frequent recalculation of the IS weights and construction of the approximation, potentially negating the gains from introducing parallelization. Thus, we derive theoretical results to determine the optimal weight update and replenishment frequencies, and propose a way of attaining the asymptotically optimal trade-off. We show that, for a certain class of models, allowing the gaps between update and replenishment times to grow exponentially guarantees asymptotic optimality even under model misspecification, resulting in a method that controls the asymptotic cost and degeneracy of the procedure while minimizing latency. For the aforementioned reasons, we refer to the resulting approach as recursive adaptive importance sampling with optimal replenishment (RAISOR).

We begin with a review of basic concepts to introduce the notation and present our proposed approach in Section \ref{sec:methods}. Section \ref{sec:theory} contains our main theoretical contribution, which is the proof of the asymptotic optimality of our proposed replenishment heuristic. Section \ref{sec:applications} shows the effectiveness of our method in simulated experiments and in the problem of sea surface temperature estimation using in-situ data from ships, moored buoys, and Argo floats. Finally, Section \ref{sec:discussion} discusses advantages and weakness of the proposed approach, and includes our perspective on future work.

\section{Methods}\label{sec:methods}

\subsection{Importance Sampling}

Our interest lies in evaluating integrals of the form
\begin{equation}\label{eq:IS goal}
    I_{n}(f) = \int f(\bs{\theta}) \ [\bs{\theta} | \bs{y}_{1:n}] \ d\bs{\theta} = \mathbb{E}\bigg(f(\bs{\theta}) \bigg| \bs{y}_{1:n} \bigg),
\end{equation}
for some measurable $f : \mathbb{R}^{d} \to \mathbb{R}$ with $\mathbb{E}(|f(\bs{\theta})|  | \bs{y}_{1:n} ) < +\infty$. If we cannot directly evaluate $I_{n}(f)$, a simple and well-known Monte Carlo estimator is given by 
\begin{equation}
    \hat{I}_{n}(f) = \frac{1}{M}\sum_{m=1}^{M}f(\bs{\theta}_{m}), \quad \text{ where } \bs{\theta}_{1}, \dots, \bs{\theta}_{M} \overset{\text{i.i.d.}}{\sim} [\bs{\theta} | \bs{y}_{1:n}],
\end{equation}
which is both unbiased and consistent for $I_{n}(f)$. However, this estimator requires us to sample from the posterior distribution directly, which often cannot be achieved. An alternative is to sample from an auxiliary proposal distribution $[\bs{\theta}]_{\text{prop}}$, and use the estimator
\begin{equation}
    \begin{aligned}
        \tilde{I}_{n}(f) &= \sum_{m=1}^{M} \tilde{w}(\bs{\theta}_{m}) f(\bs{\theta}_{m}), 
        &\text{ where } 
        \tilde{w}(\bs{\theta}_{m}) &= \frac{w(\bs{\theta}_{m})}{\sum_{k=1}^{M} w(\bs{\theta}_{k})}, &
        w(\bs{\theta}) &= \frac{[\bs{\theta} | \bs{y}_{1:n} ]}{[\bs{\theta}]_{\text{prop}}},
    \end{aligned}
\end{equation}
and $\bs{\theta}_{1}, \dots, \bs{\theta}_{M} \overset{\text{i.i.d.}}{\sim} [\bs{\theta}]_{\text{prop}}$. In this case, $\tilde{I}_{n}(f)$ is known as the self-normalized IS estimator, $[\bs{\theta}]_{\text{prop}}$ is the IS distribution, $\{w(\bs{\theta}_{m})\}_{m=1}^{M}$ are the IS weights, and $\{\tilde{w}(\bs{\theta}_{m})\}_{m=1}^{M}$ are the self-normalized weights. To evaluate $\tilde{w}(\bs{\theta})$ we only need to compute the posterior up to a normalizing constant, hence $\tilde{I}_{n}(f)$ can be used without knowledge of $[\bs{y}_{1:n}]$. Additionally, note that $\{(\bs{\theta}_{m}, \tilde{w}(\bs{\theta}_{m}))\}_{m=1}^{M}$ allows us to estimate $I_{n}(f)$ for any choice of $f$, which includes posterior moments and probabilities, and therefore it can be seen as a weighted sample from the posterior distribution.

Under suitable conditions, see \cite{geweke1989bayesian}, self-normalized IS estimators are known to be consistent with a bias of order $O(M^{-1})$ and satisfy a central limit theorem. The latter result can then be used to obtain asymptotically valid confidence intervals for any estimated quantity of interest, allowing for consistent estimation of the Monte Carlo error.

\subsection{Adaptive Importance Sampling}

Despite their generality, IS estimators are known to be sensitive to the IS distribution $[\bs{\theta}]_{\text{prop}}$, with poor choices leading to large variances, see for instance \cite{owen2000safe}. Originally proposed by \cite{kloek1978bayesian} and \cite{naylor1988econometric}, adaptive IS (AIS) estimators mitigate this issue by introducing a family of distributions $[\bs{\theta}]_{\text{prop}} = [\bs{\theta} | \bs{\eta}]_{\text{prop}}$, indexed by some potentially infinite dimensional quantity $\bs{\eta} \in \Xi$, and searching for the value $\bs{\eta}_{*}$ that yields the best estimator. Here, $\Xi$ denotes the space of possible values of $\bs{\eta}$ given the specific family of proposal distributions being considered (e.g., if $[\bs{\theta} | \bs{\eta}]_{\text{prop}} = \text{Normal}_{d}(\bs{\theta} | \bs{\eta}, I_{d})$, then $\bs{\eta} \in \Xi \subseteq \mathbb{R}^{d}$).

Many different criteria can be used to determine the optimal choice of $\bs{\eta}_{*}$, but one can typically represent this as an optimization problem of the form
\begin{equation}\label{eq:AIS minimization}
    \bs{\eta}_{*} = \arg\min_{\bs{\eta} \in \Xi} D_{f}\bigg( [\bs{\theta} | \bs{y}_{1:n}]  \bigg\|  [\bs{\theta} | \bs{\eta}]_{\text{prop}} \bigg),
\end{equation}
for some objective function $D_{f}(\cdot \| \cdot)$, which is a function of the IS and target distributions that can also depend on the choice of integrand $f$ in (\ref{eq:IS goal}). In the context of Bayesian inference, we are usually interested in estimating $I_{n}(f)$ for $f$ in a class of functions that might be unknown prior to model fitting, so $D_{f} = D$ is usually chosen to be agnostic with respect to $f$. Common choices are to take $D$ as a distance between moments or a divergence between the IS density and the target (e.g., \citealp{cappe2008adaptive, cornuet2012adaptive, guilmeau2024adaptive}). In this case, the Kullback-Leibler (KL) divergence is the most popular due to its balance between computational convenience and quality of the resulting approximation, but other alternatives have been considered \citep{ryu2014adaptive}.

\begin{algorithm}[tb!]
    \caption{Adaptive Importance Sampling}\label{alg:AIS}
    \begin{algorithmic}
    \State Choose an initial proposal $[\bs{\theta} | \bs{\eta}^{(1)}]_{\text{prop}}$
    \For{$s \in \{1, \dots, S\}$}
        \State Sample $\bs{\theta}_{1}^{(s)}, \dots, \bs{\theta}_{M}^{(s)} \overset{\text{i.i.d.}}{\sim} \left[ \bs{\theta} \left| \bs{\eta}^{(s)} \right. \right]_{\text{prop}}$ \Comment{in parallel for each $m$}
        \State Take $\tilde{w}(\bs{\theta}_{m}^{(s)}) \propto \frac{[\bs{y}_{1:n} | \bs{\theta}_{m}^{(s)}][\bs{\theta}_{m}^{(s)}]}{[\bs{\theta}_{m}^{(s)} | \bs{\eta}^{(s)}]_{\text{prop}}}$ \Comment{in parallel for each $m$}
        \State Obtain $D_{f}^{(s)}(\cdot \| \cdot)$, an IS estimate of the objective function
        \State Obtain $\bs{\eta}^{(s+1)} = \arg \min_{\bs{\eta} \in \Xi} D_{f}^{(s)} \bigg( [\bs{\theta} | \bs{y}_{1:n}] \bigg\| [\bs{\theta} | \bs{\eta}]_{\text{prop}} \bigg)$
    \EndFor
    \State \textbf{Return:} Weighted sample from the posterior $\{(\bs{\theta}_{m}^{(S)}, \tilde{w}(\bs{\theta}_{m}^{(S)}))\}_{m=1}^{M}$
    \end{algorithmic}
\end{algorithm}

Choosing $\bs{\eta}_{*}$ as the minimizer of a divergence may not be tractable in practice. Such a minimization typically requires one to evaluate posterior expectations, which is the goal of IS estimation in the first place. For this reason, AIS methods involve an iterative approach, with each iteration using the newly-generated samples from the IS distribution to improve the estimate of the objective function, which is then minimized to find a better proposal, as described in Algorithm \ref{alg:AIS}.

Since its initial conception, multiple variations and improvements of AIS have been considered (e.g., \citealp{zhang1996nonparametric, cappe2004population, martino2015adaptive}), including hybrid AIS and MCMC methods (e.g., \citealp{botev2013markov, martino2015adaptive}). For a more comprehensive review, see \cite{bugallo2017adaptive}. AIS follows an algorithmically simple structure, but showing the convergence of the corresponding IS estimates in the general case can be challenging due to the flexibility one has when building proposal distributions that depend on previous iterations. Initial efforts to analyze convergence considered particular cases (e.g., \citealp{oh1992adaptive, zhang1996nonparametric}), but, more recently, convergence in the more general setting was established under suitable regularity conditions \citep{delyon2018asymptotic, akyildiz2021convergence}.

\subsection{Recursive Adaptive Importance Sampling with Optimal Replenishment}\label{sub:RAISOR}

Although AIS is a flexible and parallel-friendly framework to fit general classes of Bayesian models, shortcomings hinder its widespread adoption. In what follows, we discuss these limitations and propose RAISOR as a sequence of adaptations that address them.

\subsubsection{Recursion}

In the context of AIS techniques, a common challenge is the choice of the initial IS distribution because it influences the speed of convergence and numerical stability of the procedure. Ideally, it should be close to the optimal proposal, but due to the nature of the problem this is often infeasible. In the context of fitting Bayesian models, guidance regarding the initial proposal is often vague, but a natural approach is to use the prior distribution. In principle, this can be a good option if the prior closely matches the posterior distribution, but when this is not the case (e.g., when using vague priors or when the data are highly informative), the problem persists.

To address this issue, annealed IS methods were proposed (e.g., \citealp{evans1991chaining, gramacy2008importance}), introducing a sequence of distributions that lie between the initial proposal and the target, thus smoothly bridging the gap between the distributions and increasing the stability of the procedure. From the perspective of RB methods and SMC samplers (SMCS) (e.g., \citealp{chopin2002sequential, del2006sequential, dai2022invitation}), a simple way to build a bridge of distributions is to consider the sequence $\{ [\bs{\theta} | \bs{y}_{1:n_{j}}] \}_{j=1}^{J}$, for some choice of increasing sample sizes $\{n_{j}\}_{j=1}^{J}$ with $n_{J} = n$, which changes from an initial partial posterior $[\bs{\theta} | \bs{y}_{1:n_{1}}]$ to the full posterior $[\bs{\theta} | \bs{y}_{1:n_{J}}] = [\bs{\theta} | \bs{y}_{1:n}]$ as $j$ goes from $1$ to $J$. The choice of $\{n_{j}\}$ implicitly partitions the data, thus we denote the $j\text{th}$ partition as $\tilde{\bs{y}}_{j} := \bs{y}_{(n_{j-1}+1):n_{j}}$, so that the relation $\bs{y}_{1:n_{j}} = \tilde{\bs{y}}_{1:j}$ holds for all $j > 1$. For consistency, we use the convention $\tilde{\bs{y}}_{1:0} = \bs{y}_{1:0} = \varnothing$, so that $[\bs{\theta} | \bs{y}_{1:0}] = [\bs{\theta}]$. Thus, if $\{(\bs{\theta}_{m}, \tilde{w}_{j}(\bs{\theta}_{m}))\}_{m=1}^{M}$ represents a weighted sample targeting $[\bs{\theta} | \tilde{\bs{y}}_{1:j}]$, we can use a generalization of (\ref{eq:recursion identity}) to update the weights recursively and without the need to resample by taking
\begin{equation}\label{eq:recursive weights}
    \begin{aligned}
        \tilde{w}_{j}(\bs{\theta}_{m}) 
        &= \frac{\frac{[\bs{\theta}_{m} | \tilde{\bs{y}}_{1:j}]}{[\bs{\theta}_{m}]_{\text{prop}}}}{ \sum_{k=1}^{M} \frac{[\bs{\theta}_{k} | \tilde{\bs{y}}_{1:j}]}{[\bs{\theta}_{k}]_{\text{prop}}} }
        = \frac{\frac{[\bs{\theta}_{m} | \tilde{\bs{y}}_{1:(j-1)}]}{[\bs{\theta}_{m}]_{\text{prop}}} [\tilde{\bs{y}}_{j} | \tilde{\bs{y}}_{1:(j-1)}, \bs{\theta}_{m}]}{ \sum_{k=1}^{M} \frac{[\bs{\theta}_{k} | \tilde{\bs{y}}_{1:(j-1)}]}{[\bs{\theta}_{k}]_{\text{prop}}} [\tilde{\bs{y}}_{j} | \tilde{\bs{y}}_{1:(j-1)}, \bs{\theta}_{k}] } \\
        &= \frac{\tilde{w}_{j-1}(\bs{\theta}_{m}) [\tilde{\bs{y}}_{j} | \tilde{\bs{y}}_{1:(j-1)}, \bs{\theta}_{m}]}{ \sum_{k=1}^{M} \tilde{w}_{j-1}(\bs{\theta}_{k}) [\tilde{\bs{y}}_{j} | \tilde{\bs{y}}_{1:(j-1)}, \bs{\theta}_{k}] } 
        \propto \tilde{w}_{j-1}(\bs{\theta}_{m}) [\tilde{\bs{y}}_{j} | \tilde{\bs{y}}_{1:(j-1)}, \bs{\theta}_{m}],
    \end{aligned}
\end{equation}
for all $j \in \{ 1, \dots, J\}$. Therefore, to update the weights from the initial proposal to the posterior, one only needs a sample from $[\bs{\theta} | \bs{y}_{1:n_{1}}]$ and to calculate the sequence of conditional likelihood functions evaluated at the sampled values. Due to its similarity with the prior-proposal-recursive Bayesian (PP-RB) method of \cite{hooten2021making}, we call this procedure ``importance PP-RB,'' summarized in Algorithm \ref{alg:importance PP-RB}. Importance PP-RB is equivalent to directly using $[\bs{\theta} | \bs{y}_{1:n_{1}}]$ as the proposal targeting $[\bs{\theta} | \bs{y}_{1:n}]$, with $n_{1} = 0$ corresponding to proposing from the prior and $n_{1} = n$ to directly sampling from the posterior. Taking $n_{1} \approx 0$ generally results in samples too dissimilar to the final target, and thus ill-behaved IS estimators, while taking $n_{1} \approx n$ diminishes the gains from parallelization, given that the initial sampling from $[\bs{\theta} | \bs{y}_{1:n_{1}}]$ is what we seek to optimize. To bridge this gap, we present a modified version of this approach in what follows.

\begin{algorithm}[tb!]
    \caption{Importance PP-RB}\label{alg:importance PP-RB}
    \begin{algorithmic}
    \State Sample $\bs{\theta}_{1}, \dots, \bs{\theta}_{M} \sim [\bs{\theta} | \tilde{\bs{y}}_{1}]$ and set $\tilde{w}_{1}(\bs{\theta}_{m}) = \frac{1}{M}$
    \For{$j \in \{2, \dots, J\}$}
        \State Update $\tilde{w}_{j}(\bs{\theta}_{m}) \propto \tilde{w}_{j-1}(\bs{\theta}_{m}) [\tilde{\bs{y}}_{j} | \tilde{\bs{y}}_{1:(j-1)}, \bs{\theta}_{m}]$ \Comment{in parallel for each $m$}
    \EndFor
    \State \textbf{Return:} Weighted sample from the posterior $\{(\bs{\theta}_{m}, \tilde{w}_{J}(\bs{\theta}_{m}))\}_{m=1}^{M}$
    \end{algorithmic}
\end{algorithm}

\subsubsection{Adaptation}

Considering the importance PP-RB procedure and similar to SMC methods, we treat the sample size $n$ as a time index and track how the quality of the resulting IS estimators deteriorates as it increases given a fixed $n_{1}$. Because the initial sample is generated directly from $[\bs{\theta} | \bs{y}_{1:n_{1}}]$, it does not suffer from degeneracy when taking itself as the target. As we update the weights to get closer to the posterior, we expect the weighted samples targeting their corresponding distributions to yield estimators with increasing variance over time. To improve importance PP-RB, a natural idea is to introduce replenishment to the procedure using adaptation steps, leading to a recursive AIS (RAIS) framework. This has been explored in the SMCS literature (e.g., \citealp{fearnhead2013adaptive}) with replenishment introduced using proposals based on MCMC transition kernels, but here we consider the more general class of proposals that allow for density evaluation.

Adaptation can be achieved by, after each weight update, using the resulting weighted sample to: estimate an objective function, find the estimated optimal approximation, generate from it to replenish the samples, and recalculate the IS weights considering the new samples, resulting in Algorithm \ref{alg:RAIS}. As long as the new approximation is closer to the current target distribution than the previous one, we expect the variance of the corresponding estimators to decrease after each adaptation step. This allows us to take $n_{1} \ll n$ and choose $\{n_{j}\}$ to introduce enough intermediate steps to ensure that estimators are high-quality throughout the fitting process.

\begin{algorithm}[tb!]
    \caption{Recursive Adaptive Importance Sampling}\label{alg:RAIS}
    \begin{algorithmic}
    \State Sample $\bs{\theta}_{1}^{(1)}, \dots, \bs{\theta}_{M}^{(1)} \sim [\bs{\theta} | \tilde{\bs{y}}_{1}]$ 
    \State Take $\tilde{w}_{2}(\bs{\theta}_{m}^{(1)}) \propto [\tilde{\bs{y}}_{2} | \tilde{\bs{y}}_{1}, \bs{\theta}_{m}^{(1)}]$ \Comment{in parallel for each $m$}
    \For{$j \in \{2, \dots, J-1\}$}
        \State Obtain $D_{f}^{(j)}(\cdot \| \cdot)$, an IS estimate of the objective function
        \State Obtain $\bs{\eta}^{(j)} = \arg \min_{\bs{\eta} \in \Xi} D_{f}^{(j)} \bigg( [\bs{\theta} | \tilde{\bs{y}}_{1:j}] \bigg\| [\bs{\theta} | \bs{\eta}]_{\text{prop}} \bigg)$
        \State Sample $\bs{\theta}_{1}^{(j)}, \dots, \bs{\theta}_{M}^{(j)} \overset{\text{i.i.d.}}{\sim} \left[ \bs{\theta} \left| \bs{\eta}^{(j)} \right. \right]_{\text{prop}}$ \Comment{in parallel for each $m$}
        \State Take $\tilde{w}_{j+1}(\bs{\theta}_{m}^{(j)}) \propto \frac{[\tilde{\bs{y}}_{1:(j+1)} | \bs{\theta}_{m}^{(j)}][\bs{\theta}_{m}^{(j)}]}{[\bs{\theta}_{m}^{(j)} | \bs{\eta}^{(j)}]_{\text{prop}}}$ \Comment{in parallel for each $m$}
    \EndFor
    \State \textbf{Return:} Weighted sample from the posterior $\{(\bs{\theta}_{m}^{(J-1)}, \tilde{w}_{J}(\bs{\theta}_{m}^{(J-1)}))\}_{m=1}^{M}$
    \end{algorithmic}
\end{algorithm}

Despite the fact that taking $n_{1} \ll n$ in RAIS reduces the initial cost of sampling from $[\bs{\theta} | \tilde{\bs{y}}_{1}]$, because additional computation is required for each adaptation step, the method can scale poorly with the sample size $n$ depending on the number of intermediate steps introduced. To show this, we assume that the computational cost of evaluating an expression proportional to $[\bs{\theta} | \bs{y}_{1:n}]$ is $O(n^{p})$, for some constant $p \geq 1$, which we refer to as the base cost of the model. Note that the only operation that scales with the partial sample size $n_{j}$ in Algorithm \ref{alg:RAIS} is the evaluation of the likelihood, which is the dominating cost of the procedure for large $n$. The worst case scenario has $J = n$ and $n_{j} = j$, and results in a cost per Monte Carlo sample $C$ of recalculating the weights throughout the procedure of $C(n) = O\left( \sum_{j=1}^{n} j^{p} \right) = O(n^{p+1})$, which becomes prohibitively large as $n$ grows relative to the base cost of the model.

\subsubsection{Optimal Replenishment}

We seek to introduce as many adaptation steps as possible to ensure high quality estimation, while minimizing the cost. Considering an idealized implementation of Algorithm \ref{alg:RAIS}, which we discuss below, and under i.i.d. assumptions with additional regularity conditions (see Section \ref{sec:theory}), Theorem \ref{thm:asymptotic distirbution} establishes that an asymptotically optimal trade-off can be achieved considering a simple choice of $\{n_{j}\}$, for some notion of sample quality that we define below. Despite only holding under idealized conditions, this result can then be used as a reference to propose effective heuristic choices of $\{n_{j}\}$ in general.

Taking $[\bs{\theta} | \bs{y}_{1:n}] \propto [\bs{y}_{1:n} | \bs{\theta}] [\bs{\theta}]$ to represent the usual posterior obtained via Bayes' theorem, we define the quality of $[\bs{\theta} | \bs{y}_{1:n_{1}}]$ as a proposal to generate samples targeting $[\bs{\theta} | \bs{y}_{1:n}]$ as
\begin{equation}\label{eq:RESS}
    \begin{aligned}
        Q(n | n_{1}) = Q\left([\bs{\theta} | \bs{y}_{1:n}] \left| \vphantom{\frac{1}{1}} \right. [\bs{\theta} | \bs{y}_{1:n_{1}}] \right) := \left\{ 1 + D_{\chi^{2}} \left( [\bs{\theta} | \bs{y}_{1:n}] \left\| \vphantom{\frac{1}{1}} \right. [\bs{\theta} | \bs{y}_{1:n_{1}}] \right) \right\}^{-1},
    \end{aligned}
\end{equation}
for $n \geq n_{1}$, where $D_{\chi^{2}}(f(\bs{\theta}) \| g(\bs{\theta})) = \int \frac{(f(\bs{\theta}) - g(\bs{\theta}))^{2}}{g(\bs{\theta})}  d\bs{\theta}$ represents the $\chi^{2}$-divergence between two distributions with corresponding densities $f$ and $g$. This quantity always lies in $[0,1]$, is closely related to the effective sample size (ESS, commonly used to track degeneration in particle filters, e.g., \citealp{del2006sequential, del2012adaptive}), and can be seen as the proportion of information retained when using samples from $[\bs{\theta}|\bs{y}_{1:n_{1}}]$ to approximate $[\bs{\theta} | \bs{y}_{1:n}]$. For details about these connections, see Section C of the Supplementary Material \citep{barreto2026recursive}.

Theorem \ref{thm:asymptotic distirbution} implies that, assuming perfect replenishment (i.e., sampling directly from $[\bs{\theta} | \tilde{\bs{y}}_{1:j}]$ instead at $[\bs{\theta} | \bs{\eta}^{(j)}]_{\text{prop}}$ at every iteration of Algorithm \ref{alg:RAIS}) and under the conditions stated in Section \ref{sub:notation and assumptions}, if one takes an exponentially increasing sequence $\{n_{j} \}$ with $n_{j} = \lceil \alpha^{-(j-1)} \rceil$ for some constant $\alpha \in (0,1)$, then 
\begin{equation}\label{eq:quality convergence}
    Q(n_{j+1} | n_{j}) \overset{d}{\to} Q_{\theta^{*},\alpha}(\bs{z}),
\end{equation}
as $j$ goes $+\infty$, where $Q_{\theta^{*},\alpha}(\bs{z}) \in (0,1)$ is a non-degenerate random variable with $Q_{\theta^{*},\alpha}(\bs{z}) \to 0$ as $\alpha$ goes to $0$. Therefore, for sufficiently small values of $\alpha$, the quality eventually becomes arbitrarily close to $0$, indicating that the gaps between replenishment times cannot grow larger than exponentially if one wants to avoid degeneration (see Section \ref{sec:theory} for details). But, when analyzing the computational cost in Algorithm \ref{alg:RAIS}, taking $n_{j} = \lceil \alpha^{-(j-1)} \rceil$ for $j \in \{1, \dots, J_{n}\}$ implies that $J_{n} = \left\lfloor -\frac{\log(n)}{\log(\alpha)} + 1 \right\rfloor$, resulting in
\begin{equation}\label{eq:cost}
    \begin{aligned}
        &C(n) 
        = \sum_{j=1}^{J_{n}} \lceil \alpha^{-(j-1)} \rceil^{p}
        \leq \frac{1+\alpha}{\alpha^{-p}} \sum_{j=1}^{J_{n}} \alpha^{-jp}
        = \left( \frac{1+\alpha}{\alpha^{-p}} \right) \frac{\alpha^{-J_{n}p}-1}{1-\alpha^{p}}
        \leq \frac{1 + \alpha}{1-\alpha^{p}} \left( n^{p} -\alpha^{p} \right) \\
        &\text{and}\quad C(n) 
        = \sum_{j=1}^{J_{n}} \lceil \alpha^{-(j-1)} \rceil^{p} 
        \geq \alpha^{p} \sum_{k=1}^{J_{n}} \alpha^{-jp}
        = \alpha^{p}\frac{\alpha^{-J_{n}p}-1}{1-\alpha^{p}}
        \geq \frac{1}{1-\alpha^{p}}(n^{p} - \alpha^{p}),
    \end{aligned}    
\end{equation}
therefore $C(n) = O\left( n^{p} \right)$. This implies that the time complexity of Algorithm \ref{alg:RAIS} with exponential replenishment is $O(n^{p})$, which is of the same order as other well-established methods (e.g., the Metropolis-Hastings algorithm), although methods attaining super-efficient scaling properties for certain models have been considered (e.g., \citealp{bierkens2019zig}).

\begin{table}[tb!]
    \centering
    \begin{tabular}{lcl}
        \toprule
        Replenishment Times & Computational Cost & Estimation Quality \\
        \midrule
        $\alpha \to 1$ (sub-exponential) & $C_{\text{SUB}}(n) > O(n^{p} )$ & $Q_{\theta^{*},\alpha}(\bs{z}) \overset{\mathbb{P}}{\to} 1$ \\
        $\alpha \in (0,1)$ (exponential) & $C_{\text{EXP}}(n) = O(n^{p})$ & $Q_{\theta^{*},\alpha}(\bs{z}) \in (0,1)$ \\ 
        $\alpha \to 0$ (super-exponential) & $C_{\text{SUP}}(n) = O(n^{p})$ & $Q_{\theta^{*},\alpha}(\bs{z}) \overset{\mathbb{P}}{\to} 0$ \\ 
        \bottomrule
    \end{tabular}
    \caption{Table comparing the asymptotic quality and computational cost for a model with base cost $O(n^p)$ when considering $\{n_{j}\}$ with $n_{j} = \lceil \alpha^{-(j-1)} \rceil$.}
    \label{tab:trade-off}
\end{table}

Table \ref{tab:trade-off} summarizes the role $\alpha$ plays in the RAIS algorithm, and makes it clear that, when aiming for both computational efficiency and sample quality, replenishment times should grow exponentially to achieve an asymptotically optimal trade-off, resulting in the RAIS with optimal replenishment (RAISOR) framework.

\subsubsection{RAISOR}

In what follows, we show why choosing an exponentially increasing sequence $\{n_{j}\}$ is not enough to obtain a stable procedure, and present a better way to leverage optimal replenishment in the implementation of RAIS methods. The proposed approach determines when replenishment should be introduced in a way that is more robust to the choice of $\{n_{j}\}$, while still using the theoretical results of Section \ref{sec:theory} to control the computational cost and mitigate latency.

When using an asymptotically optimal choice of $\{n_{j}\}$, even under the assumptions of Theorem \ref{thm:asymptotic distirbution}, Algorithm \ref{alg:RAIS} can still present instability. To see this, we first introduce the notation
\begin{equation}
    Q(n_{j} | \bs{\eta}) 
    := Q\left( [\bs{\theta} | \bs{y}_{1:n_{j}}] \left| \vphantom{\frac{1}{1}} \right. [\bs{\theta} | \bs{\eta}]_{\text{prop}} \right)
    = \frac{\left\{ \int \tilde{w}_{j}(\bs{\theta}) [\bs{\theta} | \bs{\eta}]_{\text{prop}} \ d\bs{\theta} \right\}^{2}}{\int \left\{ \tilde{w}_{j}(\bs{\theta}) \right\}^{2} [\bs{\theta} | \bs{\eta}]_{\text{prop}} \ d\bs{\theta}},
\end{equation}
for all $j \geq 1$ and all $\bs{\eta}$, where $\bs{\theta}_{1}, \dots, \bs{\theta}_{M} \overset{\text{i.i.d.}}{\sim} [\bs{\theta} | \bs{\eta}]_{\text{prop}}$. Analyzing Algorithm \ref{alg:RAIS}, instabilities arise when the variance of the IS estimator $D_{f}^{(j)}(\cdot \| \cdot)$ of the objective function is too high at the $j\text{th}$ iteration due to sample degeneration, that is, when $Q(n_{j+1} | \bs{\eta}^{(j)} ) \approx 0$. By the definition of $\bs{\eta}^{(j)}$ and assuming that the class of proposals is rich enough to accurately approximate the current partial posterior, we expect $Q(n_{j+1} | \bs{\eta}^{(j)}) \approx 0$ to occur only when $Q(n_{j+1} | n_{j}) \approx 0$. Therefore, we can mitigate degeneracy issues by conservatively taking $n_{j+1}$ close to $n_{j}$, which can always be achieved by sufficiently increasing $\alpha$ in the context of optimal replenishment. However, from (\ref{eq:cost}) we know the cost of model fitting significantly increases as $\alpha \to 1$, resulting in a trade-off between algorithmic stability and computational cost.

Even though taking a larger value of $\alpha$ is typically enough to prevent instability, in practice it often introduces more replenishment steps than necessary. A natural way of controlling this excess is to use Importance PP-RB while tracking an estimate of $Q(n_{j+1} | \bs{\eta}^{(j)})$ and to introduce replenishment steps only when it falls below a certain threshold $q$ (see Section B of the Supplementary Material for guidance regarding the choice of $q$). This motivates the introduction of the estimator
\begin{equation}\label{eq:quality estimate}
    \hat{Q}(n_{j} | \bs{\eta})
    := \frac{ \left\{ \frac{1}{M} \sum_{m=1}^{M} \tilde{w}_{j}(\bs{\theta}_{m}) \right\}^{2}}{\frac{1}{M}\sum_{m=1}^{M} \left\{ \tilde{w}_{j}(\bs{\theta}_{m}) \right\}^{2}},
\end{equation}
where $\bs{\theta}_{1}, \dots, \bs{\theta}_{M} \overset{i.i.d.}{\sim} [\bs{\theta} | \bs{\eta}]_{\text{prop}}$, which we can use to track the quality of the approximation, resulting in Algorithm \ref{alg:RAISOR}. This estimator can be computed in $O(M)$ operations regardless of $n_{j}$, making it an attractive option for large $n$, although it requires communication between processors. A similar approach to replenishment is often used in SMC methods to fit state space models to streaming data (e.g., \citealp{del2012adaptive}) or in the SMCS approach for static models (e.g., \citealp{del2006sequential, dai2022invitation}), but we consider an optimal weight update schedule that maintains stability while minimizing latency, with the number of observations incorporated growing exponentially with each update. For a numerical assessment of the impact of this optimal schedule, see the simulation examples in Section \ref{sec:applications}.

\begin{algorithm}[tb!]
    \caption{Recursive Adaptive Importance Sampling with Optimal Replenishment}\label{alg:RAISOR}
    \begin{algorithmic}
    \State \textbf{Require:} Exponentially increasing sequence $\{n_{j}\}$ to generate the partition $\{\tilde{\bs{y}}_{j}\}$
    \State Sample $\bs{\theta}_{1}^{(1)}, \dots, \bs{\theta}_{M}^{(1)} \sim [\bs{\theta} | \tilde{\bs{y}}_{1}]$, set $\tilde{w}_{1}(\bs{\theta}_{m}^{(1)}) = \frac{1}{M}$, and $s \gets 1$
    \For{$j \in \{2, \dots, J\}$}
        \State Update $\tilde{w}_{j}(\bs{\theta}^{(s)}) \propto \tilde{w}_{j-1}(\bs{\theta}^{(s)}) [\tilde{\bs{y}}_{j} | \tilde{\bs{y}}_{1:(j-1)}, \bs{\theta}_{m}^{(s)}]$ \Comment{in parallel for each $m$}
        \State Calculate $\hat{Q}(n_{j} | \bs{\eta}^{(s)})$ using (\ref{eq:quality estimate})${}^{(*)}$
        \If{$\hat{Q}(n_{j} | \bs{\eta}^{(s)}) < q$}{ $s \gets s+1$}
            \State Obtain $D_{f}^{(s)}(\cdot \| \cdot)$, an IS estimate of the objective function
            \State Obtain $\bs{\eta}^{(s)} = \arg \min_{\bs{\eta} \in \Xi} D_{f}^{(s)} \bigg( [\bs{\theta} | \tilde{\bs{y}}_{1:j}] \bigg\| [\bs{\theta} | \bs{\eta}]_{\text{prop}} \bigg)$
            \State Sample $\bs{\theta}_{1}^{(s)}, \dots, \bs{\theta}_{M}^{(s)} \overset{\text{i.i.d.}}{\sim} \left[ \bs{\theta} \left| \bs{\eta}^{(s)} \right. \right]_{\text{prop}}$ \Comment{in parallel for each $m$}
            \State Take $\tilde{w}_{j}(\bs{\theta}_{m}^{(s)}) \propto \frac{[\tilde{\bs{y}}_{1:j} | \bs{\theta}_{m}^{(s)}][\bs{\theta}_{m}^{(s)}]}{[\bs{\theta}_{m}^{(s)} | \bs{\eta}^{(s)}]_{\text{prop}}}$ \Comment{in parallel for each $m$}
        \EndIf
    \EndFor
    \State \textbf{Return:} Weighted sample from the posterior $\{(\bs{\theta}_{m}^{(s)}, \tilde{w}_{J}(\bs{\theta}_{m}^{(s)}))\}_{m=1}^{M}$
    \State
    \State ${}^{(*)}$here we use the convention that $[\bs{\theta} | \bs{\eta}^{(1)}]_{\text{prop}} = [\bs{\theta} | \tilde{\bs{y}}_{1}]$
    \end{algorithmic}
\end{algorithm}

Regarding the computational cost of Algorithm \ref{alg:RAISOR}, assuming that $\{n_{j}\}$ grows exponentially as before, it remains the same cost as of Algorithm \ref{alg:RAIS}. To see this, note that even when replenishment is introduced after every weight update, its asymptotic cost must still follow (\ref{eq:cost}). For the weight updates, given that $[\tilde{\bs{y}}_{j} | \tilde{\bs{y}}_{1:(j-1)}, \bs{\theta}] = \frac{[\tilde{\bs{y}}_{1:j} | \bs{\theta}]}{[\tilde{\bs{y}}_{1:(j-1)} | \bs{\theta}]}$, the additional cost of evaluating the conditional likelihoods is no more than $O(n_{j}^{p})$ so, even though for certain classes of models it can be reduced to $O(n_{j}^{p-1})$, it still follows (\ref{eq:cost}) in the worst case scenario. Given that these are the only steps that scale with $n$, the cost of Algorithm \ref{alg:RAISOR} is $O(n^{p})$.

By construction, Algorithm \ref{alg:RAISOR} only ``skips'' unnecessary replenishment steps, so we expect that a result analogous to equation (\ref{eq:quality convergence}) should still hold in this case. However, the inherent stochasticity of $\hat{Q}(n_{j} | \bs{\eta})$ and the introduction of imperfect replenishment makes it challenging to establish such a result even under the assumptions of Theorem \ref{thm:asymptotic distirbution}, therefore this is the subject of ongoing research.

Although Algorithm \ref{alg:RAISOR} is significantly more reliable when compared to Algorithm \ref{alg:RAIS}, more advanced strategies can be considered for further improvement. We discuss such strategies as well as implementation details regarding how to heuristically choose the objective function, the family of approximation distributions, and the sequence $\{n_{j}\}$ in Section B of the Supplementary Material.

\section{Theory}\label{sec:theory}

\subsection{Notation and Assumptions}\label{sub:notation and assumptions}

Following the particular case of i.i.d. observations from \cite{kleijn2012bernstein}, let $\{\bs{y}_{i}\}_{i=1}^{+\infty}$ be a sequence of i.i.d. random vectors in a probability space $(\Omega, \mathcal{F}, \mathbb{P}_{*})$ and generated from an unknown distribution $F_{*}$. Consider the potentially misspecified Bayesian hierarchical model
\begin{equation}
    \begin{aligned}
        \bs{y}_{1}, \dots, \bs{y}_{n} | \bs{\theta} &\overset{\text{iid}}{\sim} F(\cdot | \bs{\theta}), \\
        \bs{\theta} &\sim F_{\theta},
    \end{aligned}
\end{equation}
for all $n \geq 1$, where $F(\cdot | \bs{\theta})$ is a parametric family of distributions indexed by $\bs{\theta} \in \Theta$, with $\Theta \subseteq \mathbb{R}^{d}$ open, and $F_{\theta}$ is the prior distribution for $\bs{\theta}$. We denote the log-likelihood function associated with observations $\bs{y}_{(n_{0}+1):n_{1}}$ as $\ell_{n_{0}+1}^{n_{1}}(\bs{\theta}) = \log [\bs{y}_{(n_{0}+1):n_{1}} | \bs{\theta}]$.

Note that we do \textit{not} assume the parametric family of distributions $F(\cdot | \bs{\theta})$ contains the true $F_{*}$, but instead that there exists a unique $\bs{\theta}^{*}$ in the interior of $\Theta$ satisfying
\begin{equation}\label{eq:true theta}
    \bs{\theta}^{*} = \arg\min_{\bs{\theta} \in \Theta} \mathbb{E}_{*}\left( \log \frac{[\bs{y}_{1}]_{*}}{[\bs{y}_{1} | \bs{\theta}]} \right),
\end{equation}
where $[\bs{y}]_{*}$ is the true density of $\bs{y}_{i}$ under $\mathbb{P}_{*}$, and we use the subscript ``$*$'' to indicate that the corresponding object (e.g., expectation) is taken with respect to $\mathbb{P}_{*}$. To ensure (\ref{eq:true theta}) is well-defined, we assume there exists a single measure that dominates $F_{*}$ and all $F(\cdot | \bs{\theta})$. Importantly, even though our model includes a prior distribution over $\bs{\theta}$, we treat $\bs{\theta}$ as an indexing quantity.

We assume that the prior distribution $F_{\theta}$ has a Lebesgue-density $[\bs{\theta}]$, that is continuous and positive in a neighborhood $B_{\theta^{*}}$ of $\bs{\theta}^{*}$. We assume that the likelihood function of the model is such that $\ell(\bs{\theta}) := \log [\bs{y} | \bs{\theta}]$ is differentiable at $\bs{\theta}^{*}$ with derivative $\dot{\ell}(\bs{\theta}) = \frac{\partial}{\partial\bs{\theta}} \log [\bs{y} | \bs{\theta}]$, there exists a neighborhood $U$ of $\bs{\theta}^{*}$ and a square integrable function $m_{\theta^{*}}(\bs{y})$ satisfying the inequality
\begin{equation}
    |\log [\bs{y}|\bs{\theta}_{1}] - \log[\bs{y} | \bs{\theta}_{2}] | \leq m_{\theta^{*}}(\bs{y}) \| \bs{\theta}_{1} - \bs{\theta}_{2} \|_{2}
\end{equation}
for all $\bs{\theta}_{1}, \bs{\theta}_{2} \in U$, $\mathbb{P}_{*}$-almost surely, and that there exists a positive-definite $d \times d$ matrix $\bs{V}_{\theta^{*}}$ such that
\begin{equation}\label{eq:taylor series approximation}
    -\mathbb{E}_{*}\left( \log\frac{[\bs{y} | \bs{\theta}]}{[\bs{y} | \bs{\theta}^{*}]} \right) = \frac{1}{2}(\bs{\theta} - \bs{\theta}^{*})' \bs{V}_{\theta^{*}} (\bs{\theta} - \bs{\theta}^{*})  + o(\| \bs{\theta} - \bs{\theta}^{*} \|_{2}^{2}),
\end{equation}
for $\bs{\theta} \to \bs{\theta}^{*}$. Although not strictly necessary, for simplicity we also assume that for all $n \geq 1$, the maximum likelihood estimator (MLE) $\hat{\bs{\theta}}_{1}^{n}$ based on $\bs{y}_{1:n}$ exists, lies in the interior of $\Theta$, $\mathbb{P}_{*}$-almost surely, and that regularity conditions are satisfied to ensure that $\hat{\bs{\theta}}_{1}^{n} \overset{\mathbb{P}_{*}}{\to} \bs{\theta}^{*}$, see for instance \citet[chap. 5]{van2000asymptotic} for sufficient conditions for the consistency of the MLE under model misspecification based on the theory of M-estimators.

\subsection{Convergence Result}\label{sub:main result}

As we discussed in Section \ref{sec:methods}, RAISOR relies on the estimated quality to determine when replenishment should take place throughout the fitting process, and therefore, analyzing the behavior of this statistic is essential to understand how well the procedure is expected to perform. Theorem \ref{thm:asymptotic distirbution} establishes the asymptotic distribution of the $Q(n|n_{1})$ in the case of i.i.d. observations while accounting for model misspecification.

\begin{theorem}\label{thm:asymptotic distirbution}
    Let $Q(n | n_{1})$ be as defined in (\ref{eq:RESS}) and $n_{1} = \alpha_{n} n$ for all $n \geq 1$ and some sequence of constants $\{ \alpha_{n} \}$ such that $\alpha_{n} \in \left\{\frac{1}{n}, \dots, \frac{n}{n} \right\}$ and $\alpha_{n} \to \alpha \in (0,1)$. Assume that for every sequence of constants $m_{n} \to +\infty$
    \begin{equation}\label{eq:posterior concentration}
        \begin{aligned}
            \mathbb{E}_{*}\left( \int_{B^{c}\left( \bs{\theta}^{*}, n^{-\frac{1}{2}}m_{n} \right)} [\bs{\theta} | \bs{y}_{1:n}] \, d\bs{\theta} \right) \to 0,
        \end{aligned}
    \end{equation}
    as $n \to +\infty$, where $B(\bs{\theta}^{*}, r)$ represents a $L_{2}$ neighborhood of radius $r$ around $\bs{\theta}^{*}$, and $B^{c}(\bs{\theta}^{*}, r) = \mathbb{R}^{d} \setminus B(\bs{\theta}^{*}, r)$. Then, under the assumptions stated in Section \ref{sub:notation and assumptions},
    \begin{equation}\label{eq:limit RESS}
        Q(n|n_{1}) \overset{d}{\to} Q_{\theta^{*},\alpha}(\bs{z}) := \left\{\alpha(2-\alpha)\right\}^{\frac{d}{2}} \exp\left\{-\tfrac{1-\alpha}{2-\alpha} \bs{z}' \bs{M}_{\theta^{*}} \bs{z} \right\},
    \end{equation}
    as $n$ goes to $+\infty$, where $\bs{z} \sim \textnormal{Normal}_{d}(\bs{0}, \bs{I}_{d})$ and $\bs{M}_{\theta^{*}} = (\bs{W}_{\theta^{*}}^{\frac{1}{2}})'\bs{V}_{\theta^{*}}^{-1} (\bs{W}_{\theta^{*}}^{\frac{1}{2}})$, with
    \begin{equation}
        \bs{W}_{\theta^{*}} = \mathbb{E}_{*}\left(\left. \left\{ \tfrac{\partial}{\partial \bs{\theta}} \log [\bs{y} | \bs{\theta}] \tfrac{\partial}{\partial \bs{\theta}} \log [\bs{y} | \bs{\theta}]' \right\} \right|_{\bs{\theta} = \bs{\theta}^{*}} \right),
    \end{equation}
    $\bs{V}_{\theta^{*}}$ as in (\ref{eq:taylor series approximation}), and $\bs{W}_{\theta^{*}}^{\frac{1}{2}}$ such that $\bs{W}_{\theta^{*}} = (\bs{W}_{\theta^{*}}^{\frac{1}{2}})(\bs{W}_{\theta^{*}}^{\frac{1}{2}})'$.
\end{theorem}

The technical details of the proof of Theorem \ref{thm:asymptotic distirbution} are presented in Section A of the Supplementary Material, but the proof is conceptually straightforward. We can analytically characterize the distribution of $Q(n|n_{1})$ when considering the normal case for both the data and posterior distributions (see Section C of the Supplementary Material) and, under regularity conditions, the posterior density and likelihood function will converge to a normal density by the Bernstein-von-Mises theorem and classical results. Therefore, the asymptotic behavior of $Q(n|n_{1})$ for general i.i.d. models reduces to the normal case, yielding the desired result. The posterior concentration condition in (\ref{eq:posterior concentration}), even though often impractical to directly verify, is required to ensure that the Berstein-von-Mises theorem holds, but sufficient conditions in the i.i.d. setting can be found in Theorems 3.1 and 3.2 of \cite{kleijn2012bernstein}.

We can decompose $Q_{\theta^{*},\alpha}(\bs{z})$ in (\ref{eq:limit RESS}) into the factors $u_{1}(\alpha) = \left\{ \alpha (2-\alpha )\right\}^{\frac{d}{2}}$ and $u_{2}(\alpha, \bs{z}) = \exp\left\{-\frac{1-\alpha}{2-\alpha} \bs{z}' \bs{M}_{\theta^{*}} \bs{z}\right\}$, which represent two separate aspects of the limiting behavior of $Q(n|n_{1})$ (see Section C of the Supplementary Material for details). The factor $u_{1}(\alpha)$ gives us a deterministic upper bound and, assuming for simplicity that $n_{1} = \lceil \alpha n \rceil$, we can use it to show that
\begin{equation}\label{eq:RESS upper bound}
    \begin{aligned}
        u_{1}(\alpha) &\geq q_{\text{min}} &  &\Longrightarrow & n &\leq c(q_{\text{min}},d) \cdot n_{1}, & c(q_{\text{min}},d) = q_{\text{min}}^{-\frac{2}{d}} \left\{ 1 + \sqrt{1-q_{\text{min}}^{\frac{2}{d}}} \right\},
    \end{aligned}
\end{equation}
for all $q_{\text{min}} \in (0,1)$. This implies that, if one uses weighted samples from $[\bs{\theta} | \bs{y}_{1:n_{1}}]$ targeting $[\bs{\theta} | \bs{y}_{1:n}]$, the quality of the approximation measured in terms of the $Q(n|n_{1})$ is guaranteed to be below $q_{\text{min}}$ for sufficiently large values of $n$ and $n_{1}$ satisfying $n > c(q_{\text{min}},d) \cdot n_{1}$. Note that even though taking $n \leq c(r_{\text{min}}, d) \cdot n_{1}$ does not guarantee an asymptotically high-quality approximation, it can still be used to derive computational heuristics (see Section B of the Supplementary Material) that can be used when applying RAISOR in practice.

\section{Applications}\label{sec:applications}

We applied our method to analyze simulated data and demonstrate both the efficiency and numerical stability of the proposed approach. We also applied our approach to analyze sea surface temperature data using a geostatistical model with Mat\'ern covariance function.

\subsection{Simulation Experiments}

We considered two simulation experiments: mean estimation using a conjugate normal model with i.i.d. observations, to show the asymptotic behavior of $Q(n | n_{1})$ under the assumptions of Theorem \ref{thm:asymptotic distirbution}, and function estimation using a Gaussian process (GP) regression, as a way of evaluating how our method behaves in a non-trivial example outside the scope of the theorem.

\subsubsection{Mean Estimation}

We let $y_{1}, \dots, y_{n} \overset{\text{i.i.d.}}{\sim} \text{Normal}(0, \sigma^{2})$, with $\sigma^{2}$ known, and consider the Bayesian model 
\begin{equation}
    \begin{aligned}
        y_{i} | \mu &\overset{\text{ind}.}{\sim} \text{Normal}(\mu, \sigma^{2}), \\
        \mu &\sim \text{Normal}(\mu_{0}, \sigma^{2}_{0}),
    \end{aligned}
\end{equation}
for $i \in \{1, \dots, n\}$. Well-known algebraic manipulations show that the posterior distribution is given by $\mu | \bs{y}_{1:n} \sim \text{Normal}(\mu_{n}, \sigma^{2}_{n})$, where $\sigma^{2}_{n} = \left\{ \sigma^{-2}_{0} + n \sigma^{-2} \right\}^{-1}$ and $\mu_{n} = \sigma^{2}_{n} \{ \sigma^{-2}_{0}\mu_{0} + \sigma^{-2} \sum_{i=1}^{n} y_{i} \}$. 

\begin{figure}[tb!]
    \centering
    \includegraphics[width=0.95\textwidth]{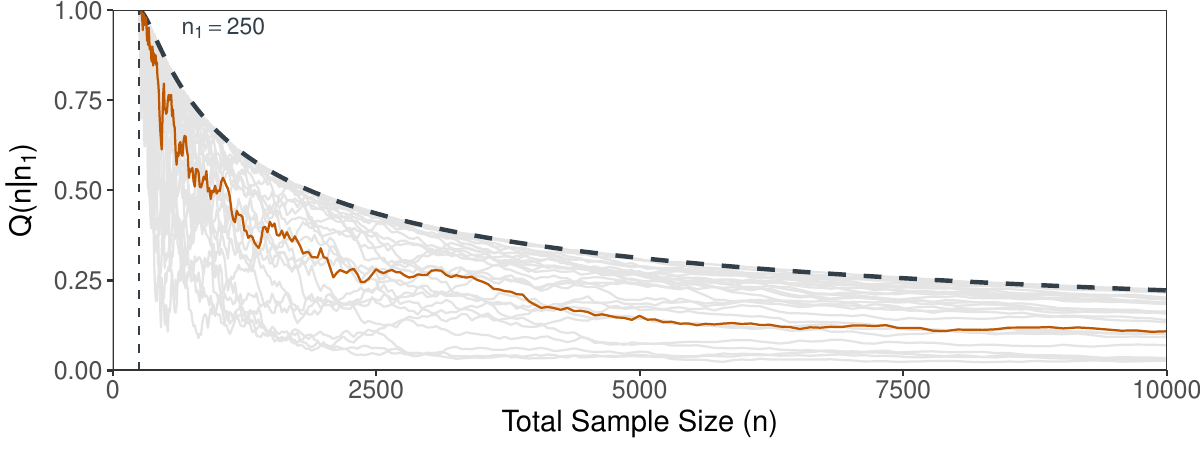}
    \caption{Solid lines shows the evolution of $Q(n|n_{1})$ as $n$ increases considering $n_{1} = 250$, $\mu_{0} = 0$, $\sigma^{2}_{0} = 10^{4}$, and $\sigma^{2} = 1$, across $30$ independent replicates of the experiment. One trajectory was highlighted for ease of visualization. Dashed line denotes the theoretical asymptotic upper bound.}
    \label{fig:RESS}
\end{figure}

Considering a sample size of $n = 10^{4}$, Figure \ref{fig:RESS} shows 30 independent replications of the evolution of the quality trajectory calculated with $M = 50000$ samples from an initial posterior distribution $[\mu | \bs{y}_{1:n_{1}}]$ with $n_{1} = 250$. The dashed line corresponds to the asymptotic theoretical upper bound $u_{1}(\alpha)$, derived in Section \ref{sec:theory}, calculated with $\alpha = \frac{n_{1}}{n}$. Considering that no sample replenishment is present, the trajectories in Figure \ref{fig:RESS} follow a noisy trajectory toward zero, although the speed of decay varies significantly across replicates. It is worth pointing out that, as indicated in Section \ref{sec:theory} and in Section C of the Supplementary Material, the speed of decay is also expected to vary with the parameter dimension and under model misspecification.

\begin{figure}[tb!]
    \centering
    \includegraphics[width=0.95\textwidth]{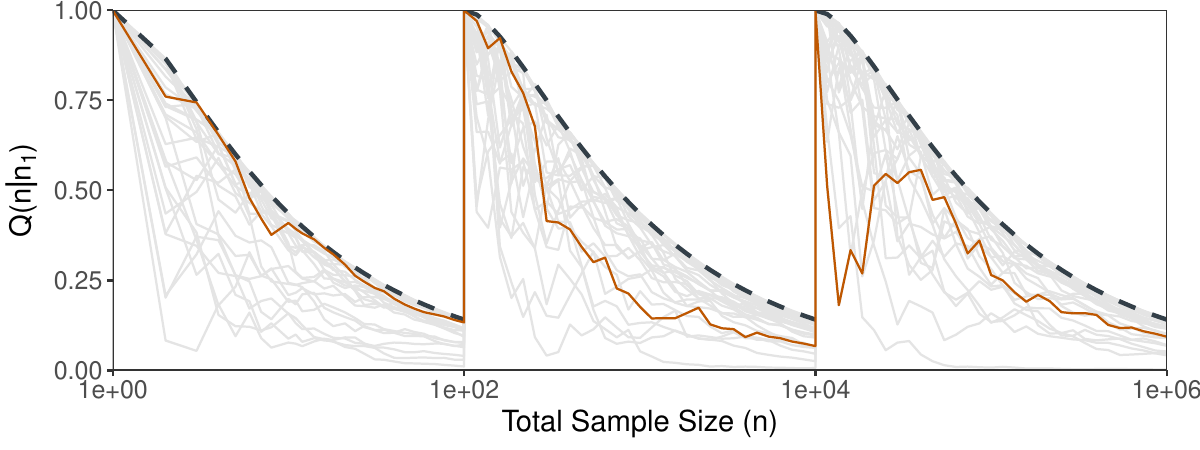}
    \caption{Solid lines shows the evolution of $Q(n|n_{1})$ as $n$ increases with replenishment introduced at $10^{0}$, $10^{2}$, and $10^{4}$, with $\mu_{0} = 0$, $\sigma^{2}_{0} = 10^{4}$, and $\sigma^{2} = 1$, across $30$ independent replicates of the experiment. One trajectory was highlighted for ease of visualization. Dashed line denotes the theoretical asymptotic upper bound when accounting for replenishment.}
    \label{fig:RESS Exponential}
\end{figure}

Following Algorithm \ref{alg:RAIS}, to show the effect of replenishment, we performed the same simulation experiment with $n = 10^{6}$, but introducing sample replenishment at $10^{0}$, $10^{2}$, and $10^{4}$. Because the posterior distribution has a known form for all $n$, we can replenish our samples by directly generating new values from the partial posterior distribution without the need to introduce approximations, resulting in perfect replenishment. Figure \ref{fig:RESS Exponential} shows the resulting $Q(n|n_{1})$ trajectories.

In Figure \ref{fig:RESS Exponential}, the replenishment naturally divides the plot in three regions, with the quality returning to 1 after each replenishment time. Noting that the $x$-axis is on a logarithmic scale and that replenishment times were taken to grow exponentially, when comparing each region there is evidence of the convergence in distribution stated in Theorem \ref{thm:asymptotic distirbution}. Although Theorem \ref{thm:asymptotic distirbution} does not guarantee the joint convergence in distribution of the trajectories, Figure \ref{fig:RESS Exponential} indicates that the $Q(n|n_{1})$ may also converge to a continuous stochastic process indexed by $\alpha = \lim_{n \to +\infty} \frac{n_{1}}{n} \in [0,1]$. Additionally, Figure \ref{fig:RESS Exponential} suggests that, when accounting for the randomness of the data generation, the quality trajectories can get arbitrarily close to 0 at any deterministic replenishment time, indicating that Algorithm \ref{alg:RAISOR} may be required to ensure robustness.

\subsubsection{Gaussian Process Regression}

We consider the problem of function estimation using a geostatistcial model (i.e., Gaussian process regression) of the form
\begin{equation}\label{eq:GP data process}
    \begin{aligned}
        \bs{y}_{1:n} | \bs{\beta}, \sigma^{2}, \tau^{2}, \phi &\sim \text{Normal}_{n}(\bs{X}\bs{\beta}, \bs{\Sigma}(\sigma^{2}, \tau^{2}, \phi)), \\
        \bs{\beta} &\sim \text{Normal}_{p}(\bs{\mu}_{\beta}, \bs{\Sigma}_{\beta}), \\
        \sigma^{2} &\sim \text{Inverse-Gamma}\left( \tfrac{\alpha_{1}}{2}, \tfrac{\alpha_{2}}{2} \right), \\
        \tau^{2} &\sim \text{Uniform}(0, 1), \\
        \phi &\sim \text{Half-Normal}(0, \gamma^{2}),
    \end{aligned}
\end{equation}
where $\bs{y}_{1:n} = (y(\bs{s}_{1}), \dots, y(\bs{s}_{n}))'$ is a vector of (noisy) observations of the function of interest at the spatial locations $\bs{s}_{1:n} = (\bs{s}_{1}, \dots, \bs{s}_{n})'$, $\bs{X}$ is a matrix of covariates, $\bs{\beta}$ is the vector of corresponding regression coefficients, and 
\begin{equation}\label{eq:covariance}
    \begin{aligned}
        \bs{\Sigma}(\sigma^{2}, \tau^{2}, \phi) &= \sigma^{2} \{(1-\tau^{2}) \bs{R}(\phi) + \tau^{2}\bs{I}_{n}\}, &
        \text{with } \bs{R}(\phi) &= \tfrac{2^{1-\nu}}{\Gamma(\nu)}\left( \sqrt{2\nu} \tfrac{\bs{D}}{\phi} \right)^{\nu} K_{\nu}\left( \sqrt{2\nu} \tfrac{\bs{D}}{\phi} \right),
    \end{aligned}
\end{equation}
corresponds to the Mat\'ern covariance matrix \citep{matern1986spatial, stein1999interpolation} with a nugget effect, $\bs{D}$ is the $n \times n$ matrix of pairwise distances between the locations $\bs{s}_{1:n}$, $K_{\nu}$ is the modified Bessel function of the second kind (see \citealp[sec. 9 and 10]{abramowitz1965handbook}), and $\sigma^{2}$, $\tau^{2}$, and $\phi$ are spatial parameters to be estimated.

The geostatistical model in (\ref{eq:GP data process}) is notoriously challenging to fit even to moderately large data sets (e.g., for $n \approx 5000$). The cost of likelihood evaluations is $O(n^{3})$, due to the need to invert a dense $n \times n$ covariance matrix, while the memory requirement is $O(n^{2})$. To mitigate this, many approximations of GPs have been proposed in the literature (e.g., \citealp{heaton2019case, liu2020gaussian}), so we consider the Vecchia approximation \citep{vecchia1988estimation, katzfuss2021general}, and in particular the Nearest-Neighbor GP (NNGP) approximation \citep{datta2016hierarchical}, to fit the GP regression model.

In our simulations, we generated six data sets for different values of $n$, with spatial locations $s_{ij} \overset{\text{i.i.d.}}{\sim} \text{Uniform}(0,1)$ for $i \in \{1, \dots, n\}$ and $j \in \{1, 2\}$, $\bs{X}$ consisting of a column of ones and the two spatial coordinates, $\bs{\beta} = (8, 4, 16)'$, $\sigma^{2} = 4$, $\tau^{2} = 0.05$, $\phi = 0.05$, $\nu = \frac{3}{2}$, $\{D\}_{ij} = \| \bs{s}_{i} - \bs{s}_{j} \|_{2}$, and the vector of observations $\bs{y}_{1:n}$ generated according to the exact GP regression model in (\ref{eq:GP data process}). We fitted the NNGP approximation of the model with $k_{n} = \left\lceil 1.2 \log_{10}^{2} n \right\rceil$ neighbors and random ordering to each simulated data set and compared five approaches: MCMC using a single core, parallel MCMC (pMCMC) chains using all available cores, AIS, annealed AIS (AAIS), RAIS (using Algorithm \ref{alg:RAISOR} with a naive update schedule given by $n_{j} = n_{1} + 20j$), and our proposed approach. To ensure a fair comparison between IS-based methods, all of them considered the same adaptation mechanism and initial proposal, given by the subposterior with $n_{1} = 10$ sampled via pMCMC. Implementation details are provided in Section B of the Supplementary Material.

\begin{table}[tb!]
    \centering
    \begin{tabular}{lccccc}
        \toprule
        & \multicolumn{5}{c}{Sample Size ($n$)} \\
        \midrule
        Method & 80 & 320 & 1280 & 5120 & 20480 \\
        \midrule
        MCMC   & 3.67 & 9.15 & 33.12 & 146.13 & 704.53 \\ 
        pMCMC  & \textbf{0.78} & \textbf{1.92} &  \textbf{7.43} &  34.25 & 161.05 \\ 
        AIS    & 2.14 &  $-$ &   $-$ &    $-$ &    $-$ \\
        AAIS   & 2.09 & 4.01 & 12.45 &  49.69 & 282.14 \\ 
        RAIS & 3.17 & 6.55 & 24.50 & 119.56 & 1122.26 \\
        RAISOR & 3.42 & 5.47 & 10.76 &  \textbf{32.28} & \textbf{133.64} \\
        \bottomrule
    \end{tabular}
    \caption{Runtime in minutes to fit the GP with Vecchia approximation to simulated data sets considering different sample sizes. Empty entries correspond to failure due to numerical instability.}
    \label{tab:time comparison}
\end{table}

We generated $M = 50000$ samples using R \citep{team2024r} with all approaches, discarding the initial $5000$ as burn-in for all MCMC chains, with all approaches but the single core MCMC making use of twenty-four parallel cores. In our experiments, AIS failed to run for all sample sizes greater than $n = 80$ due to numerical instability (i.e., because $Q(n|n_{1}) \approx 0$). Table \ref{tab:time comparison} shows the runtime of each approach and sample size considered, with pMCMC having the faster execution up to moderate $n$, being surpassed by RAISOR for large $n$.

\begin{figure}[tb!]
    \centering
    \includegraphics[width=0.95\textwidth]{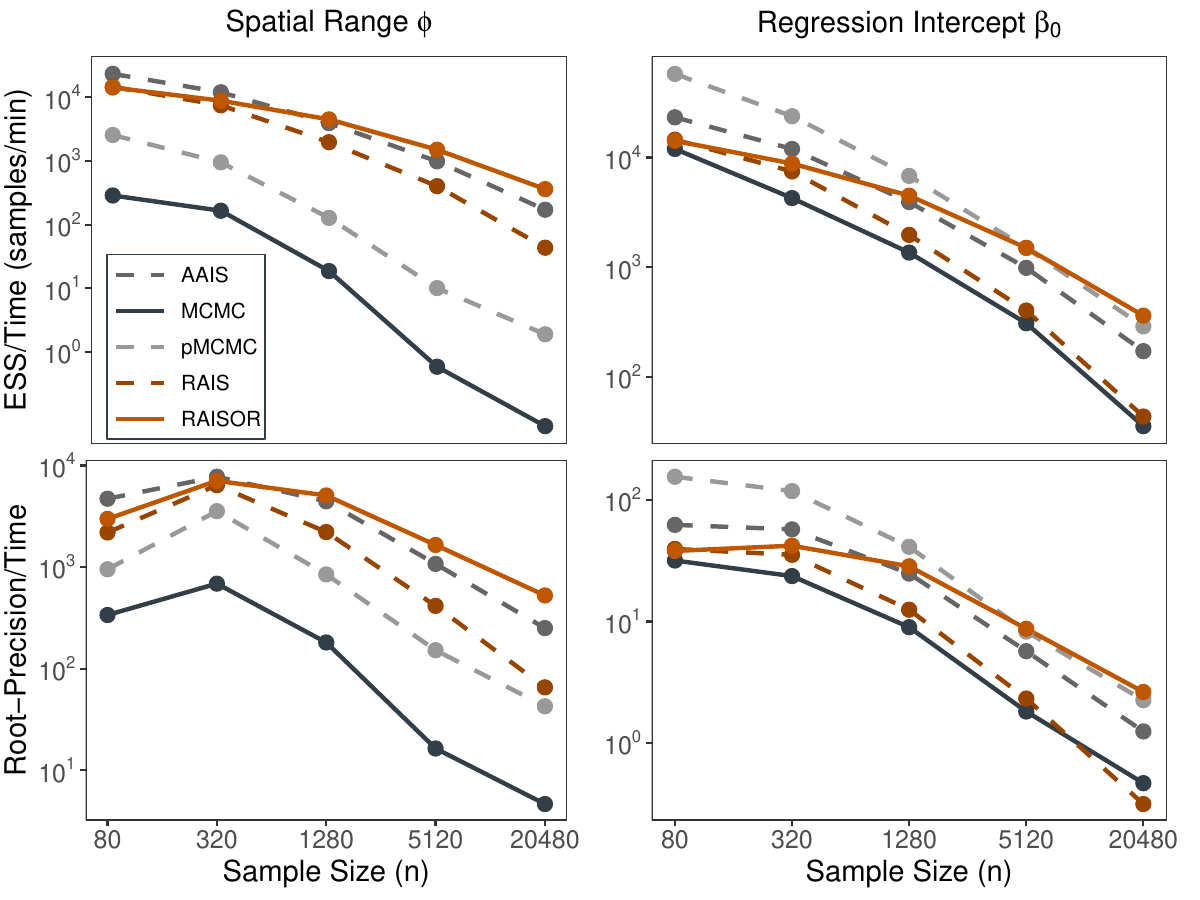}
    \caption{Comparison of the quality of the generated samples per minute for all stable methods considering two parameters. Top plots use the ESS as measure of sample quality, while bottom plots consider the estimated square root of the precision of the Monte Carlo estimates.}
    \label{fig:MCMC vs RAISOR}
\end{figure}

We also compared the quality of the generated samples. Figure \ref{fig:MCMC vs RAISOR} shows the effective sample size (ESS) and the reciprocal of the standard deviation (i.e., the root-precision) of the Monte Carlo estimators per time unit considering $\beta_{0}$ and $\phi$. For the MCMC approaches, the ESS was calculated following \cite{gamerman2006markov}, while the IS-based ESS used $M$ times the estimated quality (see Section C of the Supplementary Material). Comparing the curves, we see that across all parameters and metrics RAISOR yields greater quality per time unit for large sample sizes relative to competing approaches. Notably, comparing the performance of the RAIS and RAISOR implementations we see that the adoption an optimal weight update schedule significantly reduced the runtime of the procedure, improving the efficiency of the algorithm by nearly one order of magnitude. Results for the remaining parameters were analogous, and therefore omitted.

\begin{figure}[tb!]
    \centering
    \includegraphics[width=0.95\textwidth]{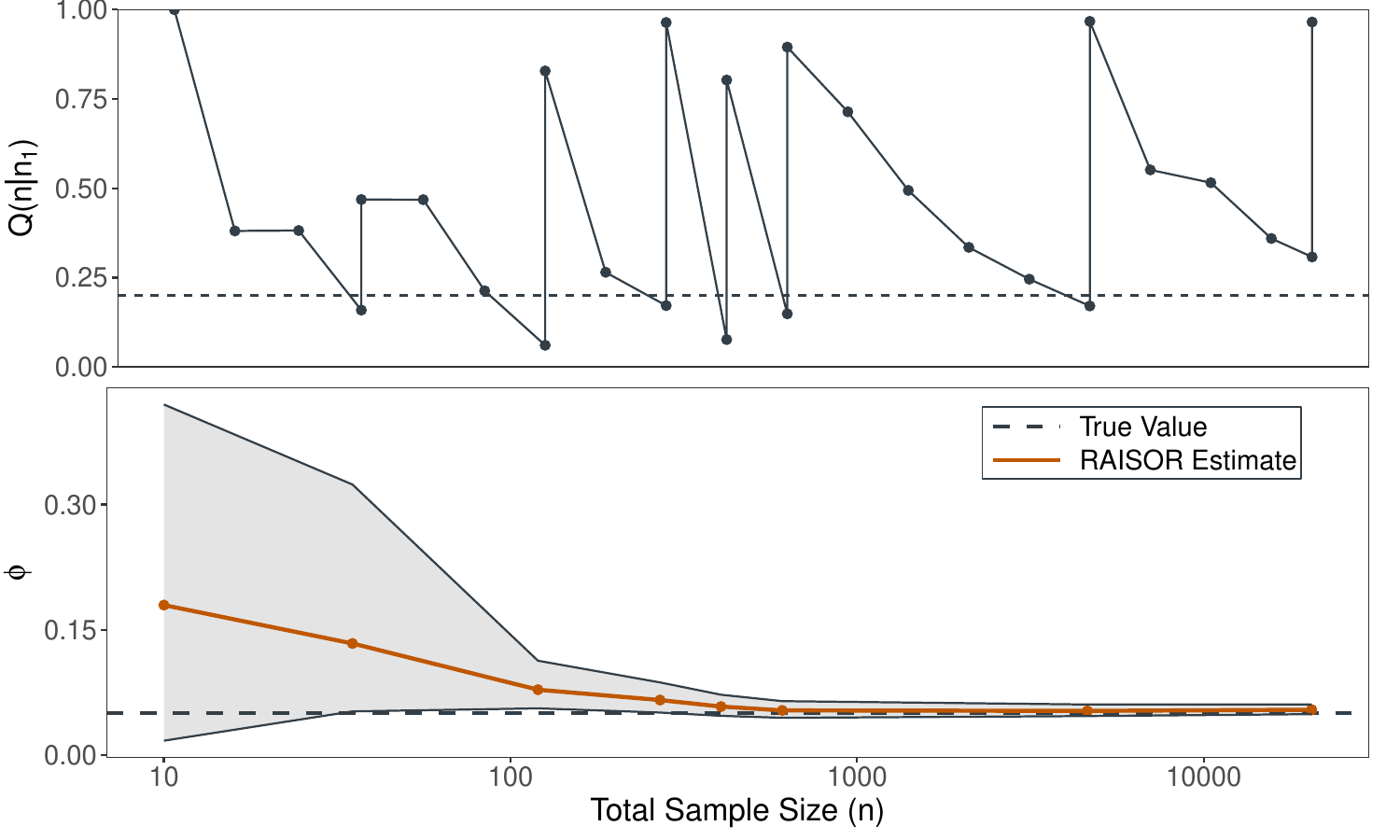}
    \caption{Top plot shows the quality trajectory obtained while fitting the GP regression model with $n = 20480$. The dashed line corresponds to $q = 0.2$. Bottom plot shows the evolution of posterior estimates of $\phi$ as the thick solid line, with the thin solid lines representing $95\%$ credible intervals and the dashed line corresponding to the true value $\phi = 0.05$.}
    \label{fig:RESS Evolution}
\end{figure}

To visualize how the RAISOR approach fits the model, considering the data set with $n = 20480$, Figure \ref{fig:RESS Evolution} shows on top the quality trajectory evaluated after each weight update or recalculation step. The dashed line represents the replenishment threshold $q = 0.2$, so the method always introduces replenishment after each weight update that reduces the quality below this line. Comparing it with Figure \ref{fig:RESS Exponential}, it is clear that the RESS does not reach one after each replenishment time, which occurs due to approximation error. The bottom plot of Figure \ref{fig:RESS Evolution} shows how the evolution of the estimate of the spatial range parameter $\phi$ as the fitting process progresses. The uncertainty associated with our estimate shrinks as the sample size increases, eventually approaching the truth.

\subsection{Application: Sea Surface Temperature Estimation}

\begin{figure}[tb!]
    \centering
    \includegraphics[width=0.5\textwidth]{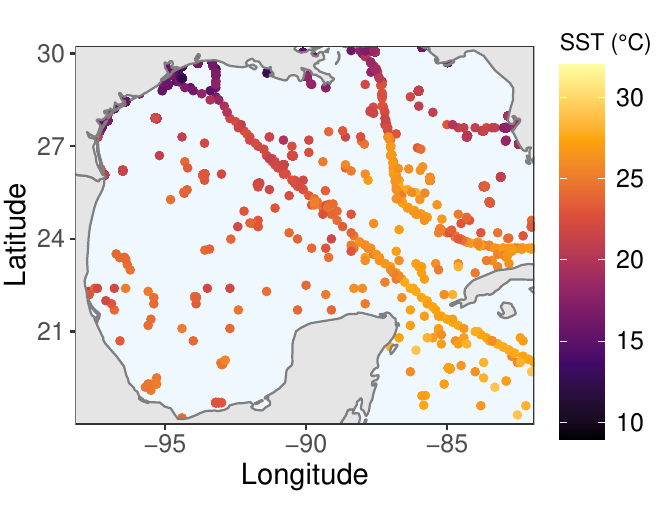}
    \caption{Sea surface temperature measurements (°C) in the Gulf of Mexico during January of 2021, obtained from the in-situ SST Quality Monitor system by the National Oceanic and Atmospheric Administration.}
    \label{fig:SST data}
\end{figure}

We consider sea surface temperature (SST) measurements in the Gulf of Mexico during January of 2021. The data were collected as a part of the in-situ SST Quality Monitor (iQuam) system developed by the National Oceanic and Atmospheric Administration (NOAA) \citep{xu2014situ}, compiling on-site measurements from multiple sources (e.g., ships, buoys, and Argo floats). These data can be used for quality control of SST obtained from other sources, such as satellite measurements, to make high resolution products or in ecological, environmental, and climatological applications. Figure \ref{fig:SST data} shows $n = 6014$ daytime observations, with $4951$ of those arising from ships, $63$ from Argo floats, and $1000$ from moored buoy measurements.

\begin{figure}[tb!]
    \centering
    \includegraphics[width=0.95\textwidth]{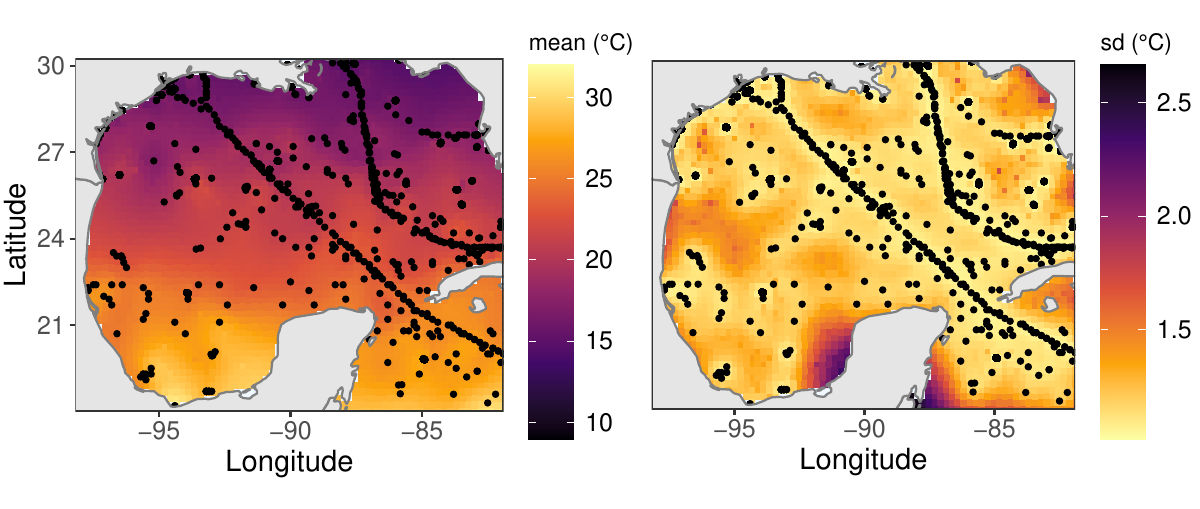}
    \caption{Estimated posterior predictive mean and marginal standard deviation of the SST at a grid of locations obtained with the proposed approach. Points represent the observed locations.}
    \label{fig:prediction}
\end{figure}

We fitted the Vecchia approximation of the GP regression model in (\ref{eq:GP data process}) to the SST data using $k = 18$ neighbors, considering both the MCMC and RAISOR approaches as described previously. We used the geodesic distance $\{D\}_{ij} = d_{\text{GEO}}(\bs{s}_{i}, \bs{s}_{j})$ in the covariance model in (\ref{eq:covariance}) and included $p = 3$ covariates in model, with the first corresponding to an intercept and the remaining two to the spacial coordinates. The total runtime was 2 hours 56 minutes for a single core MCMC and 46 minutes for the RAISOR approach with twenty-four cores, while it would require weeks to fit the exact geostatistical model. Figure \ref{fig:prediction} shows the posterior predictive mean and standard deviation of the SST at a grid of locations obtained using the proposed approach. These resulting maps can be used in downstream analyses that facilitate ecological and environmental research. For instance, \cite{liu2025rapid} studied the effect of sea temperature on the formation of Hurricane Ian, that caused extensive damage after reaching the state of Florida in 2022, and they established a causal link between anomalies in the sea temperature in the Gulf of Mexico and Ian's rapid intensification from a Category 3 to a Category 5 hurricane.

\section{Discussion}\label{sec:discussion}

In the context of distributed computing environments, RAISOR combines the strengths of RB and AIS methods. When compared to the importance PP-RB algorithm, it introduces replenishment to mitigate the sample degeneracy issue, and relative to AIS approaches, it targets a sequence of partial posteriors, bridging the gap between the initial proposal and the final target and thus circumventing the initialization problem. Our approach is conceptually similar to AAIS methods, but with the key caveat that, while AAIS typically involves evaluating the target distribution at every iteration (e.g., \citealp{neal2001annealed, liu2014adaptive, goshtasbpour2023adaptive, zhang2025annealed}), RAISOR only requires evaluation of its final target in the last iteration, significantly reducing the computational cost for large $n$. To build on previous works that have also considered the posterior recursive formulation to aid computation (e.g., \citealp{chopin2002sequential, hooten2021making, dai2022invitation}), RAISOR leverages the natural posterior contraction rate to introduce weight update and replenish steps at an optimal rate, significantly mitigating the effect of latency.

When considering the effect of the dimension $d$ of $\bs{\theta}$ on the asymptotic distribution of $Q(n|n_{1})$, Theorem \ref{thm:asymptotic distirbution} establishes that sample degeneration is expected to accelerate significantly as $d$ increases (see also \citealp{bengtsson2008curse, beskos2014stability, agapiou2017importance, chatterjee2018sample, dai2022invitation}). A similar problem also occurs when using non-parametric density estimators to approximate the current target distribution because the approximation error increases exponentially with $d$. Thus, RAISOR is less effective when used to sample from high-dimensional posterior distributions, but will yield appreciable gains in low-dimensional parameter models with $d \leq 10$. Another aspect worth considering is that RAISOR, as proposed, does not address posterior distributions with dimension increasing with the sample size, such as generalized linear mixed models, latent Markov random fields, and state-space models, although extensions could be proposed similar to \cite{hooten2021making}. These points raise the question of whether RAISOR could be combined with other approaches, such as MCMC or SMC, to generate methods better equipped to address these issues while still allowing for efficient parallelization with minimal communication cost.

Our results in Section \ref{sec:theory} are based on the Bernstein-von Mises theorem under model misspecification of \cite{kleijn2012bernstein}, which holds under the local asymptotic normality (LAN) conditions and encompasses classes of models beyond the i.i.d. case. Given that our proof of Theorem \ref{thm:asymptotic distirbution} relies on the independence assumption for the true data generating process and the assumed Bayesian model, our result cannot be immediately extended to this setting, although we conjecture that under suitable additional conditions a similar result should hold under LAN assumptions. This opens the question of how RAISOR is expected to behave when fitting models with strong dependence (e.g., the geostatistical model). We present in Section C of the Supplement Material informal arguments suggesting that our approach should still work reliably, which is in concordance with the empirical results presented in Section \ref{sec:applications}, but a formal treatment is the subject of ongoing research.

\if1\anon
{
    \section{Acknowledgments}\label{sec:acknowledgment}

  The authors thank Alex Barth, Berkeley Ho, Justin Van Ee, Michael Schwob, Myungsoo Yoo, Nikunj Goel, and Rachael Ren for helpful discussions and insights. Similarly, the authors thank the anonymous associated editor and reviewers for their comments and suggestions to improve this work.
} \fi

\section{Disclosure statement}\label{sec:disclosure}

The authors have no knowledge of any conflict of interests to declare.

\section{Data Availability Statement}\label{sec:data availability}

The SST data used in this project are available at the following URL:

https://www.star.nesdis.noaa.gov/socd/sst/iquam/.

\bibliography{bibliography.bib}

\appendix

\appendix

\section{Technical Proofs}

For organizational purposes, we split the proof of Theorem 3.1 into lemmas that are stated and proven in this Supplementary Material. The first lemma allows us to rewrite $Q(n | n_{1})$ as the ratio of two expectations with respect to a normal approximation of the posterior $[\bs{\theta} | \bs{y}_{1:n_{1}}]$ using the Bernstein-von Mises theorem under model misspecification of \cite{kleijn2012bernstein}.

\begin{lemma}\label{lm:normal swap}
    Let $Q(n | n_{1})$ be as defined in Section 2.3 and $n_{1} = \alpha_{n} n$, for all $n \geq 1$ and some sequence of constants $\{\alpha_{n}\}_{n}$ such that $\alpha_{n} \in \left\{ \frac{1}{n}, \dots, \frac{n}{n}\right\}$ and $\alpha_{n} \to \alpha \in (0,1)$. Then, under assumptions stated in Section 3.1 and the posterior concentration condition in equation (16), 
    \begin{equation}
        \left| Q(n|n_{1}) - Q_{\phi}(n|n_{1})  \right| \overset{\mathbb{P}_{*}}{ \to} 0,
    \end{equation}
    as $n$ goes to $+\infty$, where 
    \begin{equation}\label{eq:RESS normal}
        Q(n | n_{1}) = \frac{ \left\{ \int R_{(n_{1}+1):n}(\bs{\theta}, \hat{\bs{\theta}}_{n_{1}+1}^{n}) \ \phi(\bs{\theta} | \hat{\bs{\theta}}_{1}^{n_{1}}, n_{1}^{-1}\bs{V}_{\theta^{*}}^{-1}) \ d\bs{\theta} \right\}^{2}}{ \int R_{(n_{1}+1):n}^{2}(\bs{\theta}, \hat{\bs{\theta}}_{n_{1}+1}^{n}) \ \phi(\bs{\theta} | \hat{\bs{\theta}}_{1}^{n_{1}}, n_{1}^{-1}\bs{V}_{\theta^{*}}^{-1}) \ d\bs{\theta}},
    \end{equation}
    $R_{(n_{1}+1):n}(\bs{\theta}, \tilde{\bs{\theta}}) = \exp\left\{ \ell_{n_{1}+1}^{n}(\bs{\theta}) - \ell_{n_{1}+1}^{n}(\tilde{\bs{\theta}}) \right\}$ is the likelihood ratio, and $\phi(\cdot | \bs{\mu}, \bs{\Sigma})$ represents the density of a normal random vector with mean $\bs{\mu}$ and covariance matrix $\bs{\Sigma}$.
\end{lemma}

\begin{proof}
    To avoid unnecessary complications, note that $\alpha_{n} \to \alpha \in (0,1)$, which implies that $\alpha_{n} < 1$ for all large $n$, and so we assume without loss of generality that $\alpha_{n} < 1$ for all $n \geq 2$. Additionally, given that $\bs{y}_{1:n}$ are i.i.d., we have
    \begin{equation}
        \begin{aligned}
            \hat{\bs{\theta}}_{n_{1}+1}^{n} = \arg \max_{\bs{\theta} \in \Theta} \log [\bs{y}_{(n_{1}+1):n} | \bs{\theta}] \overset{d}{=} \arg \max_{\bs{\theta} \in \Theta} \log [\bs{y}_{1:(n-n_{1})} | \bs{\theta}] = \hat{\bs{\theta}}_{1}^{n-n_{1}},
        \end{aligned}
    \end{equation}
    therefore $\hat{\bs{\theta}}_{n_{1}+1}^{n}$ exists and lies in the interior of $\Theta$, $\mathbb{P}_{*}$-almost surely, for all $n \geq 2$.
    
    To make $Q(n|n_{1})$ and $Q_{\phi}(n | n_{1})$ comparable, we begin by rewriting 
    \begin{equation}
        \begin{aligned}
            Q(n|n_{1}) 
            &= \left\{ 1 + D_{\chi^{2}}( [\bs{\theta} | \bs{y}_{1:n}] \ \| \ [\bs{\theta} | \bs{y}_{1:n_{1}}] ) \right\}^{-1} \\
            &= \left\{ \int \frac{[\bs{\theta} | \bs{y}_{1:n}]^{2}}{[\bs{\theta} | \bs{y}_{1:n_{1}}]} \ d\bs{\theta} \right\}^{-1} 
            = \frac{[\bs{y}_{1:n}]^{2}}{[\bs{y}_{1:n_{1}}]} \left\{ \int \frac{[\bs{y}_{1:n} | \bs{\theta}]^{2}}{[\bs{y}_{1:n_{1}} | \bs{\theta}]} [\bs{\theta}] \ d\bs{\theta} \right\}^{-1} \\
            &= \frac{ \left\{ \int [\bs{y}_{1:n} | \bs{\theta}] [\bs{\theta}] \ d\bs{\theta} \right\}^{2}}{ [\bs{y}_{1:n_{1}}] \int \frac{[\bs{y}_{1:n} | \bs{\theta}]^{2}}{[\bs{y}_{1:n_{1}} | \bs{\theta}]} [\bs{\theta}] \ d\bs{\theta}}
            = \frac{ \left\{ \int [\bs{y}_{(n_{1}+1):n} | \bs{y}_{1:n_{1}}, \bs{\theta}] [\bs{\theta} | \bs{y}_{1:n_{1}}] \ d\bs{\theta} \right\}^{2}}{ \int [\bs{y}_{(n_{1}+1):n} | \bs{y}_{1:n_{1}}, \bs{\theta}]^{2} [\bs{\theta} | \bs{y}_{1:n_{1}}] \ d\bs{\theta}} 
             \\
             &\overset{\text{ind}}{=} \frac{ \left\{ \int [\bs{y}_{(n_{1}+1):n} | \bs{\theta}] [\bs{\theta} | \bs{y}_{1:n_{1}}] \ d\bs{\theta} \right\}^{2}}{ \int [\bs{y}_{(n_{1}+1):n} | \bs{\theta}]^{2} [\bs{\theta} | \bs{y}_{1:n_{1}}] \ d\bs{\theta}} 
             = \frac{ \left\{ \int R_{(n_{1}+1):n}(\bs{\theta}, \hat{\bs{\theta}}_{n_{1}+1}^{n}) \ [\bs{\theta} | \bs{y}_{1:n_{1}}] \ d\bs{\theta} \right\}^{2}}{ \int R_{(n_{1}+1):n}^{2}(\bs{\theta}, \hat{\bs{\theta}}_{n_{1}+1}^{n}) \ [\bs{\theta} | \bs{y}_{1:n_{1}}] \ d\bs{\theta}},
        \end{aligned}
    \end{equation}
    for all $n \geq 2$. Note that the above expression is only well-defined because $\alpha_{n} < 1$, which implies $n_{1} < n$. To streamline the calculations, we introduce the notation
    \begin{equation}
        \begin{aligned}
            a_{n,\gamma} &:= \int R^{\gamma}_{(n_{1}+1):n}(\bs{\theta}, \hat{\bs{\theta}}_{n_{1}+1}^{n}) \ [\bs{\theta} | \bs{y}_{1:n_{1}}] \ d\bs{\theta}, \\
            b_{n,\gamma} &:= \int R^{\gamma}_{(n_{1}+1):n}(\bs{\theta}, \hat{\bs{\theta}}_{n_{1}+1}^{n}) \ \phi(\bs{\theta} | \hat{\bs{\theta}}_{1}^{n_{1}}, n_{1}^{-1}\bs{V}_{\theta^{*}}^{-1}) \ d\bs{\theta},
        \end{aligned}
    \end{equation}
    so we have
    \begin{equation}\label{eq:bound RESS}
        \begin{aligned}
            &\left| Q(n|n_{1}) - Q_{\phi}(n|n_{1})  \right| 
            = \left| \frac{a_{n,1}^{2}}{a_{n,2}} - \frac{b_{n,1}^{2}}{b_{n,2}} \right| 
            = \left| \frac{b_{n,2} a_{n,1}^{2} - a_{n,2} b_{n,1}^{2}}{a_{n,2} b_{n,2}} \right|  \\
            &= \left| \frac{\left\{ b_{n,2} - a_{n,2} \right\} a_{n,1}^{2} - a_{n,2} \left\{ b_{n,1}^{2} - a_{n,1}^{2} \right\} }{a_{n,2} b_{n,2}} \right| 
            \leq \frac{\left| b_{n,2} - a_{n,2} \right| a_{n,1}^{2} }{a_{n,2} b_{n,2}} + \frac{ a_{n,2} \left| b_{n,1}^{2} - a_{n,1}^{2} \right| }{a_{n,2} b_{n,2}} \\
            &= \frac{\left| b_{n,2} - a_{n,2} \right| }{ b_{n,2}} \cdot \frac{a_{n,1}^{2}}{a_{n,2}} + \frac{ (b_{n,1} + a_{n,1}) \left| b_{n,1} - a_{n,1} \right| }{ b_{n,2}}
            \leq \frac{\left| b_{n,2} - a_{n,2} \right| + 2\left| b_{n,1} - a_{n,1} \right| }{ b_{n,2}},
        \end{aligned}
    \end{equation}
    where on the last inequality we used that $a_{n,1}^{2} \leq a_{n,2}$, by Jensen's inequality, and $a_{n,1} + b_{n,1} \leq 2$, because the likelihood ratio is bounded by 1. Note that
    \begin{equation}\label{eq:bound l1}
        \begin{aligned}
            \left| a_{n,\gamma} - b_{n,\gamma} \right|
            &\leq \int \underbrace{R^{\gamma}_{(n_{1}+1):n}(\bs{\theta}, \hat{\bs{\theta}}_{n_{1}+1}^{n})}_{\leq 1} \left|  [\bs{\theta} | \bs{y}_{1:n_{1}}] - \phi(\bs{\theta} | \hat{\bs{\theta}}_{1}^{n_{1}}, n_{1}^{-1}\bs{V}_{\theta^{*}}^{-1}) \right| \ d\bs{\theta} \\
            &\leq \int \left|  [\bs{\theta} | \bs{y}_{1:n_{1}}] - \phi(\bs{\theta} | \hat{\bs{\theta}}_{1}^{n_{1}}, n_{1}^{-1}\bs{V}_{\theta^{*}}^{-1}) \right| \ d\bs{\theta},
        \end{aligned}
    \end{equation}
    for $\gamma > 0$, so combining (\ref{eq:bound RESS}) and (\ref{eq:bound l1}) yields
    \begin{equation}
        \begin{aligned}
            \left| Q(n|n_{1}) - Q_{\phi}(n|n_{1})  \right| 
            \leq 3\frac{\int \left|  [\bs{\theta} | \bs{y}_{1:n_{1}}] - \phi(\bs{\theta} | \hat{\bs{\theta}}_{1}^{n_{1}}, n_{1}^{-1}\bs{V}_{\theta^{*}}^{-1}) \right| \ d\bs{\theta}}{\int R^{2}_{(n_{1}+1):n}(\bs{\theta}, \hat{\bs{\theta}}_{n_{1}+1}^{n}) \ \phi(\bs{\theta} | \hat{\bs{\theta}}_{1}^{n_{1}}, n_{1}^{-1}\bs{V}_{\theta^{*}}^{-1}) \ d\bs{\theta}}.
        \end{aligned}
    \end{equation}
    Because $\alpha_{n} \to \alpha > 0$ as $n$ goes to $+\infty$, we also have $n_{1} \to +\infty$, so, by Lemma \ref{lm:integral convergence}, the denominator converges in distribution to a positive random variable and, by Theorem 2.1 of \cite{kleijn2012bernstein}, the numerator converges in $\mathbb{P}_{*}$-probability to 0. Combining those with the continuous mapping theorem yields the desired result.
\end{proof}

In the next lemma, we establish the limiting distribution of $Q_{\phi}(n|n_{1})$, defined in equation (\ref{eq:RESS normal}), which forms the foundation of the proof of Theorem 3.1.

\begin{lemma}\label{lm:integral convergence}
    Let $n_{1} = \alpha_{n} n$ for all $n \geq 1$ and some sequence of constants $\{ \alpha_{n} \}_{n}$ such that $\alpha_{n} \in \left\{\frac{1}{n}, \dots, \frac{n}{n} \right\}$ and $\alpha_{n} \to \alpha \in (0,1)$. Additionally, let $f:(0,1)^{k} \to \mathbb{R}^{m}$ be a continuous function $\lambda$-almost everywhere in $(0,1)^{k}$, where $\lambda$ represents the Lebesgue measure, and
    \begin{equation}
        \mathbb{I}_{n,\gamma}(\bs{\theta}^{*}) = \int R^{\gamma}_{(n_{1}+1):n}(\bs{\theta}, \hat{\bs{\theta}}_{n_{1}+1}^{n}) \ \phi(\bs{\theta} | \hat{\bs{\theta}}_{1}^{n_{1}}, n_{1}^{-1}\bs{V}_{\theta^{*}}^{-1}) \ d\bs{\theta}
    \end{equation}
    for all $\gamma > 0$. Then, under the assumptions stated in Section 3.1,
    \begin{equation}
        f\left( \vphantom{\frac{1}{1}} \mathbb{I}_{n,\gamma_{1}}(\bs{\theta}^{*}), \dots, \mathbb{I}_{n,\gamma_{k}}(\bs{\theta}^{*}) \right) \overset{d}{\to} f\left( \vphantom{\frac{1}{1}} \mathbb{I}_{\theta^{*},\gamma_{1}}(\bs{z}), \dots,  \mathbb{I}_{\theta^{*},\gamma_{k}}(\bs{z})\right),
    \end{equation}
    where $\bs{z} \sim \textnormal{Normal}_{d}(\bs{0}, \bs{I}_{d})$, $\gamma_{1}, \dots, \gamma_{k}$ are positive constants, and
    \begin{equation}
        \begin{aligned}
            \log \mathbb{I}_{\theta^{*},\gamma}(\bs{z}) := \tfrac{d}{2} \log\left( \tfrac{\alpha}{\gamma + (1-\gamma)\alpha} \right) - \tfrac{\gamma}{2(\gamma + (1-\gamma) \alpha)} \bs{z}' \bs{M}_{\theta^{*}} \bs{z}.
        \end{aligned}
    \end{equation}
    for all $\gamma > 0$ and $\bs{z} \in \mathbb{R}^{d}$.
\end{lemma}

\begin{proof}
    To avoid excessive repetition, all limits are taken as $n$ goes to $+\infty$ unless otherwise stated. Additionally, for the same reasons considered in the proof of Lemma \ref{lm:normal swap}, because $\alpha_{n} \to \alpha \in (0,1)$, we can assume without loss of generality that $\alpha_{n} < 1$ for all $n \geq 2$, and we have that $\hat{\bs{\theta}}_{n_{1}+1}^{n}$ exists and lies in the interior of $\Theta$, $\mathbb{P}_{*}$-almost surely, ensuring that $\mathbb{I}_{n,\gamma}(\bs{\theta}^{*})$ is well-defined.
    
    To show that $\mathbb{I}_{n,\gamma}(\bs{\theta}^{*})$ converges in distribution, our strategy is to first obtain an asymptotically tight bound of the likelihood ratio when restricting the integral over fixed compact sets. Then, following the proof of Theorem 2.1 in \cite{kleijn2012bernstein}, we let the compact sets grow in a controlled manner to obtain the desired result. With that in mind, we separate $\mathbb{I}_{n,\gamma}(\bs{\theta}^{*})$ into two parts, obtaining
    \begin{equation}\label{eq:decomposition}
        \begin{aligned}
            \int R^{\gamma}_{(n_{1}+1):n}(\bs{\theta}, \hat{\bs{\theta}}_{n_{1}+1}^{n}) \ \phi(\bs{\theta} | \hat{\bs{\theta}}_{1}^{n_{1}}, n_{1}^{-1}\bs{V}_{\theta^{*}}^{-1}) \ d\bs{\theta}
            &= \mathbb{I}_{n,\gamma}(\bs{\theta}^{*}, r_{n-n_{1}}) + \mathbb{I}^{c}_{n,\gamma}(\bs{\theta}^{*}, r_{n-n_{1}})
        \end{aligned}
    \end{equation}
    where
    \begin{equation}
        \begin{aligned}
            \mathbb{I}_{n,\gamma}(\bs{\theta}^{*}, r_{n-n_{1}}) &= \int_{B(\bs{\theta}^{*},r_{n-n_{1}})} R^{\gamma}_{(n_{1}+1):n}(\bs{\theta}, \hat{\bs{\theta}}_{n_{1}+1}^{n}) \ \phi(\bs{\theta} | \hat{\bs{\theta}}_{1}^{n_{1}}, n_{1}^{-1}\bs{V}_{\theta^{*}}^{-1}) \ d\bs{\theta}, \\
            \mathbb{I}^{c}_{n,\gamma}(\bs{\theta}^{*}, r_{n-n_{1}}) &= \int_{B^{c}(\bs{\theta}^{*},r_{n-n_{1}})} R^{\gamma}_{(n_{1}+1):n}(\bs{\theta}, \hat{\bs{\theta}}_{n_{1}+1}^{n}) \ \phi(\bs{\theta} | \hat{\bs{\theta}}_{1}^{n_{1}}, n_{1}^{-1}\bs{V}_{\theta^{*}}^{-1}) \ d\bs{\theta},
        \end{aligned}
    \end{equation}
    with $B(\bs{\theta}^{*},r)$ an $L_{2}$ neighborhood of radius $r$ around $\bs{\theta}^{*}$ (i.e., $B(\bs{\theta}^{*},r) = \{ \bs{\theta} \in \Theta : \| \bs{\theta} - \bs{\theta}^{*} \|_{2} \leq r \}$), and $r_{x} = r(x)$ can be any continuous function in $\mathbb{R}_{+}$ satisfying 
    \begin{equation}\label{eq:radii condition 1}
        \begin{aligned}
            \lim_{x \to +\infty} r(x) &= 0 && \text{and} & \lim_{x \to +\infty} x r^{2}(x) = +\infty.
        \end{aligned}
    \end{equation}
    Considering the two terms of the decomposition in (\ref{eq:decomposition}), we can bound the second one and use a change of variables to write
    \begin{equation}\label{eq:term 1 bound}
        \begin{aligned}
            0 \leq \mathbb{I}^{c}_{n,\gamma}(\bs{\theta}^{*}, r_{n-n_{1}})
            &= \int_{B^{c}(\bs{\theta}^{*}, r_{n-n_{1}})} \underbrace{R^{\gamma}_{(n_{1}+1):n}(\bs{\theta}, \hat{\bs{\theta}}_{n_{1}+1}^{n})}_{\leq 1} \ \phi(\bs{\theta} | \hat{\bs{\theta}}_{1}^{n_{1}}, n_{1}^{-1}\bs{V}_{\theta^{*}}^{-1}) \ d\bs{\theta} \\
            &\leq \int_{B^{c}(\bs{\theta}^{*},r_{n-n_{1}})} \phi(\bs{\theta} | \hat{\bs{\theta}}_{1}^{n_{1}}, n_{1}^{-1}\bs{V}_{\theta^{*}}^{-1}) \ d\bs{\theta} \\
            &= \int_{B^{c}(\bs{\theta}^{*},1)} \phi(\bs{\theta} | r_{n-n_{1}}^{-1}\{ \hat{\bs{\theta}}_{1}^{n_{1}} - \bs{\theta}^{*} \} +\bs{\theta^{*}}, \{n_{1} r_{n-n_{1}}^{2}\}^{-1}\bs{V}_{\theta^{*}}^{-1}) \ d\bs{\theta}.
        \end{aligned}
    \end{equation}
    Using that $n_{1} \to +\infty$ because $\alpha_{n} \to \alpha \in (0,1)$, and noting that the MLE is consistent by assumption and asymptotically normal by Lemma 2.2 of \cite{kleijn2012bernstein}, we get
    \begin{equation}\label{eq:MLE normality}
        \sqrt{n_{1}} (\hat{\bs{\theta}}_{1}^{n_{1}} - \bs{\theta}^{*}) \overset{d}{\to} \text{Normal}_{d}(\bs{0}, \bs{\Sigma}_{\theta^{*}})
    \end{equation}
    where $\bs{\Sigma}_{\theta^{*}} = \bs{V}_{\theta^{*}}^{-1} \bs{W}_{\theta^{*}} \bs{V}_{\theta^{*}}^{-1}$ is the sandwich matrix. Furthermore, with $n\{ 1-\alpha_{n}\} \to +\infty$, $\frac{\alpha_{n}}{1-\alpha_{n}} \to \frac{\alpha}{1-\alpha} \in (0, +\infty)$, and $r$ continuous we have
    \begin{equation}
        \begin{aligned}
            n_{1}r^{2}_{n-n_{1}} 
            &= \alpha_{n} n  r^{2}\left( n \left\{ 1 - \alpha_{n} \right\} \right)
            = \frac{\alpha_{n}}{1-\alpha_{n}} n\{1-\alpha_{n}\} r^{2}(n\{1-\alpha_{n}\}) \to +\infty,
        \end{aligned}
    \end{equation}
    which we can combine with (\ref{eq:MLE normality}) to obtain
    \begin{equation}
        r_{n-n_{1}}^{-1} (\hat{\bs{\theta}}_{1}^{n_{1}} - \bs{\theta}^{*}) +\bs{\theta}^{*} = \underbrace{ \{ n_{1}r_{n-n_{1}}^{2} \}^{-\frac{1}{2}} }_{\to 0} \underbrace{ \sqrt{n_{1}} ( \hat{\bs{\theta}}_{1}^{n_{1}} - \bs{\theta}^{*} )}_{= O_{P}(1)} +\bs{\theta}^{*} \overset{\mathbb{P}_{*}}{\to} \bs{\theta}^{*}
    \end{equation}
    by Slutsky's theorem. Note that the normal density in the upper bound of (\ref{eq:term 1 bound}) converges to a Dirac delta at $\bs{\theta}^{*}$, while the region of integration remains constant and bounded away from $\bs{\theta}^{*}$. Hence, the integral converges to $0$ and we have
    \begin{equation}\label{eq:term 1 convergence}
        \mathbb{I}^{c}_{n,\gamma}(\bs{\theta}^{*}, r_{n-n_{1}}) =\int_{B^{c}(\bs{\theta}^{*}, r_{n-n_{1}})} R^{\gamma}_{(n_{1}+1):n}(\bs{\theta}, \hat{\bs{\theta}}_{n_{1}+1}^{n}) \ \phi(\bs{\theta} | \hat{\bs{\theta}}_{1}^{n_{1}}, n_{1}^{-1}\bs{V}_{\theta^{*}}^{-1}) \ d\bs{\theta} 
        \overset{\mathbb{P}_{*}}{\to} 0.
    \end{equation}
    
    Next, to address the first term of the decomposition in equation (\ref{eq:decomposition}), we rewrite $r(x) = x^{-\frac{1}{2}} r^{*}(x)$ for some positive function $r^{*}(x) = r_{x}^{*}$ in $\mathbb{R}_{+}$, denote the convergence rate by $\delta_{x} = x^{-\frac{1}{2}}$, and use a change of variables to write
    \begin{equation}\label{eq:term 2 split}
        \begin{aligned}
            &\mathbb{I}_{n,\gamma}(\bs{\theta}^{*}, r_{n-n_{1}})
            =\int_{B(\bs{\theta}^{*},r_{n-n_{1}})} R^{\gamma}_{(n_{1}+1):n}(\bs{\theta}, \hat{\bs{\theta}}_{n_{1}+1}^{n}) \ \phi(\bs{\theta} | \hat{\bs{\theta}}_{1}^{n_{1}}, n_{1}^{-1}\bs{V}_{\theta^{*}}^{-1}) \ d\bs{\theta} \\
            &= \int_{B(\bs{\theta}^{*}, r_{n-n_{1}})} R^{\gamma}_{(n_{1}+1):n}(\bs{\theta}, \bs{\theta}^{*}) \ \phi(\bs{\theta} | \hat{\bs{\theta}}_{1}^{n_{1}}, n_{1}^{-1}\bs{V}_{\theta^{*}}^{-1}) \ d\bs{\theta} \ \left\{ R^{\gamma}_{(n_{1}+1):n}(\bs{\theta}^{*}, \hat{\bs{\theta}}_{n_{1}+1}^{n}) \right\} \\
            &= \underbrace{ \int_{B(\bs{0},r^{*}_{n-n_{1}})} R^{\gamma}_{(n_{1}+1):n}\left( \bs{\theta}^{*} + \delta_{n-n_{1}}\bs{\theta} , \bs{\theta}^{*} \right) \phi\left( \bs{\theta} \left| \delta_{n-n_{1}}^{-1}(\hat{\bs{\theta}}_{1}^{n_{1}} - \bs{\theta}^{*}), \tfrac{n-n_{1}}{n_{1}} \bs{V}_{\theta^{*}}^{-1} \right. \right) \ d\bs{\theta} }_{:= \mathbb{J}_{n,\gamma}(\bs{\theta}^{*}, r^{*}_{n-n_{1}})} \\ 
            &\quad \times  R^{\gamma}_{(n_{1}+1):n}(\bs{\theta}^{*}, \hat{\bs{\theta}}_{n_{1}+1}^{n}) 
            = \mathbb{J}_{n,\gamma}(\bs{\theta}^{*}, r^{*}_{n-n_{1}}) \ R^{\gamma}_{(n_{1}+1):n}(\bs{\theta}^{*}, \hat{\bs{\theta}}_{n_{1}+1}^{n}).
        \end{aligned}
    \end{equation}
    Before proceeding, we define the auxiliary quantities
    \begin{equation}
        \begin{aligned}
            a_{n_{1}+1}^{n}(\bs{\theta}^{*}, \bs{\theta}) &:= \log R_{(n_{1}+1):n}\left( \bs{\theta}^{*} + \delta_{n-n_{1}} \bs{\theta} , \bs{\theta}^{*} \right) - \bs{\theta}' \delta_{n-n_{1}}\dot{\ell}_{n_{1}+1}^{n}(\bs{\theta}^{*}) + \tfrac{1}{2}\bs{\theta}' \bs{V}_{\theta^{*}} \bs{\theta} \\
            a_{n_{1}+1}^{n}(\bs{\theta}^{*}, r^{*}) &:= \sup_{\bs{\theta} \in B(\bs{0}, r^{*})} \left| a_{n_{1}+1}^{n}(\bs{\theta}^{*}, \bs{\theta}) \right|,
        \end{aligned}
    \end{equation}
    so, for all $\bs{\theta} \in B(\bs{0}, r^{*}_{n-n_{1}})$, we can write
    \begin{equation}\label{eq:ratio upper bound}
        \begin{aligned}
            \log R_{(n_{1}+1):n}\left( \bs{\theta}^{*} + \delta_{n-n_{1}}\bs{\theta} , \bs{\theta}^{*} \right)
            &= a_{n_{1}+1}^{n}(\bs{\theta}^{*}, \bs{\theta}) + \bs{\theta}'\delta_{n-n_{1}}\dot{\ell}_{n_{1}+1}^{n}(\bs{\theta}^{*}) - \tfrac{1}{2}\bs{\theta}' \bs{V}_{\theta^{*}} \bs{\theta} \\
            &\leq \left| a_{n_{1}+1}^{n}(\bs{\theta}^{*}, \bs{\theta}) \right| + \bs{\theta}'\delta_{n-n_{1}}\dot{\ell}_{n_{1}+1}^{n}(\bs{\theta}^{*}) - \tfrac{1}{2}\bs{\theta}' \bs{V}_{\theta^{*}} \bs{\theta} \\
            &\leq a_{n_{1}+1}^{n}(\bs{\theta}^{*}, r^{*}_{n-n_{1}}) + \bs{\theta}'\delta_{n-n_{1}}\dot{\ell}_{n_{1}+1}^{n}(\bs{\theta}^{*}) - \tfrac{1}{2}\bs{\theta}' \bs{V}_{\theta^{*}} \bs{\theta},
        \end{aligned}
    \end{equation}
    and analogously, we have
    \begin{equation}\label{eq:ratio lower bound}
        \log R_{(n_{1}+1):n}\left( \bs{\theta}^{*} + \delta_{n-n_{1}}\bs{\theta} , \bs{\theta}^{*} \right) \geq -a_{n_{1}+1}^{n}(\bs{\theta}^{*}, r^{*}_{n-n_{1}}) + \bs{\theta}'\delta_{n-n_{1}}\dot{\ell}_{n_{1}+1}^{n}(\bs{\theta}^{*}) - \tfrac{1}{2}\bs{\theta}' \bs{V}_{\theta^{*}} \bs{\theta}.
    \end{equation}
    In what follows, we use inequalities (\ref{eq:ratio upper bound}) and (\ref{eq:ratio lower bound}), and Lemma 2.1 of \cite{kleijn2012bernstein} to find asymptotically tight upper and lower bounds of $\mathbb{J}_{n,\gamma}(\bs{\theta}^{*}, r^{*}_{n-n_{1}})$. Starting with the upper bound, we write
    \begin{equation}\label{eq:integral upper bound}
        \begin{aligned}
            &\mathbb{J}_{n,\gamma}(\bs{\theta}^{*}, r^{*}_{n-n_{1}})
            \leq \exp\left\{ \gamma a_{n_{1}+1}^{n}(\bs{\theta}^{*}, r^{*}_{n-n_{1}})\right\} \\
            &\quad \times \int_{B(\bs{0},r^{*}_{n-n_{1}})} e^{ \gamma\bs{\theta}'\delta_{n-n_{1}}\dot{\ell}_{n_{1}+1}^{n}(\bs{\theta}^{*}) - \tfrac{1}{2}\bs{\theta}'\gamma \bs{V}_{\theta^{*}} \bs{\theta} }  \phi\left( \bs{\theta} \left| \delta_{n-n_{1}}^{-1}(\hat{\bs{\theta}}_{1}^{n_{1}} - \bs{\theta}^{*}), \tfrac{n-n_{1}}{n_{1}} \bs{V}_{\theta^{*}}^{-1} \right. \right) \ d\bs{\theta} \\
            &= \left(\tfrac{n_{1}}{\gamma n + (1-\gamma) n_{1}}\right)^{\frac{d}{2}} e^{b_{n, \gamma}(\bs{\theta}^{*}) + \gamma a_{n_{1}+1}^{n}(\bs{\theta}^{*}, r^{*}_{n-n_{1}})} \int_{B(\bs{0},r^{*}_{n-n_{1}})} \phi\left( \bs{\theta} \left| \bs{\mu}_{n,\gamma}(\bs{\theta}^{*}), \bs{\Omega}^{-1}_{n,\gamma} (\bs{\theta}^{*}) \right. \right) \ d\bs{\theta},
        \end{aligned}
    \end{equation}
    where the quantities
    \begin{equation}
        \begin{aligned}
            \bs{\Omega}_{n,\gamma} (\bs{\theta}^{*}) &= \tfrac{\gamma n + (1-\gamma)n_{1}}{n-n_{1}}\bs{V}_{\theta^{*}}, \\ 
            \bs{\mu}_{n,\gamma}(\bs{\theta}^{*}) &= \bs{\Omega}^{-1}_{n,\gamma} (\bs{\theta}^{*}) \left\{ \gamma \delta_{n-n_{1}} \dot{\ell}_{n_{1}+1}^{n}(\bs{\theta}^{*}) + \tfrac{n_{1}}{(n-n_{1})^{\frac{1}{2}}} \bs{V}_{\theta^{*}} (\hat{\bs{\theta}}_{1}^{n_{1}} - \bs{\theta}^{*}) \right\},
        \end{aligned}
    \end{equation}
    were identified by the usual quadratic form for combining Gaussian kernels, and the notation
    \begin{equation}
        \begin{aligned}
            b_{n, \gamma}(\bs{\theta}^{*}) 
            &= -\tfrac{1}{2}\left\{ (\hat{\bs{\theta}}_{1}^{n_{1}} - \bs{\theta}^{*})' n_{1}
             \bs{V}_{\theta^{*}}(\hat{\bs{\theta}}_{1}^{n_{1}} -  \bs{\theta}^{*}) - \bs{\mu}'_{n,\gamma}(\bs{\theta}^{*}) \bs{\Omega}_{n,\gamma} (\bs{\theta}^{*}) \bs{\mu}_{n,\gamma}(\bs{\theta}^{*}) \right\},
        \end{aligned}
    \end{equation}
    was introduced to simplify the expressions that follow. Analogously, the lower bound is of the form
    \begin{multline}
        \mathbb{J}_{n,\gamma}(\bs{\theta}^{*}, r^{*}_{n-n_{1}})
        \geq \left(\tfrac{n_{1}}{\gamma n + (1-\gamma) n_{1}}\right)^{\frac{d}{2}} \exp\left\{ b_{n, \gamma}(\bs{\theta}^{*}) - \gamma a_{n_{1}+1}^{n}(\bs{\theta}^{*}, r^{*}_{n-n_{1}}) \right\} \\ 
        \int_{B(\bs{0},r^{*}_{n-n_{1}})} \phi\left( \bs{\theta} \left| \bs{\mu}_{n,\gamma}(\bs{\theta}^{*}), \bs{\Omega}^{-1}_{n,\gamma} (\bs{\theta}^{*}) \right. \right) \ d\bs{\theta},
    \end{multline}
    which differs from the upper bound in equation (\ref{eq:integral upper bound}) only by the presence of a minus sign in front of $\gamma a_{n_{1}+1}^{n}(\bs{\theta}^{*}, r^{*}_{n-n_{1}})$. Leveraging this last fact allows us to write
    \begin{equation}\label{eq:bound term 2}
        \begin{aligned}
            \left| \log\left\{ \frac{ \mathbb{J}_{n,\gamma}(\bs{\theta}^{*}, r^{*}_{n-n_{1}}) e^{-b_{n,\gamma}(\bs{\theta}^{*})} \left( \tfrac{n_{1}}{\gamma n + (1-\gamma)n_{1}} \right)^{-\frac{d}{2}} }{ \int_{B(\bs{0}, r^{*}_{n-n_{1}})} \phi \left(\bs{\theta} | \bs{\mu}_{n,\gamma}(\bs{\theta}^{*}), \bs{\Omega}^{-1}_{n, \gamma}(\bs{\theta}^{*}) \right) d\bs{\theta} } \right\} \right| \leq \gamma a_{n_{1}+1}^{n}(\bs{\theta}^{*}, r^{*}_{n-n_{1}}),
        \end{aligned}
    \end{equation}
    which we can use along with the triangle inequality to obtain the inequality
    \begin{multline}\label{eq:bound term 2 final}
        \left| \log \mathbb{J}_{n,\gamma}(\bs{\theta}^{*}, r^{*}_{n-n_{1}}) - \log \mathbb{J}_{n,\gamma}(\bs{\theta}^{*}) \right| \\
        \leq \underbrace{ \gamma a_{n_{1}+1}^{n}(\bs{\theta}^{*}, r^{*}_{n-n_{1}})}_{(\text{I})} + \underbrace{ \left| \log \int_{B(\bs{0}, r^{*}_{n-n_{1}})} \phi \left(\bs{\theta} | \bs{\mu}_{n,\gamma}(\bs{\theta}^{*}), \bs{\Omega}^{-1}_{n, \gamma}(\bs{\theta}^{*}) \right) \ d\bs{\theta} \right|}_{(\text{II})},
    \end{multline}
    for all $r^{*}_{n-n_{1}} > 0$, where 
    \begin{equation}
        \mathbb{J}_{n,\gamma}(\bs{\theta}^{*}) = \exp\left\{ b_{n,\gamma}(\bs{\theta}^{*}) + \tfrac{d}{2} \log\left( \tfrac{n_{1}}{\gamma n + (1-\gamma)n_{1}} \right) \right\}.
    \end{equation} 
    Therefore, establishing the limiting distribution of $\mathbb{J}_{n,\gamma}(\bs{\theta}^{*}, r^{*}_{n-n_{1}})$ becomes easier if we show that the upper bound above converges to $0$ in $\mathbb{P}_{*}$-probability for some choice of function $r^{*}(x) = r^{*}_{x}$. Note that any choice $r^{*}(x)$ such that $r(x) = \delta(x) r^{*}(x)$ is continuous and still satisfies (\ref{eq:radii condition 1}), or equivalently, such that
    \begin{equation}\label{eq:radii condition 2}
        \begin{aligned}
            \lim_{x \to +\infty} r_{x}^{*} &= +\infty, & 
            \lim_{x \to +\infty} x^{-\frac{1}{2}}r_{x}^{*} &= 0,
        \end{aligned}
    \end{equation}
    and $r^{*}$ is continuous, guarantees that (\ref{eq:term 1 convergence}) remains valid. Starting with term (I) of equation (\ref{eq:bound term 2 final}), due to the i.i.d. assumption we have
    \begin{equation}
        \begin{aligned}
            a_{n+1}^{n}(\bs{\theta}^{*}, \bs{\theta}) 
            &= \log R_{(n_{1}+1):n} (\bs{\theta}^{*} + \delta_{n-n_{1}} \bs{\theta}, \bs{\theta}^{*}) - \bs{\theta}'\delta_{n-n_{1}} \dot\ell_{n_{1}+1}^{n}(\bs{\theta}^{*}) + \tfrac{1}{2} \bs{\theta}'\bs{V}_{\theta^{*}} \bs{\theta} \\
            &\overset{d}{=} \log R_{1:(n-n_{1})} (\bs{\theta}^{*} + \delta_{n-n_{1}} \bs{\theta}, \bs{\theta}^{*}) - \bs{\theta}'\delta_{n-n_{1}} \dot\ell_{1}^{n-n_{1}}(\bs{\theta}^{*}) + \tfrac{1}{2} \bs{\theta}'\bs{V}_{\theta^{*}} \bs{\theta} \\
            &= a_{1}^{n-n_{1}}(\bs{\theta}^{*}, \bs{\theta}),
        \end{aligned}
    \end{equation}
    therefore, by Lemma 2.1 of \cite{kleijn2012bernstein} and because $n-n_{1} = (1-\alpha_{n})n \to +\infty$,
    \begin{equation}
        \begin{aligned}
            a_{n_{1}+1}^{n}(\bs{\theta}^{*}, r^{*}) 
            = \sup_{\bs{\theta} \in B(\bs{0}, r^{*})} \left| a_{n_{1}+1}^{n}(\bs{\theta}^{*}, \bs{\theta}) \right|
            \overset{d}{=} \sup_{\bs{\theta} \in B(\bs{0}, r^{*})} \left| a_{1}^{n-n_{1}}(\bs{\theta}^{*}, \bs{\theta}) \right| \overset{\mathbb{P}_{*}}{\to} 0,
         \end{aligned}
    \end{equation}
    so 
    \begin{equation}
        \begin{aligned}
            a_{n_{1}+1}^{n}(\bs{\theta}^{*}, r^{*}) \overset{\mathbb{P}_{*}}{\to} 0
         \end{aligned}
    \end{equation}
    for any fixed $r^{*}$. Thus, if $\{ \tilde{r}_{m} \}_{m=1}^{+\infty}$ is a sequence of positive radii satisfying limiting properties analogous to equation (\ref{eq:radii condition 2}), then the convergence of $a_{n_{1}+1}^{n}(\bs{\theta}^{*}, \tilde{r}_{m}) \to 0$ in $\mathbb{P}_{*}$-probability still holds for all fixed $m \geq 1$, but may fail if we directly take the limit of $a_{n_{1}+1}^{n}(\bs{\theta}^{*}, \tilde{r}_{n-n_{1}})$. To address this, similar to the proof of Theorem 2.1 of \cite{kleijn2012bernstein}, it is always possible to construct a monotonic subsequence $\{ \tilde{r}_{m_{n}} \}_{n=1}^{+\infty}$ that grows sufficiently slowly to ensure that 
    \begin{equation}
        0 \leq a_{n_{1}+1}^{n}(\bs{\theta}^{*}, r^{*}_{m_{n-n_{1}}}) 
        \leq a_{n_{1}+1}^{n}(\bs{\theta}^{*}, r^{*}_{m_{n}}) 
        \overset{\mathbb{P}_{*}}{\to} 0
    \end{equation}
    and with $r^{*}_{m_{n}} \leq r_{n}^{*}$ for all $n$, while maintaining $\lim_{n \to +\infty} r^{*}_{m_{n-n_{1}}} = +\infty$. Extending such a sequence to a continuous function on the positive real numbers (e.g., with linear interpolation) results in a function $r^{*}_{x} = r^{*}(x)$ satisfying equation (\ref{eq:radii condition 2}) and such that 
    \begin{equation}\label{eq:bound convergence 1}
        a_{n_{1}+1}^{n}(\bs{\theta}^{*}, r^{*}_{n-n_{1}}) \overset{\mathbb{P}_{*}}{\to} 0,
    \end{equation}
    as desired. Before addressing term (II) of equation (\ref{eq:bound term 2 final}), note that by the i.i.d. assumption and the central limit theorem we have
    \begin{equation}\label{eq:score normality}
        \begin{aligned}
            \delta_{n-n_{1}}\dot{\ell}_{n_{1}+1}^{n}( \bs{\theta}^{*}) 
            &= \frac{1}{\sqrt{n-n_{1}}}\sum_{i=n_{1}+1}^{n} \frac{\partial}{\partial\bs{\theta}} \log [\bs{y}_{i} | \bs{\theta}]\Bigg|_{\theta = \theta^{*}} \\
            &\overset{d}{=} \frac{1}{\sqrt{n-n_{1}}}\sum_{i=1}^{n-n_{1}} \frac{\partial}{\partial\bs{\theta}} \log [\bs{y}_{i} | \bs{\theta}]\Bigg|_{\theta = \theta^{*}}
            \overset{d}{\to} \text{Normal}_{d}\left( \bs{0}, \bs{W}_{\theta^{*}} \right),
        \end{aligned}
    \end{equation}
    so, from equations (\ref{eq:MLE normality}) and (\ref{eq:score normality}), we known
    \begin{equation}\label{eq:mean and variance convergence}
        \begin{aligned}
            \bs{\Omega}_{n,\gamma} (\bs{\theta}^{*}) 
            &= \tfrac{\gamma n + (1-\gamma)n_{1}}{n-n_{1}}\bs{V}_{\theta^{*}}
            = \tfrac{\gamma n + (1-\gamma) \alpha_{n} n }{n- \alpha_{n} n }\bs{V}_{\theta^{*}}
            \overset{\mathbb{P}_{*}}{\to} \tfrac{\gamma + (1-\gamma)\alpha}{1-\alpha }\bs{V}_{\theta^{*}}, \\
             \bs{\mu}_{n,\gamma} (\bs{\theta}^{*})
            &= \bs{\Omega}^{-1}_{n,\gamma} (\bs{\theta}^{*}) \left\{ \gamma \delta_{n-n_{1}} \dot{\ell}_{n_{1}+1}^{n}(\bs{\theta}^{*}) + \tfrac{n_{1}}{(n-n_{1})^{\frac{1}{2}}} \bs{V}_{\theta^{*}} (\hat{\bs{\theta}}_{1}^{n_{1}} - \bs{\theta}^{*}) \right\} \\
            &= \bs{\Omega}^{-1}_{n,\gamma} (\bs{\theta}^{*}) \bigg\{ \left( \tfrac{n_{1}}{n-n_{1}} \right)^{\frac{1}{2}} \underbrace{ \bs{V}_{\theta^{*}} \sqrt{n_{1}} (\hat{\bs{\theta}}_{1}^{n_{1}} - \bs{\theta}^{*})}_{\to \bs{W}_{\theta^{*}}^{\frac{1}{2}} \bs{z}_{0}} + \gamma \underbrace{\delta_{n-n_{1}} \dot{\ell}_{n_{1}+1}^{n}(\bs{\theta}^{*})}_{\to \bs{W}_{\theta^{*}}^{\frac{1}{2}}\bs{z}_{1}} \bigg\} \\
            &\overset{d}{\to} \tfrac{1-\alpha }{\gamma + (1-\gamma)\alpha}\bs{V}_{\theta^{*}}^{-1} \bs{W}_{\theta^{*}}^{\frac{1}{2}} \bigg\{ \left(\tfrac{\alpha}{1-\alpha}\right)^{\frac{1}{2}} \bs{z}_{0} + \gamma \bs{z}_{1} \bigg\},
        \end{aligned}
    \end{equation}
    where $\bs{z}_{0}, \bs{z}_{1} \overset{\text{iid}}{\sim} \text{Normal}_{d}(\bs{0}, \bs{I}_{d})$, and $\bs{W}_{\theta^{*}}^{\frac{1}{2}}$ is the left factor of the Cholesky decomposition of $\bs{W}_{\theta^{*}}$. Note that the independence between $\bs{z}_{0}$ and $\bs{z}_{1}$ is not immediate, because we only established convergence in distribution, but it results from the fact that $\delta_{n-n_{1}} \dot{\ell}_{n_{1}+1}^{n}(\bs{\theta}^{*})$ and $\bs{V}_{\theta^{*}} \sqrt{n_{1}} (\hat{\bs{\theta}}_{1}^{n_{1}} - \bs{\theta}^{*})$ are independent for all $n$, given that each one of them only depends on a disjoint subset of $\bs{y}_{1:n}$. Next, considering term (II) of equation (\ref{eq:bound term 2 final}), we use a change of variables to write
    \begin{multline}\label{eq:annoying integral}
        \int_{B(\bs{0}, r^{*}_{n-n_{1}})} \phi \left(\bs{\theta} | \bs{\mu}_{n,\gamma}(\bs{\theta}^{*}), \bs{\Omega}^{-1}_{n, \gamma}(\bs{\theta}^{*}) \right) \ d\bs{\theta} \\
        = \int_{B(\bs{0}, 1)} \phi \left( \bs{\theta} \left| \{r^{*}_{n-n_{1}}\}^{-1} \bs{\mu}_{n,\gamma}(\bs{\theta}^{*}), \{r^{*}_{n-n_{1}}\}^{-2} \bs{\Omega}^{-1}_{n, \gamma}(\bs{\theta}^{*}) \right. \right) \ d\bs{\theta},
    \end{multline}
    but we have $r^{*}_{n-n_{1}} \to +\infty$, so
    \begin{equation}
        \begin{aligned}
            \{r^{*}_{n-n_{1}}\}^{-1} \bs{\mu}_{n,\gamma}(\bs{\theta}^{*}) &\overset{\mathbb{P}_{*}}{\to} \bs{0} &&\text{and}&
            \{r^{*}_{n-n_{1}}\}^{-2} \bs{\Omega}^{-1}_{n, \gamma}(\bs{\theta}^{*}) &\overset{\mathbb{P}_{*}}{\to} \bs{0},
        \end{aligned}
    \end{equation}
    which imply that the sequence of densities on the r.h.s. of equation (\ref{eq:annoying integral}) are converging to a point mass at $\bs{0}$, while the region of integration is a fixed neighborhood around $\bs{0}$. Therefore the integral must converge to 1, and we have
    \begin{equation}\label{eq:bound convergence 2}
        \left| \log \int_{B(\bs{0}, r^{*}_{n-n_{1}})} \phi \left(\bs{\theta} | \bs{\mu}_{n,\gamma}(\bs{\theta}^{*}), \bs{\Omega}^{-1}_{n, \gamma}(\bs{\theta}^{*}) \right) \ d\bs{\theta} \right| \overset{\mathbb{P}_{*}}{\to} 0,
    \end{equation}
    which, when considering (\ref{eq:bound term 2 final}) and (\ref{eq:bound convergence 1}), implies 
    \begin{equation}\label{eq:bound convergence final}
        \left| \log \mathbb{J}_{n,\gamma}(\bs{\theta}^{*}, r^{*}_{n-n_{1}}) - \log \mathbb{J}_{n,\gamma}(\bs{\theta}^{*}) \right| \overset{\mathbb{P}_{*}}{\to} 0.
    \end{equation}
    Next, due to the i.i.d. assumption, by Lemma 2.2 of \cite{kleijn2012bernstein} and because $n-n_{1}= (1-\alpha_{n})n \to +\infty$, we have
    \begin{multline}\label{eq:MLE and score convergence}
        \left\| \sqrt{n-n_{1}} (\hat{\bs{\theta}}_{n_{1}+1}^{n} - \bs{\theta}^{*}) - \tfrac{1}{\sqrt{n-n_{1}}} \bs{V}_{\theta^{*}}^{-1} \dot\ell_{n_{1}+1}^{n}(\bs{\theta}^{*}) \right\|_{2} \\ 
        \overset{d}{=} \left\| \sqrt{n-n_{1}} (\hat{\bs{\theta}}_{1}^{n-n_{1}} - \bs{\theta}^{*}) - \tfrac{1}{\sqrt{n-n_{1}}} \bs{V}_{\theta^{*}}^{-1} \dot\ell_{1}^{n-n_{1}}(\bs{\theta}^{*}) \right\|_{2} \overset{\mathbb{P}_{*}}{\to} 0,
    \end{multline}
    and, analogously to equations (\ref{eq:MLE normality}) and (\ref{eq:score normality}), we have
    \begin{equation}\label{eq:MLE normality 2}
        \begin{aligned}
            \tfrac{1}{\sqrt{n-n_{1}}} \bs{V}_{\theta^{*}}^{-1} \dot\ell_{n_{1}+1}^{n}(\bs{\theta}^{*}) &\overset{d}{\to} \text{Normal}_{d}(\bs{0}, \bs{\Sigma}_{\theta^{*}}), \\
            \sqrt{n-n_{1}} (\hat{\bs{\theta}}_{n_{1}+1}^{n} - \bs{\theta}^{*}) &\overset{d}{\to} \text{Normal}_{d}(\bs{0}, \bs{\Sigma}_{\theta^{*}}),
        \end{aligned}
    \end{equation}
    and consequently $\hat{\bs{\theta}}_{n_{1}+1}^{n} \overset{\mathbb{P}_{*}}{\to} \bs{\theta}^{*}$. Following the proof of Lemma 2.1 of \cite{kleijn2012bernstein}, we have by Lemma 19.31 of \cite{van2000asymptotic} that for any sequence $\{\bs{h}_{n}\}$ bounded in $\mathbb{P}_{*}$-probability
    \begin{multline}
        \log R_{1:(n-n_{1})}(\bs{\theta}^{*} + \delta_{n-n_{1}}\bs{h}_{n}, \bs{\theta}^{*}) - (n-n_{1}) \mathbb{E}_{*}\left( \log \frac{[\bs{y} | \bs{\theta}^{*} + \delta_{n-n_{1}}\bs{h}_{n}]}{ [\bs{y} | \bs{\theta}^{*}]} \right) \\
        - \tfrac{1}{\sqrt{n-n_{1}}} \bs{h}_{n}'\dot\ell_{1}^{n-n_{1}}(\bs{\theta}^{*}) + \sqrt{n-n_{1}} \underbrace{\mathbb{E}_{*}\left( h_{n}'\dot\ell(\bs{\theta}^{*}) \right)}_{= 0} \overset{\mathbb{P}_{*}}{\to} 0,
    \end{multline}
    and as assumed in equation (15) of Section 3.1 we have that
    \begin{equation}
        \begin{aligned}
            (n-n_{1})\mathbb{E}_{*}\left( \log \frac{[\bs{y} | \bs{\theta}^{*} + \delta_{n-n_{1}}\bs{h}_{n}]}{ [\bs{y} | \bs{\theta}^{*}]} \right) - \tfrac{1}{2} \bs{h}_{n}' \bs{V}_{\theta^{*}} \bs{h}_{n} \overset{\mathbb{P}_{*}}{\to} 0,
        \end{aligned}
    \end{equation}
    therefore
    \begin{equation}
        \begin{aligned}
            \log R_{1:(n-n_{1})}(\bs{\theta}^{*} + \delta_{n-n_{1}}\bs{h}_{n}, \bs{\theta}^{*}) - \tfrac{1}{\sqrt{n-n_{1}}} \bs{h}_{n}'\dot\ell_{1}^{n-n_{1}}(\bs{\theta}^{*}) + \tfrac{1}{2} \bs{h}_{n}' \bs{V}_{\theta^{*}} \bs{h}_{n} \overset{\mathbb{P}_{*}}{\to} 0,
        \end{aligned}
    \end{equation}
    and by the i.i.d. assumption it must be the case that for all sequences $\{\bs{h}_{n}\}$ bounded in $\mathbb{P}_{*}$-probability we have that
    \begin{equation}
        \begin{aligned}
            \log R_{(n_{1}+1):n}(\bs{\theta}^{*} + \delta_{n-n_{1}}\bs{h}_{n}, \bs{\theta}^{*}) - \tfrac{1}{\sqrt{n-n_{1}}} \bs{h}_{n}'\dot\ell_{n_{1}+1}^{n}(\bs{\theta}^{*}) + \tfrac{1}{2} \bs{h}_{n}' \bs{V}_{\theta^{*}} \bs{h}_{n} \overset{\mathbb{P}_{*}}{\to} 0.
        \end{aligned}
    \end{equation}
    In particular, taking $\bs{h}_{n} = \sqrt{n-n_{1}} (\hat{\bs{\theta}}_{n_{1}+1}^{n} - \bs{\theta}^{*} )$, which is bounded in probability due to (\ref{eq:MLE normality 2}), results in
    \begin{multline}\label{eq:log-likelihood convergence}
        \log R_{(n_{1}+1):n}(\hat{\bs{\theta}}_{n_{1}+1}^{n}, \bs{\theta}^{*}) - (\hat{\bs{\theta}}_{n_{1}+1}^{n} - \bs{\theta}^{*})' \dot\ell_{n_{1}+1}^{n}(\bs{\theta}^{*}) \\
        + \tfrac{n-n_{1}}{2} (\hat{\bs{\theta}}_{n_{1}+1}^{n} - \bs{\theta}^{*})'\bs{V}_{\theta^{*}} (\hat{\bs{\theta}}_{n_{1}+1}^{n} - \bs{\theta}^{*}) \overset{\mathbb{P}_{*}}{\to} 0.
    \end{multline}
    The limits in (\ref{eq:MLE normality 2}) also imply that
    \begin{equation}\label{eq:log-likelihood limit}
        \begin{aligned}
            &(\hat{\bs{\theta}}_{n_{1}+1}^{n} - \bs{\theta}^{*})' \dot\ell_{n_{1}+1}^{n}(\bs{\theta}^{*}) - \tfrac{n-n_{1}}{2} (\hat{\bs{\theta}}_{n_{1}+1}^{n} - \bs{\theta}^{*})'\bs{V}_{\theta^{*}} (\hat{\bs{\theta}}_{n_{1}+1}^{n} - \bs{\theta}^{*}) \\
            &= \sqrt{n-n_{1}} (\hat{\bs{\theta}}_{n_{1}+1}^{n} - \bs{\theta}^{*})' \bs{V}_{\theta^{*}} \tfrac{1}{\sqrt{n-n_{1}}} \bs{V}_{\theta^{*}}^{-1} \dot\ell_{n_{1}+1}^{n}(\bs{\theta}^{*}) \\
            &\quad - \tfrac{1}{2} \sqrt{n-n_{1}} (\hat{\bs{\theta}}_{n_{1}+1}^{n} - \bs{\theta}^{*})'\bs{V}_{\theta^{*}} \sqrt{n-n_{1}}(\hat{\bs{\theta}}_{n_{1}+1}^{n} - \bs{\theta}^{*}) \\
            &= \underbrace{\sqrt{n-n_{1}} (\hat{\bs{\theta}}_{n_{1}+1}^{n} - \bs{\theta}^{*})'}_{\to \bs{V}_{\theta^{*}}^{-1} \bs{W}_{\theta^{*}}^{\frac{1}{2}} \tilde{\bs{z}}_{1}} \bs{V}_{\theta^{*}} \Big\{ \underbrace{\tfrac{1}{\sqrt{n-n_{1}}} \bs{V}_{\theta^{*}}^{-1} \dot\ell_{n_{1}+1}^{n}(\bs{\theta}^{*})}_{\to \bs{V}_{\theta^{*}}^{-1} \bs{W}_{\theta^{*}}^{\frac{1}{2}} \bs{z}_{1}} -\tfrac{1}{2} \underbrace{\sqrt{n-n_{1}} (\hat{\bs{\theta}}_{n_{1}+1}^{n} - \bs{\theta}^{*}}_{\to \bs{V}_{\theta^{*}}^{-1} \bs{W}_{\theta^{*}}^{\frac{1}{2}} \tilde{\bs{z}}_{1}}) \Big\} \\
            &\overset{d}{\to} \tilde{\bs{z}}_{1}' \underbrace{ \bs{W}_{\theta^{*}}^{\frac{1}{2}} \bs{V}_{\theta^{*}}^{-1} \bs{W}_{\theta^{*}}^{\frac{1}{2}} }_{:= \bs{M}_{\theta^{*}}} \left\{ \bs{z}_{1} - \tfrac{1}{2} \tilde{\bs{z}}_{1} \right\} = \tilde{\bs{z}}_{1}' \bs{M}_{\theta^{*}} \left\{ \bs{z}_{1} - \tfrac{1}{2} \tilde{\bs{z}}_{1} \right\},
        \end{aligned}
    \end{equation}
    where $\bs{z}_{1}, \tilde{\bs{z}}_{1} \overset{\text{id}}{\sim} \text{Normal}_{d}(\bs{0}, \bs{I}_{d})$, and $\bs{z}_{1}$ is the same as in (\ref{eq:mean and variance convergence}). Note that because of (\ref{eq:MLE and score convergence}), the random variables $\bs{z}_{1}$ and $\tilde{z}_{1}$ must be almost surely identical, so for simplicity we denote both of them as $\bs{z}_{1}$. Combining this last fact with (\ref{eq:log-likelihood convergence}) and (\ref{eq:log-likelihood limit}) results in
    \begin{equation}\label{eq:log-likelihood final limit}
        \begin{aligned}
            \log R_{(n_{1}+1):n}(\bs{\theta}^{*}, \hat{\bs{\theta}}_{n_{1}+1}^{n}) = -\log R_{(n_{1}+1):n}(\hat{\bs{\theta}}_{n_{1}+1}^{n}, \bs{\theta}^{*})  \overset{d}{\to} -\tfrac{1}{2} \bs{z}_{1}' \bs{M}_{\theta^{*}} \bs{z}_{1},
        \end{aligned}
    \end{equation}
    and similarly, from (\ref{eq:MLE normality}) and (\ref{eq:mean and variance convergence}), we get
    \begin{equation}\label{eq:integral convergence}
        \begin{aligned}
            \log \mathbb{J}_{n,\gamma}(\bs{\theta}^{*}) 
            &= - \tfrac{n_{1}}{2}(\hat{\bs{\theta}}_{1}^{n_{1}} - \bs{\theta}^{*})' \bs{V}_{\theta^{*}} (\hat{\bs{\theta}}_{1}^{n_{1}} -  \bs{\theta}^{*}) + \tfrac{1}{2}\bs{\mu}'_{n,\gamma}(\bs{\theta}^{*}) \bs{\Omega}_{n,\gamma} (\bs{\theta}^{*}) \bs{\mu}_{n,\gamma}(\bs{\theta}^{*}) \\
            &\quad + \tfrac{d}{2} \log\left( \tfrac{n_{1}}{\gamma n + (1-\gamma) n_{1}} \right) \\
            &\overset{d}{\to} \tfrac{1-\alpha }{2(\gamma + (1-\gamma)\alpha)} \left\{ \left(\tfrac{\alpha}{1-\alpha}\right)^{\frac{1}{2}} \bs{z}_{0} + \gamma \bs{z}_{1} \right\}' ( \bs{W}_{\theta^{*}}^{\frac{1}{2}} )' \bs{V}_{\theta^{*}}^{-1} \bs{W}_{\theta^{*}}^{\frac{1}{2}} \left\{ \left(\tfrac{\alpha}{1-\alpha}\right)^{\frac{1}{2}} \bs{z}_{0} + \gamma \bs{z}_{1} \right\} \\
            & \quad -\tfrac{1}{2}\bs{z}_{0} ( \bs{W}_{\theta^{*}}^{\frac{1}{2}} )' \bs{V}_{\theta^{*}}^{-1} \bs{W}_{\theta^{*}}^{\frac{1}{2}} \bs{z}_{0} + \tfrac{d}{2} \log\left( \tfrac{\alpha}{\gamma + (1-\gamma)\alpha} \right) \\
            &= \tfrac{1-\alpha }{2(\gamma + (1-\gamma)\alpha)} \left\{ \left(\tfrac{\alpha}{1-\alpha}\right)^{\frac{1}{2}} \bs{z}_{0} + \gamma \bs{z}_{1} \right\}' \bs{M}_{\theta^{*}} \left\{ \left(\tfrac{\alpha}{1-\alpha}\right)^{\frac{1}{2}} \bs{z}_{0} + \gamma \bs{z}_{1} \right\} \\ 
            & \quad -\tfrac{1}{2} \bs{z}_{0} \bs{M}_{\theta^{*}} \bs{z}_{0} + \tfrac{d}{2} \log\left( \tfrac{\alpha}{\gamma + (1-\gamma)\alpha} \right) 
            =: \log \mathbb{J}_{\theta^{*},\gamma} (\bs{z}_{0}, \bs{z}_{1}),
        \end{aligned}
    \end{equation}
    where $\bs{z}_{0},\bs{z}_{1} \overset{\text{ind}}{\sim} \text{Normal}_{d}(\bs{0}, \bs{I}_{d})$. From (\ref{eq:bound convergence final}) and (\ref{eq:integral convergence}) we obtain that
    \begin{equation}\label{eq:annoying integral convergence}
        \log \mathbb{J}_{n,\gamma}(\bs{\theta}^{*}, r^{*}_{n-n_{1}}) \overset{d}{\to} \log \mathbb{J}_{\theta^{*},\gamma}(\bs{z}_{0}, \bs{z}_{1}),
    \end{equation}
    and using (\ref{eq:log-likelihood final limit}) and (\ref{eq:annoying integral convergence}) we have
    \begin{equation}\label{eq:term 2 convergence}
        \begin{aligned}
            &\log \mathbb{I}_{n,\gamma}(\bs{\theta}^{*}, r^{*}_{n-n_{1}}) = \gamma \log R_{(n_{1}+1):n}(\bs{\theta}^{*}, \hat{\bs{\theta}}_{n_{1}+1}^{n}) + \log \mathbb{J}_{n,\gamma}(\bs{\theta}^{*}, r^{*}_{n-n_{1}}) \\
            &\overset{d}{\to} \tfrac{d}{2} \log\left( \tfrac{\alpha}{\gamma + (1-\gamma)\alpha} \right) -\tfrac{1}{2} \bs{z}_{0}' \bs{M}_{\theta^{*}} \bs{z}_{0} -\tfrac{\gamma}{2} \bs{z}_{1}' \bs{M}_{\theta^{*}} \bs{z}_{1} \\ 
            & \quad +\tfrac{1-\alpha }{2(\gamma + (1-\gamma)\alpha)} \left\{ \left( \tfrac{\alpha}{1-\alpha} \right)^{\frac{1}{2}} \bs{z}_{0} + \gamma \bs{z}_{1} \right\}' \bs{M}_{\theta^{*}} \left\{ \left(\tfrac{\alpha}{1-\alpha}\right)^{\frac{1}{2}} \bs{z}_{0} + \gamma \bs{z}_{1} \right\} \\
            &= \tfrac{d}{2} \log\left( \tfrac{\alpha}{\gamma + (1-\gamma)\alpha} \right) - \left\{ \tfrac{\gamma (1-\alpha)}{2(\gamma + (1-\gamma) \alpha)} \right\} \bs{z}_{0}'\bs{M}_{\theta^{*}}\bs{z}_{0} - \left\{ \tfrac{\gamma \alpha}{2(\gamma + (1-\gamma) \alpha)} \right\} \bs{z}_{1}' \bs{M}_{\theta^{*}} \bs{z}_{1} \\
            & \quad +\left\{ \tfrac{\gamma \alpha^{\frac{1}{2}} (1-\alpha)^{\frac{1}{2}}}{\gamma + (1-\gamma) \alpha} \right\} \bs{z}_{0}'\bs{M}_{\theta^{*}} \bs{z}_{1} \\
            &= \tfrac{d}{2} \log\left( \tfrac{\alpha}{\gamma + (1-\gamma)\alpha} \right) - \tfrac{\gamma}{2(\gamma + (1-\gamma) \alpha)} \left\{ (1-\alpha)^{\frac{1}{2}} \bs{z}_{0} - \alpha^{\frac{1}{2}} \bs{z}_{1} \right\}' \bs{M}_{\theta^{*}} \left\{ (1-\alpha)^{\frac{1}{2}} \bs{z}_{0} - \alpha^{\frac{1}{2}} \bs{z}_{1} \right\} \\
            &= \tfrac{d}{2} \log\left( \tfrac{\alpha}{\gamma + (1-\gamma)\alpha} \right) - \tfrac{\gamma}{2(\gamma + (1-\gamma) \alpha)} \bs{z}' \bs{M}_{\theta^{*}} \bs{z}
            = \log \mathbb{I}_{\theta^{*},\gamma}(\bs{z}),
        \end{aligned}
    \end{equation}
    where $\bs{z} = (1-\alpha)^{\frac{1}{2}}\bs{z_{0}} - \alpha^{\frac{1}{2}}\bs{z}_{1} \sim \text{Normal}_{d}(\bs{0}, \bs{I}_{d})$. Finally, the triangle inequality, the continuous mapping theorem, and equations (\ref{eq:decomposition}), (\ref{eq:term 1 convergence}), and (\ref{eq:term 2 convergence}), imply that
    \begin{equation}
        \begin{aligned}
            \int R^{\gamma}_{(n_{1}+1):n}(\bs{\theta}, \hat{\bs{\theta}}_{n_{1}+1}^{n}) \ \phi(\bs{\theta} | \hat{\bs{\theta}}_{1}^{n_{1}}, n_{1}^{-1}\bs{V}_{\theta^{*}}^{-1}) \ d\bs{\theta} \overset{d}{\to} \mathbb{I}_{\theta^{*},\gamma}(\bs{z}),
        \end{aligned}
    \end{equation}
    for all $\gamma > 0$. Importantly, when choosing a set of constants $\gamma_{1}, \dots, \gamma_{k} > 0$ and accounting for the dependence between the corresponding random variables 
    \begin{equation}
        \mathbb{I}_{\theta^{*},\gamma_{1}}(\bs{z}^{(1)}), \dots, \mathbb{I}_{\theta^{*},\gamma_{k}}(\bs{z}^{(k)}),
    \end{equation}
    we must have $\bs{z}^{(i)} = \bs{z}^{(j)}$ $\mathbb{P}_{*}$-almost surely for all $i,j \in \{1, \dots, k\}$. Hence, considering that the limiting distribution $\mathbb{I}_{\theta^{*},\gamma}(\bs{z})$ is absolutely continuous with respect to the Lebesgue measure $\lambda$ (and therefore does not contain atoms at points of discontinuity of $f$), an application of the continuous mapping theorem concludes the proof.
\end{proof}

\begin{proof}[Proof of Theorem 3.1]
    Choosing $f:(0,1)^{2} \to \mathbb{R}$ with $f(x,y) = \frac{x^{2}}{y}$ and applying Lemma \ref{lm:integral convergence} yields
    \begin{equation}\label{eq:convergence surrogate}
        \begin{aligned}
            Q_{\phi}(n | n_{1}) 
            = \frac{\left\{ \mathbb{I}_{n,1}(\bs{\theta}^{*}) \right\}^{2}}{ \mathbb{I}_{n,2}(\bs{\theta}^{*}) }
            \overset{d}{\to} \frac{\{ \mathbb{I}_{\theta^{*},1}(\bs{z}) \}^{2}}{\mathbb{I}_{\theta^{*},2}(\bs{z})},
        \end{aligned}
    \end{equation}
    where
    \begin{equation}
        \begin{aligned}
            \frac{\{ \mathbb{I}_{\theta^{*},1}(\bs{z}) \}^{2}}{\mathbb{I}_{\theta^{*},2}(\bs{z})}
            &= \frac{ \left( \frac{\alpha}{1+(1-1)\alpha} \right)^{2\frac{d}{2}} \exp\left\{ -\frac{2}{2(1+(1-1)\alpha)} \bs{z}'\bs{M}_{\theta^{*}} \bs{z} \right\} }{\left( \frac{\alpha}{2+(1-2)\alpha} \right)^{\frac{d}{2}} \exp\left\{ -\frac{2}{2(2+(1-2)\alpha)} \bs{z}'\bs{M}_{\theta^{*}} \bs{z} \right\}} \\
            &= \left\{ \alpha (2-\alpha) \right\}^{\frac{d}{2}} \exp\left\{- \tfrac{1-\alpha}{2-\alpha} \bs{z}'\bs{M}_{\theta^{*}} \bs{z} \right\},
        \end{aligned}
    \end{equation}
    and $\bs{z} \sim \text{Normal}_{d}(\bs{0}, \bs{I}_{d})$. Next, note that from Lemma \ref{lm:normal swap} we have
    \begin{equation}
        \left| Q(n|n_{1}) - Q_{\phi}(n|n_{1}) \right| \overset{\mathbb{P}_{*}}{\to} 0,
    \end{equation}
    which, combined with (\ref{eq:convergence surrogate}), implies
    \begin{equation}
        Q(n|n_{1}) \overset{d}{\to} \left\{ \alpha (2-\alpha) \right\}^{\frac{d}{2}} \exp\left\{- \tfrac{1-\alpha}{2-\alpha} \bs{z}'\bs{M}_{\theta^{*}} \bs{z} \right\},
    \end{equation}
    for all $\alpha \in (0, 1)$, where $\bs{z} \sim \text{Normal}_{d}(\bs{0}, \bs{I}_{d})$.
\end{proof}

\section{Implementation Details}

For reproducibility of our results, in what follows, we present details of the implementations mentioned in Section 4. Section \ref{sub:MCMC details} presents the MCMC sampler used and Section \ref{sub:RAISOR details} provides the RAISOR sampler used along with general guidance regarding implementation choices, and Section \ref{sub:AIS details} discusses the implementation of the other IS-based approaches.

For all methods we fit the NNGP approximation of the geostatistical model given by
\begin{equation}
    \begin{aligned}
        \bs{y}_{1:n} | \bs{\beta}, \sigma^{2}, \tau^{2}, \phi &\sim \text{Normal}_{n}(\bs{X}\bs{\beta}, \bs{\Sigma}(\sigma^{2}, \tau^{2}, \phi)), \\
        \bs{\beta} &\sim \text{Normal}_{p}(\bs{\mu}_{\beta}, \bs{\Sigma}_{\beta}), \\
        \sigma^{2} &\sim \text{Inverse-Gamma}\left( \tfrac{\alpha_{1}}{2}, \tfrac{\alpha_{2}}{2} \right), \\
        \tau^{2} &\sim \text{Uniform}(0, 1), \\
        \phi &\sim \text{Half-Normal}(0, \gamma^{2}),
    \end{aligned}
\end{equation}
with the resulting posterior distribution of the form
\begin{equation}
    [\bs{\beta}, \sigma^{2}, \tau^{2}, \phi | \bs{y}_{1:n}] \propto [\bs{y}_{1:n} | \bs{\beta}, \sigma^{2}, \tau^{2}, \phi] [\bs{\beta}] [\sigma^{2}] [\tau^{2}] [\phi].
\end{equation}
To use the approximation effectively, we write $\bs{\Sigma}^{-1} = \frac{1}{\sigma^{2}}\bs{L}'\bs{L}$, where $\bs{L}$ is a sparse lower triangular matrix (see \citealp{katzfuss2021general}), and we use $\bs{L}$ instead of $\bs{\Sigma}$ in all calculations, leveraging sparse linear algebra routines when appropriate. 

Even though $\bs{\beta}$ can be integrated out of the model using normal conjugacy, the resulting expressions cannot directly leverage the sparsity of $\bs{L}$ for efficient computation. However, for any fixed value of $\bs{\beta}$ we can write
\begin{equation}
    \begin{aligned}
        \ [\sigma^{2}, \tau^{2}, \phi | \bs{y}_{1:n}] 
        = \frac{[\bs{\beta}, \sigma^{2}, \tau^{2}, \phi | \bs{y}_{1:n}]}{[\bs{\beta} | \sigma^{2}, \tau^{2}, \phi, \bs{y}_{1:n}]}
        \propto \frac{[\bs{y}_{1:n} | \bs{\beta}, \sigma^{2}, \tau^{2}, \phi] [\bs{\beta}] [\sigma^{2}] [\tau^{2}] [\phi]}{[\bs{\beta} | \sigma^{2}, \tau^{2}, \phi, \bs{y}_{1:n}]},
    \end{aligned}
\end{equation}
allowing us to implicitly integrate $\bs{\beta}$ out of the model, and we use this relation to improve the implementation of all approaches.

\subsection{MCMC}\label{sub:MCMC details}

For the MCMC approach, we use a Metropolis-Hastings within a partially collapsed Gibbs sampler \citep{van2015metropolis}. Importantly, recall that for this type of sampler the limiting distribution of the chain is not invariant to the order of the updates, therefore we update $(\tau^{2}, \phi)$, followed by $\bs{\beta}$, and then $\sigma^{2}$ to sample from the correct target distribution.

Considering $\bs{\beta}$ and $\sigma^{2}$ we leverage conjugacy to sample from their full-conditional distributions given by
\begin{equation}
    \begin{aligned}
        \bs{\beta} | \cdot &\sim \text{Normal}_{p}( \tilde{\bs{\mu}}_{\beta}, \tilde{\bs{\Sigma}}_{\beta} ), &
        \sigma^{2} | \cdot &\sim \text{Inverse-Gamma}\left( \tfrac{\tilde{\alpha}_{1}}{2}, \tfrac{\tilde{\alpha}_{2}}{2} \right),
    \end{aligned}
\end{equation}
where $\tilde{\bs{\mu}}_{\beta} = \tilde{\bs{\Sigma}}_{\beta} \left( \frac{1}{\sigma^{2}} \bs{X}' \bs{L}'\bs{L} \bs{y} + \bs{\Sigma}^{-1}_{\beta} \bs{\mu}_{\beta}\right )$, $\tilde{\bs{\Sigma}}_{\beta} = \left( \frac{1}{\sigma^{2}}\bs{X}'\bs{L}'\bs{L}\bs{X} + \bs{\Sigma}_{\beta}^{-1} \right)^{-1}$, $\tilde{\alpha}_{1} = \alpha_{1} + n$, and 
$\tilde{\alpha}_{2} = \alpha_{2} + (\bs{y} - \bs{X} \bs{\beta})' \bs{L}'\bs{L} (\bs{y} - \bs{X}\bs{\beta})$.

For $\tau$ and $\phi$ we consider a joint Metropolis update targeting the marginalized posterior $[\sigma^{2}, \tau^{2}, \phi | \bs{y}_{1:n}]$, which reduces the number of times we need to construct $\bs{L}$ and evaluate the likelihood function (note that even with the NNGP approximation those operations remain the most costly for moderate $n$). For convenience, we work with the transformed parameters $(\tilde{\tau}^{2}, \tilde{\phi}) = \left(\log \frac{\tau^{2}}{1-\tau^{2}}, \log \phi \right)$, and use a random walk proposal with noise arising from a $\text{Normal}_{2} (\bs{0}, \sigma^{2}_{\text{tune}} \bs{I}_{2} )$, where $\sigma^{2}_{\text{tune},n}$ is a hyperparameter that needs to be tuned for each sample size $n$ to avoid poor mixing. Using a few test runs for guidance, we took $$\log \sigma^{2}_{\text{tune},n_{1:6}} = (0.93, 0.34, -0.86, -1.91, -2.68, -3.35)'.$$ Additionally, for each chain we adapted $\sigma^{2}_{\text{tune},n}$ during the burn-in iterations to improve the mixing.

\subsection{RAISOR}\label{sub:RAISOR details}

Similar to other importance sampling-based techniques, RAISOR requires user choices during implementation. In what follows, we describe those and provide general heuristic guidance. Note that because RAISOR can be used to sample from different families of models, optimal choices are likely to be model-specific, and therefore the heuristics provided are meant to be initial values in a tuning process, rather than default choices. We later provide details on how the RAISOR used in Section 4 was implemented.

\subsubsection{Family of Approximating Distributions and Notion of Optimality}

The first choice to consider is that of the approximating family of distributions $[\bs{\theta} | \bs{\eta}]_{\text{prop}}$. As with any other approximation, a suitable choice has to meet basic general requirements, such as having the same support as the $j$th target partial posterior $[\bs{\theta}|\tilde{\bs{y}}_{1:j}]$, however, because $[\bs{\theta} | \bs{\eta}]_{\text{prop}}$ fulfills the role of an IS distribution, a theoretically good choice should also satisfy the following additional criteria:
\begin{enumerate}
    \item (sampling) one can efficiently generate $\bs{\theta}_{1}, \dots, \bs{\theta}_{M} \overset{\text{iid}}{\sim} [\bs{\theta} | \bs{\eta}]_{\text{prop}}$ for any choice of $\bs{\eta} \in \Xi$;
    \item (tractability) $[\bs{\theta} | \bs{\eta}]_{\text{prop}}$ is analytically available up to a proportionally constant and cheap to evaluate;
    \item (closeness) for all $\varepsilon > 0$, there is $\bs{\eta}$ such that $D_{\chi^{2}}([\bs{\theta} | \tilde{\bs{y}}_{1:j}] \| [\bs{\theta} | \bs{\eta}]_{\text{prop}} ) < \varepsilon$.
\end{enumerate}
The sampling condition ensures fast replenishment, tractability makes it possible to calculate exact IS weights (assuming $[\bs{\theta} | \tilde{\bs{y}}_{1:j}]$ is also tractable), and closeness guarantees that one can always increase $Q(n_{j} | \bs{\eta})$ after replenishment for some choice of $\bs{\eta} \in \Xi$. Taking $\bs{\theta}$ to be continuous, the first two can be easily satisfied by common density estimators such as kernel density estimation (KDE) and infinite mixture of normals applied to some transformation of the original parameter $g(\bs{\theta})$, while mixtures of discrete distributions can generally address the case of $\bs{\theta}$ discrete. Closeness requires consideration because it also depends on the posterior, but is usually guaranteed if $[\bs{\theta}|\bs{\eta}]_{\text{prop}}$ is a family of non-parametric density (or probability mass) estimators with heavy tails (e.g., mixture of $t$ distributions or KDE with a $t$ kernel). This is related to the robustness of the IS estimator, see for instance \cite{hesterberg1995weighted} and \cite{owen2000safe}, and, although relevant from a theoretical point of view, its practical implications may depend on the problem at hand, as discussed in \cite{delyon2021safe}.

Regarding the measure of discrepancy in Algorithm 2, used to determine which element from the family of approximations is best, one convenient possibility is to take
\begin{equation}
    D_{f}\bigg( [\bs{\theta} | \bs{y}_{1:n}] \bigg\| [\bs{\theta} | \bs{\eta}]_{\text{prop}} \bigg) = \mathbb{E}_{\bs{\theta} \sim [\bs{\theta} | \bs{y}_{1:n}]} \left\{ \phi\left( \frac{[\bs{\theta} | \bs{\eta}]_{\text{prop}}}{[\bs{\theta} | \bs{y}_{1:n}]} \right) \right\}, 
\end{equation}
where $\phi : [0,+\infty) \to (-\infty, +\infty]$ is a convex function such that $\phi(x) < +\infty$ for all $x > 0$, $\phi(1) = 0$, and $\phi(0) = \lim_{x \to 0^{+}} \phi(x)$. This family of divergences was introduced by \cite{csiszar1967information} and, given a weighted sample $\{\bs{\theta}_{m}, w( \bs{\theta}_{m}) \}_{m=1}^{M}$ from $[\bs{\theta} | \bs{y}_{1:n}]$, can be consistently estimated using weighted averages, resulting in an optimization with form
\begin{equation}\label{eq:optimal approximation}
    \hat{\bs{\eta}} 
    = \arg\min_{\bs{\eta} \in \Xi} \hat{D}_{f}\bigg( [\bs{\theta} | \bs{y}_{1:n}] \bigg\| [\bs{\theta} | \bs{\eta}]_{\text{prop}} \bigg) 
    = \arg\min_{\bs{\eta} \in \Xi} \sum_{m=1}^{M} \tilde{w}(\bs{\theta}_{m}) \ \phi\left( \frac{[\bs{\theta}_{m} | \bs{\eta}]_{\text{prop}}}{[\bs{\theta}_{m} | \bs{y}_{1:n}]} \right).
\end{equation}
A few common instances occur when taking $\phi(x) = \frac{(x-1)^{2}}{x}$, $\phi(x) = -\log x$, and $\phi(x) = \frac{1}{2}|x-1|$, corresponds to the $\chi^{2}$-divergence, KL-divergence, and total variation distance respectively.

Although the $\chi^{2}$-divergence is theoretically attractive due to its direct connection to the tracked quality of the samples (see Section \ref{sec:RESS}), when making this choice, one must also account for the computational cost of obtaining $\hat{\bs{\eta}}$. For this reason, the KL is significantly more popular in the context of AIS because it achieves a good trade-off between computational efficiency and quality of the approximation empirically. In this case, the optimization reduces to 
\begin{equation}
    \hat{\bs{\eta}} = \arg\max_{\bs{\eta} \in \Xi} \sum_{m=1}^{M} \tilde{w}(\bs{\theta}_{m}) \log [\bs{\theta}_{m} | \bs{\eta}]_{\text{prop}},
\end{equation}
which is equivalent to weighted maximum likelihood estimation. Hence, depending on the choice of $[\bs{\theta} | \bs{\eta}]_{\text{prop}}$, solutions can be obtained analytically or with the use of efficient algorithms. For instance, when choosing $[\bs{\theta} | \bs{\eta}]_{\text{prop}}$ to be a mixture of normals, the optimization can be done using a weighted version of the Expectation Maximization (EM) algorithm \citep{dempster1977maximum}, for which many efficient implementations are readily available.

Additionally, the cost of the optimization in (\ref{eq:optimal approximation}) depends on the size $M$ of the weighted sample $\{ \bs{\theta}_{m}, w_{m} \}_{m=1}^{M}$ targeting $[\bs{\theta} | \bs{y}_{1:n}]$, which typically contains a relatively small ESS at replenishment times. Therefore, we can use sample reduction techniques to obtain a smaller weighted sample $\{ \bs{\theta}^{*}_{m}, w^{*}_{m} \}_{m=1}^{N}$ that also targets the posterior and has a similar ESS. For instance, one can use the sampling-importance-resampling (SIR) technique \citep{rubin1988using} to obtain a smaller unweighted sample, although the introduction of potentially repeated samples tends to result in a reduction in ESS. To circumvent this, similarly to \cite{delyon2021safe}, we suggest using a weighted version of SIR, described in Algorithm \ref{alg:WSIR}, which corresponds to running SIR until $N$ unique values are sampled, then grouping repeated values using weights proportional to the number of replicates.

\begin{algorithm}[tb!]
    \caption{Weighted Sampling-Importance-Resampling}\label{alg:WSIR}
    \begin{algorithmic}
    \State \textbf{Require:} Weighted sample $\{ \bs{\theta}_{m}, w_{m} \}_{m=1}^{M}$ and target size $N < M$
    \State Get $\bs{i}_{1:N} = (i_{1}, \dots, i_{N})$ by sampling from $\{1, \dots, M\}$ without replacement with weights $\{ w_{m} \}_{m=1}^{M}$
    \State Set $\bs{\theta}_{1:N} = (\bs{\theta}_{1}, \dots, \bs{\theta}_{N}) \gets (\bs{\theta}_{i_{1}}, \dots, \bs{\theta}_{i_{N}})$
    \State Set $\bs{w}_{1:N} = (w_{1}, \dots, w_{N}) \gets (w_{i_{1}}, \dots, w_{i_{N}})$
    \State Set $\bs{w}_{1:N}^{*} = (w_{1}^{*}, \dots, w_{N}^{*}) \gets (1, \dots, 1)$
    \State Set $p \gets 1$
    \For{$j \in \{1, \dots, N-1\}$}
        \State Set $p \gets p - w_{j}$
        \State Sample $k \sim \text{Geometric}(p)$
        \If{$k > 0$}
            \State Sample $\bs{p}_{1:j} \sim \text{Multinomial}(k, \bs{w}_{1:j}) $
            \State Set $\bs{w}_{1:j}^{*} \gets \bs{w}_{1:j}^{*} + \bs{p}_{1:j} $
        \EndIf
    \EndFor
    \State Set $\bs{w}^{*}_{1:N} \gets \bs{w}^{*}_{1:N}/\sum_{j=1}^{N}w_{j}^{*}$
    \State \textbf{Return:} Weighted sample $\{\bs{\theta}_{m}, w^{*}_{m}\}_{m=1}^{N}$
    \end{algorithmic}
\end{algorithm}

\subsubsection{Initial Sample Size and Growth Rate}

When considering the initialization of RAISOR, one has two choices: to take $n_{1} = 0$, corresponding to proposing from $[\bs{\theta}]$, or to take $n_{1} > 0$, corresponding to proposing from $[\bs{\theta} | \bs{y}_{1:n_{1}}]$. Choosing the prior can work well if the distribution is reasonably informative and can be sampled from. Otherwise, it is generally better to bypass potentially ill-behaved initial iterations by taking proposing from a partial posterior. In either case, sampling can be done in a variety of ways (e.g., direct sampling, MCMC, or IS), as long as it results in a (potentially weighted) sample targeting the corresponding distribution. The optimal choice of $n_{1}$ depends on multiple factors, but our experiments indicate that $n_{1} \geq 10$ is sufficient to ensure numerical stability of the updates for posterior distributions with weakly informative priors.

As mentioned in Section 3, we can use the upper bound on $Q_{\theta^{*}, \alpha}$ to narrow down the range of reasonable values of $\alpha$, as
\begin{equation}
    \begin{aligned}
        n_{j+1} &> c(q_{\text{min}}, d) \cdot n_{j} &&\Longrightarrow &\text{RESS}(n_{j+1} | n_{j}) &< q_{\text{min}}, 
    \end{aligned}
\end{equation}
for large values of $n_{j}$, where $c(q_{\text{min}}, d)$ is as defined in Section 3 and $q_{\text{min}}$ represents the smallest quality tolerable during model fitting, see below for more details. Therefore, heuristically we can take
\begin{equation}\label{eq:growth limit}
    \alpha \in \left( q_{\text{min}}^{\frac{2}{d}} \left\{ 1+ \sqrt{1 - q_{\text{min}}^{\frac{2}{d}}} \right\}^{-1},  1 \right),
\end{equation}
but note that the lower bound above is optimistic in the sense that it assumes perfect replenishment, so setting $\alpha$ between those values is a starting point (e.g., picking the geometric mean of the end points of the interval above). Alternative heuristics may consider the entire limiting distribution rather than just its upper bound, but we leave this for future work.

Even when taking $\{n_{j}\}$ such that $n_{j+1} = \lceil \alpha^{-1} n_{j} \rceil$ with $\alpha$ satisfying (\ref{eq:growth limit}), note that due to the stochasticity inherent to the limiting distribution, it is not possible to guarantee that $\hat{Q}(n_{j+1}|\bs{\eta})$ in Algorithm 4 of the paper will remain above the threshold $q_{\text{min}}$. As discussed in Section 2, This can be problematic because a low value of $\hat{Q}(n_{j+1}|\bs{\eta}^{(j)})$ generally leads to a poor estimator of the objective function $D_{f}$, leading to an approximation $[\bs{\theta} | \bs{\eta}^{(j+1)}]_{\text{prop}}$ that is quite distant from its target $[ \bs{\theta} |  \bs{y}_{1:n_{j+1}} ]$, and consequently an unstable procedure. To mitigate this issue we considered two options. The first is to, when using samples from $[\bs{\theta} | \bs{\eta}^{(j)}]_{\text{prop}} \approx [\bs{\theta} | \bs{y}_{1:n_{j}}]$ to approximate $[\bs{\theta} | \bs{y}_{1:n_{j+1}}]$, take a few exponentially spaced values between $n_{j}$ and $n_{j+1}$, and consider the resulting quality. Then, instead of always targeting $[\bs{\theta} | \bs{y}_{1:n_{j+1}}]$, one can choose the highest sample size $n_{j+1}^{*} \in (n_{j}, n_{j+1}]$ (among the ones considered) that yields $\hat{Q}(n_{j+1}^{*} | \bs{\eta}^{(j)}) \geq q_{\text{min}}$. Alternatively, one can use an annealing step to bridge the current approximation and the next partial posterior with controlled stability until the resulting quality of the samples is above $q_{\text{min}}$ (see Subsection \ref{sub:AIS details} for details).

\subsubsection{Monte Carlo Sample Size and Quality Thresholds}

We provide guidance regarding the Monte Carlo sample size $M$, the replenishment threshold $q$ and the minimal threshold $q_{\text{min}}$, based on a reference ESS $N$. We define $N$ as the number of independent samples that would be needed to satisfactorily estimate the densities (or p.m.f.s) of the partial posterior distributions $\{ [ \bs{\theta} | \bs{y}_{1:n_{j}} ] \}_{j=1}^{J}$ and to obtain posterior estimates or predictions with satisfactory accuracy. Because $N$ is case dependent and varies with the model in question, the chosen family of approximations, the dimension of $\bs{\theta}$, and how the final posterior samples will be used, we do not provide specific guidance on how to make this choice, but note that a similar assessment is often implicitly done when using other sampling techniques, such as MCMC and AIS.

For a weighted sample of size $M$ with quality $q$, the ESS can be estimated by the product $M q$, so considering a specified value of $N$, a natural idea is to impose the constraint $M q_{\text{min}} \geq N$ because it ensures that the quality remains high enough throughout the procedure to reliably construct approximations of the partial posteriors and complete downstream inferential and prediction tasks. Additionally, we include the constraint $q > q_{\text{min}}$ because replenishment is necessary before the quality drops below $q_{\text{min}}$ to avoid numerical instability. 

As we presented in Section 2, replenishment steps are relatively expensive due to the cost of recomputing IS weights. Considering that an increase in $q_{\text{min}}$ indirectly results in more replenishment steps, we choose $M$ to minimize $q_{\text{min}}$ under the aforementioned constraints (i.e., $M$ should be taken as large as possible), which has the added benefit of increasing the compatibility of the method with parallelization. Considering the weighted SIR procedure to reduce the cost of obtaining the approximation $[\bs{\theta} |\bs{\eta}]_{\text{prop}}$, discussed previously, the main limiting factor for the choice of $M$ in this case is storage, because the memory requirement of the procedure will be at least $O(Md)$ while the cost of optimizing $\hat{D}_{f}$ will scale with the reduced sample size.

When choosing $q$, two relevant aspects need to be considered. The first one is the ratio $q/q_{\text{min}} > 1$, which creates a region of quality values where replenishment can be performed without numerical instability. Increasing $q$ can help mitigate the need for potentially expensive interventions, however, the another consideration is that it also increases the frequency of replenishment steps. For well-specified models with low-dimensional parameters $d \leq 6$, our experiments indicate that a ratio around 2 (i.e., $q = 2 q_{\min}$) yields good results.

\subsubsection{RAISOR Used in the Applications}

Considering the applications presented in Section 4, we made similar implementation choices for all data sets. We assessed that $N = 5000$ independent samples would be sufficient to estimate the posterior density and to perform all tasks of interest (in our case prediction). Therefore, to make the results compatible with MCMC, we used a total of $M = 50000$ IS samples with $q_{\text{min}} = 0.1$ and $q = 2q_{\text{min}} = 0.2$.

For convenience, we considered the reparametrization $\tilde{\bs{\theta}} = \left(\bs{\beta}', \log \sigma^{2}, \log \frac{\tau^2}{1-\tau^{2}}, \log \phi \right)'$ to ensure that the support was $\mathbb{R}^{p+3}$, and chose the family of approximating distributions $[\bs{\theta} | \bs{\eta}]_{\text{prop}}$ to be a mixture of multivariate normals for the spatial parameters and the full-conditional distribution for $\bs{\beta}$, so $\Xi$ in this case corresponds to the set of all possible mean vectors, weight vectors, and covariance matrices used to determine the mixture. Despite suspecting that the posterior density contains regions with heavy tails (at least when compared to a normal), there were no signs of strong instability during model fitting, implying that even if those regions exist their influence is negligible for all practical purposes. For computational efficiency, we chose the KL-divergence between our approximation and the target as our objective function, and used the weighted EM algorithm implemented in the R package \texttt{mclust} \citep{scrucca2023model} to do the optimization. Knowing that our posterior distribution does not present strong multi-modality but that the covariance parameters are highly dependent, we conservatively used 10 components in the mixture to account for it.

We followed the batching strategy described previously and considered the values $\alpha \in \left\{ \frac{1}{2}, \frac{2}{3}, \frac{3}{4} \right\}$. All values of $\alpha$ resulted in a similar computational cost, choosing to use $\alpha = \frac{1}{2}$ for all experiments. If the quality decreased below $q_{\text{min}}$ after a weight update, we used annealed steps until generating a sample with quality greater than $q$ (see Section \ref{sub:AIS details} for details). Regarding the initial sample size, we considered directly sampling from the prior or using MCMC to sample from a partial posterior. Because the prior was chosen to be vague for some of the parameters, using it as the initial IS distribution resulted in many annealed IS correction steps needed for stability and an overall computational cost similar to sampling from $[\bs{\theta} | \bs{y}_{1:20}]$ using MCMC. Thus, for simplicity we used the latter in all simulations.

\subsection{Other IS-based Approaches}\label{sub:AIS details}

When considering the AIS and AAIS approaches, whenever possible we used the same heuristics as described in Subsection \ref{sub:RAISOR details} for the sake of the comparability of the results. Therefore, both AIS and AAIS use as their adaptation mechanism the same normal mixture approximation, obtained via the weighted expectation-maximization algorithm used by RAISOR. Additionally, instead of running either method for a fixed number of iterations, we do it until the quality of the resulting IS sample is above the threshold $q = 0.20$ and then running one more iteration after that to maximize sample quality.

Despite having convergence guarantees, AIS methods tend to be sensitive to their initialization, potentially leading to numerical instabilities when using uninformative initial proposals. The most common way to robustify AIS implementations is to introduce annealing, however this results in an AAIS approach, which we are already considering. Alternatively, one could hybridize AIS with another sampler to first sample from a distribution that is sufficiently close to the target, then use it as proposal for AIS. Doing so requires choosing an initial proposal sufficiently close to the target, which is non-trivial task. For the aforementioned reasons, we only consider a standard implementation of AIS in our experiments, presented in Algorithm \ref{alg:AIS 2}.

\begin{algorithm}[tb!]
    \caption{Implementation of Adaptive Importance Sampling}\label{alg:AIS 2}
    \begin{algorithmic}
    \State Sample $\bs{\theta}_{1}^{(1)}, \dots, \bs{\theta}_{M}^{(1)} \sim [\bs{\theta} | \bs{\eta}^{(1)}]_{\text{prop}} := [\bs{\theta} | \bs{y}_{1:n_{1}}]$
    \State Take $\tilde{w}(\bs{\theta}_{m}^{(1)}) \propto \frac{[\bs{y}_{1:n} | \bs{\theta}_{m}^{(1)}][\bs{\theta}_{m}^{(1)}]}{[\bs{\theta}_{m}^{(1)} | \bs{\eta}^{(1)}]_{\text{prop}}}$ \Comment{in parallel for each $m$}
    \State Calculate $\hat{Q}(n | \bs{\eta}^{(1)})$ and take $s \gets 1$
    \While{$\hat{Q}(n | \bs{\eta}^{(s)}) < q$}{ $s \gets s+1$}
        \State Obtain $D_{f}^{(s)}(\cdot \| \cdot)$, an IS estimate of the objective function
        \State Obtain $\bs{\eta}^{(s)} = \arg \min_{\bs{\eta} \in \Xi} D_{f}^{(s)} \bigg( [\bs{\theta} | \bs{y}_{1:n}] \bigg\| [\bs{\theta} | \bs{\eta}]_{\text{prop}} \bigg)$
        \State Sample $\bs{\theta}_{1}^{(s)}, \dots, \bs{\theta}_{M}^{(s)} \overset{\text{i.i.d.}}{\sim} \left[ \bs{\theta} \left| \bs{\eta}^{(s)} \right. \right]_{\text{prop}}$ \Comment{in parallel for each $m$}
        \State Take $\tilde{w}(\bs{\theta}_{m}^{(s)}) \propto \frac{[\bs{y}_{1:n} | \bs{\theta}_{m}^{(s)}][\bs{\theta}_{m}^{(s)}]}{[\bs{\theta}_{m}^{(s)} | \bs{\eta}^{(s)}]_{\text{prop}}}$ \Comment{in parallel for each $m$}
        \State Calculate $\hat{Q}(n | \bs{\eta}^{(s)})$
    \EndWhile
    \State \textbf{Return:} Weighted sample from the posterior $\{(\bs{\theta}_{m}^{(s)}, \tilde{w}(\bs{\theta}_{m}^{(s)}))\}_{m=1}^{M}$
    \end{algorithmic}
\end{algorithm}

For the AAIS approach, the main consideration when it comes to implementation is the choice of how to bridge the initial proposal distribution $[\bs{\theta} | \bs{\eta}]_{\text{prop}}$ and the posterior distribution $[\bs{\theta} | \bs{y}_{1:n}]$. The most common approach in the literature is to take $\{[\bs{\theta} | \bs{y}_{1:n}, \bs{\eta}, \gamma_{j}]_{\text{pow}}\}_{j=1}^{J}$, where
\begin{equation}
    [\bs{\theta} | \bs{y}_{1:n}, \bs{\eta}, \gamma]_{\text{pow}} := \frac{[\bs{\theta} | \bs{y}_{1:n}]^{\gamma} [\bs{\theta} | \bs{\eta}]_{\text{prop}}^{1-\gamma}}{\int [\bs{\theta}' |\bs{y}_{1:n}]^{\gamma} [\bs{\theta}' | \bs{\eta}]_{\text{prop}}^{1-\gamma} \ d\bs{\theta}'},
\end{equation}
for some sequence of powers $\{\gamma^{(j)}\}_{j=1}^{J}$ with $0 = \gamma^{(1)} < \dots < \gamma^{(J)} = 1$ (e.g., \citealp{neal2001annealed, liu2014adaptive, zhang2025annealed}). Even though other ways of selecting this sequence have been proposed (e.g., \citealp{goshtasbpour2023adaptive}), we use the heuristic based on ESS independently introduced in \cite{liu2014adaptive} and Appendix A of \cite{finke2019importance} due its simplicity and effectiveness. Consider the function that measures the quality of $[\bs{\theta} | \bs{\eta}]_{\text{prop}}$ as a proposal for the bridging distribution $[\bs{\theta} | \bs{y}_{1:n}, \bs{\eta}, \gamma ]_{\text{pow}}$, given by
\begin{equation}
    \begin{aligned}
        Q(n,\gamma | \bs{\eta}) 
        &= \left\{ 1 + D_{\chi^{2}}\left( [\bs{\theta} | \bs{y}_{1:n}, \bs{\eta}, \gamma]_{\text{pow}} \| [\bs{\theta} | \bs{\eta}]_{\text{prop}} \right)\right\}^{-1} \\
        &= \left\{ \int\frac{[\bs{\theta} | \bs{y}_{1:n}, \bs{\eta}, \gamma]_{\text{pow}}^{2}}{[\bs{\theta} | \bs{\eta}]_{\text{prop}}} \ d\bs{\theta} \right\}^{-1}
        = \frac{ \left\{ \int [\bs{\theta} | \bs{y}_{1:n}]^{\gamma} [\bs{\theta} | \bs{\eta}]_{\text{prop}}^{1-\gamma} \ d\bs{\theta} \right\}^{2}}{\int [\bs{\theta} | \bs{y}_{1:n}]^{2\gamma} [\bs{\theta} | \bs{\eta}]_{\text{prop}}^{1-2\gamma} \ d\bs{\theta}} \\
        &= \frac{ \left\{ \int \left( \frac{[\bs{\theta} | \bs{y}_{1:n}]}{[\bs{\theta} | \bs{\eta}]_{\text{prop}}} \right)^{\gamma} [\bs{\theta} | \bs{\eta}]_{\text{prop}} \ d\bs{\theta} \right\}^{2}}{\int \left( \frac{[\bs{\theta} | \bs{y}_{1:n}]}{[\bs{\theta} | \bs{\eta}]_{\text{prop}}} \right)^{2\gamma} [\bs{\theta} | \bs{\eta}]_{\text{prop}} \ d\bs{\theta}}
        = \frac{ \left\{ \int \left( \frac{[\bs{y}_{1:n} | \bs{\theta}] [\bs{\theta}]}{[\bs{\theta} | \bs{\eta}]_{\text{prop}}} \right)^{\gamma} [\bs{\theta} | \bs{\eta}]_{\text{prop}} \ d\bs{\theta} \right\}^{2}}{\int \left( \frac{[\bs{y}_{1:n} | \bs{\theta}] [\bs{\theta}]}{[\bs{\theta} | \bs{\eta}]_{\text{prop}}} \right)^{2\gamma} [\bs{\theta} | \bs{\eta}]_{\text{prop}} \ d\bs{\theta}},
    \end{aligned}
\end{equation}
which can be estimated using
\begin{equation}
    \hat{Q}(n,\gamma | \bs{\eta}) = \frac{\left\{ \frac{1}{M} \sum_{m=1}^{M} \tilde{w}^{\gamma}(\bs{\theta}_{m}) \right\}^{2}}{\frac{1}{M}\sum_{m=1}^{M} \left\{ \tilde{w}^{\gamma}(\bs{\theta}_{m}) \right\}^{2}}, 
\end{equation}
where $\tilde{w}(\bs{\theta}) \propto \frac{[\bs{y}_{1:n} | \bs{\theta}] [\bs{\theta}]}{[\bs{\theta} | \bs{\eta}]_{\text{prop}}}$ and $\bs{\theta}_{1}, \dots, \bs{\theta}_{M} \overset{\text{i.i.d.}}{\sim} [\bs{\theta} | \bs{\eta}]_{\text{prop}}$. Note that $\hat{Q}(n,\gamma | \bs{\eta})$ is a strictly decreasing continuous function of $\gamma$ with $\hat{Q}(n,0 | \bs{\eta}) = 1$. Therefore, given a any proposal $[\bs{\theta} | \bs{\eta}]_{\text{prop}}$ and quality threshold $q \in (0,1)$, we can always find a sufficiently small value $\gamma \in (0,1)$ such that $\hat{Q}(n, \gamma | \bs{\eta}) \geq q$ using a root-finding algorithm. In other words, we can always find a value of $\gamma$ such that the target $[\bs{\theta} | \bs{y}_{1:n}, \bs{\eta}, \gamma]_{\text{prop}}$ is sufficiently close to $[\bs{\theta} | \bs{\eta}]_{\text{prop}}$ to avoid instabilities in the procedure, thus making AAIS fairly robust.

\begin{algorithm}[tb!]
    \caption{Implementation of Annealed Adaptive Importance Sampling}\label{alg:AAIS}
    \begin{algorithmic}
    \State Sample $\bs{\theta}_{1}^{(1)}, \dots, \bs{\theta}_{M}^{(1)} \sim [\bs{\theta} | \bs{\eta}^{(1)}]_{\text{prop}} := [\bs{\theta} | \bs{y}_{1:n_{1}}]$
    \State Take $\tilde{w}(\bs{\theta}_{m}^{(1)}) \propto \frac{[\bs{y}_{1:n} | \bs{\theta}_{m}^{(1)}][\bs{\theta}_{m}^{(1)}]}{[\bs{\theta}_{m}^{(1)} | \bs{\eta}^{(1)}]_{\text{prop}}}$ \Comment{in parallel for each $m$}
    \State Calculate $\hat{Q}(n | \bs{\eta}^{(1)})$ and take $s \gets 1$
    \While{$\hat{Q}(n | \bs{\eta}^{(s)}) < q$}{ take $s \gets s+1$}
        \State Find $\gamma^{(s-1)} \in (0,1)$ such that $\hat{Q}(n, \gamma^{(s-1)} | \bs{\eta}^{(s-1)} ) = q$
        \State Obtain $D_{f}^{(s)}(\cdot \| \cdot)$, an IS estimate of the objective function
        \State Obtain $\bs{\eta}^{(s)} = \arg \min_{\bs{\eta} \in \Xi} D_{f}^{(s)} \bigg( [\bs{\theta} | \bs{y}_{1:n}, \bs{\eta}^{(s-1)}, {\gamma}^{(s-1)}]_{\text{pow}} \bigg\| [\bs{\theta} | \bs{\eta}]_{\text{prop}} \bigg)$
        \State Sample $\bs{\theta}_{1}^{(s)}, \dots, \bs{\theta}_{M}^{(s)} \overset{\text{i.i.d.}}{\sim} \left[ \bs{\theta} \left| \bs{\eta}^{(s)} \right. \right]_{\text{prop}}$ \Comment{in parallel for each $m$}
        \State Take $\tilde{w}(\bs{\theta}_{m}^{(s)}) \propto \frac{[\bs{y}_{1:n} | \bs{\theta}_{m}^{(s)}][\bs{\theta}_{m}^{(s)}]}{[\bs{\theta}_{m}^{(s)} | \bs{\eta}^{(s)}]_{\text{prop}}}$ \Comment{in parallel for each $m$}
        \State Calculate $\hat{Q}(n | \bs{\eta}^{(s)})$
    \EndWhile
    \State \textbf{Return:} Weighted sample from the posterior $\{(\bs{\theta}_{m}^{(s)}, \tilde{w}(\bs{\theta}_{m}^{(s)}))\}_{m=1}^{M}$
    \end{algorithmic}
\end{algorithm}

After selecting a value of $\gamma$, we need to find an approximation of the objective function $D_{f}$, which will then be used to obtain $\bs{\eta}$ such that $[\bs{\theta} | \bs{\eta}]_{\text{prop}} \approx [\bs{\theta} | \bs{y}_{1:n}, \tilde{\bs{\eta}}, \gamma]_{\text{pow}}$, where $\tilde{\bs{\eta}}$ corresponds to the hyperparameter of the previous approximation. Considering the particular case of minimizing the KL-divergence, this results in
\begin{equation}
    \begin{aligned}
        \bs{\eta} 
        &= \arg\min_{\bs{\eta} \in \Xi} D_{\text{KL}}\left( [\bs{\theta} | \bs{y}_{1:n}, \tilde{\bs{\eta}}, \gamma]_{\text{pow}} \| [\bs{\theta} | \bs{\eta}]_{\text{prop}} \right) \\
        &= \arg\min_{\bs{\eta} \in \Xi} \int \log\left( \frac{[\bs{\theta} | \bs{y}_{1:n}, \tilde{\bs{\eta}}, \gamma]_{\text{pow}}}{[\bs{\theta} | \bs{\eta}]_{\text{prop}}} \right) [\bs{\theta} | \bs{y}_{1:n}, \tilde{\bs{\eta}}, \gamma]_{\text{pow}} \ d\bs{\theta} \\
        &= \arg\max_{\bs{\eta} \in \Xi} \int \log [\bs{\theta} | \bs{\eta}]_{\text{prop}} \frac{[\bs{\theta} | \bs{y}_{1:n}]^{\gamma} [\bs{\theta} | \tilde{\bs{\eta}}]_{\text{prop}}^{1-\gamma}}{\int [\bs{\theta}' |\bs{y}_{1:n}]^{\gamma} [\bs{\theta}' | \tilde{\bs{\eta}}]_{\text{prop}}^{1-\gamma} \ d\bs{\theta}'} \ d\bs{\theta} \\
        &= \arg\max_{\bs{\eta} \in \Xi} \int \left( \frac{[\bs{y}_{1:n} | \bs{\theta}] [\bs{\theta}]}{[\bs{\theta} | \tilde{\bs{\eta}}]_{\text{prop}}} \right)^{\gamma} \log [\bs{\theta} | \bs{\eta}]_{\text{prop}} [\bs{\theta} | \tilde{\bs{\eta}}]_{\text{prop}} \ d\bs{\theta} \\
    \end{aligned}
\end{equation}
which can be approximated by
\begin{equation}
    \hat{\bs{\eta}} = \arg\min_{_{\bs{\eta} \in \Xi}} \sum_{m=1}^{M} \tilde{w}^{\gamma}(\bs{\theta}_{m}) \log[\bs{\theta}_{m} | \bs{\eta}]_{\text{prop}}, 
\end{equation}
where $\tilde{w}(\bs{\theta}) \propto \frac{[\bs{y}_{1:n} | \bs{\theta}] [\bs{\theta}]}{[\bs{\theta} | \tilde{\bs{\eta}}]_{\text{prop}}}$ and $\bs{\theta}_{1}, \dots, \bs{\theta}_{M} \overset{\text{i.i.d.}}{\sim} [\bs{\theta} | \tilde{\bs{\eta}}]_{\text{prop}}$, thus reducing to a problem of weighted maximum likelihood estimation. Combining all of the aforementioned steps results in Algorithm \ref{alg:AAIS}.

\section{Sample Quality and Effective Sample Size}\label{sec:RESS}

In this section we present a well-known connection between importance sampling, the $\chi^{2}$-divergence, and the effective sample size (ESS) of a weighted Monte Carlo sample, and we provide insight into the limiting behavior of $Q(n | n_{1})$ presented in Theorem 3.1.

\subsection{Importance Sampling, Chi-Squared Divergence, and ESS}

Considering a target distribution $[\bs{\theta} | \bs{y}_{1:n}]$ and a proposal $[\bs{\theta} | \bs{\eta}]_{\text{prop}}$, we can express the $\chi^{2}$-divergence between them as
\begin{equation}\label{eq:chi-squared to var}
    \begin{aligned}
        D_{\chi^{2}}\bigg( [\bs{\theta} | \bs{y}_{1:n}]  \bigg\|  [\bs{\theta} | \bs{\eta}]_{\text{prop}} \bigg) 
        &= \int \frac{([\bs{\theta} | \bs{y}_{1:n}] - [\bs{\theta} | \bs{\eta}]_{\text{prop}})^{2}}{[\bs{\theta} | \bs{\eta}]_{\text{prop}}} \ d\bs{\theta} 
        = \int \frac{[\bs{\theta} | \bs{y}_{1:n}]^{2}}{[\bs{\theta} | \bs{\eta}]_{\text{prop}}} \ d\bs{\theta} - 1^{2} \\
        &= \mathbb{E}_{\bs{\theta} \sim [\bs{\theta} | \bs{\eta}]_{\text{prop}}}\left( w(\bs{\theta})^{2} \right) - \left\{ \mathbb{E}_{\bs{\theta} \sim [\bs{\theta} | \bs{\eta}]_{\text{prop}}}(w(\bs{\theta})) \right\}^{2} \\
        &=\text{Var}_{\bs{\theta} \sim [\bs{\theta} | \bs{\eta}]_{\text{prop}}}( w(\bs{\theta}) ),
    \end{aligned}
\end{equation}
where $w(\bs{\theta}) = \frac{[\bs{\theta | \bs{y}_{1:n}}]}{[\bs{\theta} | \bs{\eta}]_{\text{prop}}}$. This connection implies that minimizing the variance of the IS weights is equivalent to minimizing the chi-squared divergence between target and proposal, and thus we can use the weights to infer the quality of $[\bs{\theta}|\bs{\eta}]_{\text{prop}}$ as an approximation of $[\bs{\theta} | \bs{y}_{1:n}]$.

When defining the quality of $[\bs{\theta} | \bs{y}_{1:n_{1}}]$ as a proposal distribution for $[\bs{\theta} | \bs{y}_{1:n}]$ as
\begin{equation}\label{eq:RESS2}
    Q(n | n_{1}) = \left\{ 1 + D_{\chi^{2}} \left( [\bs{\theta} | \bs{y}_{1:n}] \left. \vphantom{\tfrac{0}{0}} \right\| [\bs{\theta} | \bs{y}_{1:n_{1}}] \right) \right\}^{-1},
\end{equation}
which lies in $[0,1]$. We can interpret this quantity as the proportion of information lost when using a weighted sample from $[\bs{\theta} | \bs{y}_{1:n_{1}}]$ to target $[\bs{\theta} | \bs{y}_{1:n}]$. This interpretation comes from the comparison between the variance of a simple Monte Carlo estimator $\hat{I}_{n}(f)$ and an IS estimator $\tilde{I}_{n}(f)$ with proposal $[\bs{\theta} | \bs{y}_{1:n_{1}}]$, which typically has a higher variance. Because of this, traditionally the ESS with respect to a function $f$ would be defined as
\begin{equation}
    \text{ESS}(f) = M\frac{\text{Var}_{\bs{\theta} \sim [\bs{\theta} | \bs{y}_{1:n}]}\left( \left. \hat{I}_{M}(f) \right| \bs{y}_{1:n} \right)}{\text{Var}_{\bs{\theta} \sim [\bs{\theta} | \bs{y}_{1:n_{1}}]}\left( \left. \tilde{I}_{M}(f) \right| \bs{y}_{1:n_{1}} \right)},
\end{equation}
where $M$ represents the size of the Monte Carlo sample used to calculate $\hat{I}_{M}(f)$ and $\tilde{I}_{M}(f)$, but, under certain conditions (see Appendix A of \citealp{martino2017effective}), this can be roughly approximated by
\begin{equation}\label{eq:RESS approximation}
    \text{ESS}(f) \approx M\left\{1+ \text{Var}_{\bs{\theta} \sim [\bs{\theta} | \bs{y}_{1:n_{1}}]}\left( \left. w(\bs{\theta}) \right| \bs{y}_{1:n_{1}} \right) \right\}^{-1} = M\frac{\left\{\int w(\bs{\theta}) [\bs{\theta} | \bs{y}_{1:n_{1}}] \ d\bs{\theta} \right\}^{2}}{\int w^{2}(\bs{\theta}) [\bs{\theta} | \bs{y}_{1:n_{1}}] \ d\bs{\theta}},
\end{equation}
which does not depend on the integrand $f$, and therefore can be seen as a universal measure of sample quality. For this reason, we take $Q(n|n_{1})$ as a measure of quality of a weighted sample, which we justify by writing $Q(n|n_{1})$ as a decreasing function of the $\chi^{2}$-divergence in equation (\ref{eq:RESS2}). Note that equation (\ref{eq:RESS approximation}) naturally leads to the estimator
\begin{equation}\label{eq:sample RESS}
    \hat{Q}(n | n_{1}) 
    = \frac{\left\{ \frac{1}{M}\sum_{m=1}^{M} w(\bs{\theta}_{m}) \right\}^{2}}{\frac{1}{M} \sum_{m=1}^{M} w^{2}(\bs{\theta}_{m})} 
    = \frac{\left\{ \frac{1}{M}\sum_{m=1}^{M} \tilde{w}(\bs{\theta}_{m}) \right\}^{2}}{\frac{1}{M} \sum_{m=1}^{M} \tilde{w}^{2}(\bs{\theta}_{m})},
\end{equation}
where $\{ \bs{\theta}_{m}, w^{*}(\bs{\theta}_{m}) \}_{m=1}^{M}$ is a weighted sample from $[\bs{\theta} | \bs{y}_{1:n_{1}}]$, and 
\begin{equation}
    w(\bs{\theta}) = w^{*}(\bs{\theta})[\bs{y}_{(n_{1}+1):n} | \bs{y}_{1:n_{1}}, \bs{\theta}],
\end{equation} 
which is straightforward to compute during model fitting and is the estimator we use to track the sample quality in all applications.

\subsection{Dissecting the Limiting Distribution}

Recall that the limiting distribution in Theorem 3.1, admits a decomposition of the form
\begin{equation}
    Q_{\theta^{*}, \alpha}(\bs{z}) = \left\{ \alpha (2-\alpha) \right\}^{\frac{d}{2}} \exp\left\{ -\tfrac{1-\alpha}{2-\alpha} \bs{z}'\bs{M}_{\theta^{*}}\bs{z} \right\} = u_{1}(\alpha) \cdot u_{2}(\alpha, \bs{z}),
\end{equation}
where $\bs{z} \sim \text{Normal}_{d}(\bs{0}, \bs{I}_{d})$, $u_{1}(\alpha) = \{\alpha (2-\alpha)\}^{\frac{d}{2}}$, and $u_{2}(\alpha, \bs{z}) = \exp\left\{ -\tfrac{1-\alpha}{2-\alpha}\bs{z}'\bs{M}_{\theta^{*}}\bs{z} \right\}$. In what follows, we use two examples to show that this decomposition leads to an intuitive understanding of the asymptotic behavior of $Q(n|n_{1})$.

\begin{example}\label{ex:scale}
    Let $\bs{\theta}|\bs{y}_{1:n} \sim \textnormal{Normal}_{d}(\bs{\theta}^{*}, n^{-1}\bs{V}_{\theta^{*}}^{-1})$ and $\bs{\theta}|\bs{y}_{1:n_{1}} \sim \textnormal{Normal}_{d}(\bs{\theta}^{*}, n_{1}^{-1}\bs{V}_{\theta^{*}}^{-1})$ represent posterior distributions, where $n_{1} = \alpha n  \in \mathbb{N}^{*}$ and $\phi(\cdot | \bs{\mu}, \bs{\Sigma})$ is the density of a multivariate normal random variable with mean vector $\bs{\mu}$ and covariance matrix $\bs{\Sigma}$. Then we have
    \begin{equation}\label{eq:factor 1}
        \begin{aligned}
            Q(n | n_{1})
            &=\left\{ \int \frac{ \phi^{2}(\bs{\theta} | \bs{\theta}^{*}, n^{-1}\bs{V}_{\theta^{*}}^{-1}) }{ \phi( \bs{\theta} | \bs{\theta}^{*}, n_{1}^{-1} \bs{V}_{\theta^{*}}^{-1}) } \ d\bs{\theta}  \right\}^{-1} \\
            &= \left\{ \int (2\pi\alpha)^{-\frac{d}{2}} |\bs{V}_{\theta^{*}}|^{\frac{1}{2}} \exp\left\{ -\tfrac{2-\alpha}{2} (\bs{\theta} - \bs{\theta}^{*})'\bs{V}_{\theta^{*}}(\bs{\theta} - \bs{\theta}^{*}) \right\} \ d\bs{\theta}  \right\}^{-1} \\
            &= \left\{\alpha (2-\alpha)\right\}^{\frac{d}{2}} = u_{1}(\alpha).
        \end{aligned}
    \end{equation}
\end{example}

Example \ref{ex:scale} is not a coincidence, but this connection comes from the asymptotic normality of the posterior distribution, and can be used to infer that $u_{1}(\alpha)$ describes the decrease in quality caused by the difference in scale between $[\bs{\theta} | \bs{y}_{1:n_{1}}]$ and $[\bs{\theta} | \bs{y}_{1:n}]$, because the posterior distribution concentrates around $\bs{\theta}^{*}$ with $n$ going to $+\infty$. Note that for any posterior considering i.i.d. observations the difference in scale is a consequence of the posterior concentration rate, which depends only on the sample size $n$, and thus $u_{1}(\alpha)$ is a deterministic function of $n$ and $n_{1}$. Intuitively, $u_{2}(\alpha, \bs{z})$ should appear in the limit when we also account for the difference in location between the two distributions, which we confirm in the example below.

\begin{example}\label{ex:location-scale}
    Let $\bs{y}_{1}, \dots, \bs{y}_{n} \overset{\text{iid}}{\sim} \textnormal{Normal}_{d}(\bs{\theta}^{*}, \bs{V}_{\theta^{*}}^{-1} \bs{W}_{\theta^{*}} \bs{V}_{\theta^{*}}^{-1})$, and let 
    $$\begin{aligned} \bs{\theta} | \bs{y}_{1:n} &\sim \textnormal{Normal}_{d}(\bar{\bs{y}}_{1}^{n}, n^{-1}\bs{V}_{\theta^{*}}^{-1}) & &\text{and}& \bs{\theta} | \bs{y}_{1:n_{1}} &\sim \textnormal{Normal}_{d}(\bar{\bs{y}}_{1}^{n_{1}}, n_{1}^{-1}\bs{V}_{\theta^{*}}^{-1}) \end{aligned}$$ represent posterior distributions, where $n_{1} = \alpha n \in \mathbb{N}^{*}$, $\bar{\bs{y}}_{i}^{j} = \frac{1}{j-i+1} \sum_{k=i}^{j} \bs{y}_{k}$, and $\phi(\cdot | \bs{\mu}, \bs{\Sigma})$ is the density of a multivariate normal random variable with mean vector $\bs{\mu}$ and covariance matrix $\bs{\Sigma}$. Then it can be shown that
    \begin{equation}\label{eq:factor 2}
        \begin{aligned}
            Q(n | n_{1})
            &= \left\{ \int \frac{ \phi^{2}(\bs{\theta} | \bar{\bs{y}}_{1}^{n}, n^{-1}\bs{V}_{\theta^{*}}^{-1}) }{ \phi( \bs{\theta} | \bar{\bs{y}}_{1}^{n_{1}}, n_{1}^{-1} \bs{V}_{\theta^{*}}^{-1}) } \ d\bs{\theta} \right\}^{-1} \\
            &= \left\{\tfrac{n_{1}(2n-n_{1})}{n^{2}}\right\}^{\frac{d}{2}} \exp\left\{ -\tfrac{nn_{1}}{2n-n_{1}} (\bar{\bs{y}}_{1}^{n} - \bar{\bs{y}}_{1}^{n_{1}})'\bs{V}_{\theta^{*}} (\bar{\bs{y}}_{1}^{n} - \bar{\bs{y}}_{1}^{n_{1}}) \right\} \\
            &= \left\{ \alpha(2-\alpha)\right\}^{\frac{d}{2}}\exp\left\{ -\tfrac{1-\alpha}{2-\alpha} \bs{z}' \bs{M}_{\theta^{*}}\bs{z} \right\} 
            = Q_{\theta^{*},\alpha}(\bs{z}),
        \end{aligned}
    \end{equation}
    where $\bs{z} = \left\{ \frac{n}{(n-n_{1})n_{1}} \right\}^{-\frac{1}{2}} \bs{W}_{\theta^{*}}^{-\frac{1}{2}}\bs{V}_{\theta^{*}} (\bar{\bs{y}}_{n_{1}+1}^{n} - \bar{\bs{y}}_{1}^{n_{1}} ) \sim \textnormal{Normal}_{d}(\bs{0}, \bs{I}_{d})$.
\end{example}

Example \ref{ex:location-scale} not only allows us to more clearly identify the source of randomness of the limit in Theorem 3.1, which comes from the randomness in the estimation of $\bs{\theta}^{*}$ using $\bs{y}_{1:n}$, but also provides an example we can use to understand the effect of model misspecification. Assuming a correctly specified model and under the classical regularity conditions, the matrices $\bs{V}_{\theta^{*}}$ and $\bs{W}_{\theta^{*}}$ are both equal to the Fisher information matrix at $\bs{\theta}^{*}$, and we have the simplification $\bs{M}_{\theta^{*}} = \bs{I}_{d}$. Considering now a misspecified model with $\bs{W}_{\theta^{*}} = c\bs{V}_{\theta^{*}}$ for some $c > 0$, we can rewrite (\ref{eq:factor 2}) as
\begin{equation}
    Q_{\theta^{*},\alpha}(\bs{z}) = \left\{ \alpha(2-\alpha)\right\}^{-\frac{d}{2}}\exp\left\{- \tfrac{1-\alpha}{2-\alpha} c \bs{z}'\bs{z} \right\},
\end{equation}
therefore taking $c > 1$, which would correspond to a posterior distribution that underestimates the variance of the true data generating process, results in a decrease of the $Q(n|n_{1})$, while the opposite is true for $c < 1$. This implies that underestimation of uncertainty can significantly slow model fitting, because more replenishment steps are needed throughout, while overestimating uncertainty would have the opposite effect.

These results suggest that, under the assumptions of Theorem 3.1, the degeneration in RAISOR happens either as a consequence of the concentration of the posterior distribution around $\bs{\theta}^{*}$ or due to a significant shift of the posterior mass when accounting for the gain in information from new observations. However, we argue that the same should be true in general. Intuitively, weight degeneration occurs when the targeted partial posterior differs significantly from the current proposal. But given that RAISOR (approximately) uses a partial posterior as the proposal, the difference between proposal and target must be solely a consequence of the information gain from new observations, i.e., it depends on how informative or surprising new observations are relative to what was previously accounted for. Under this reasoning, we can determine how RAISOR is expected to behave beyond the assumptions of Theorem 3.1.

When considering the stable RAISOR implementation proposed in Section 2, our quality-based replenishment heuristic can be seen as the greediest conservative approach to replenishment. It is greedy in the sense that it will delay replenishment steps as much as possible, but it is conservative because it does not allow the estimated quality to drop below a certain threshold. Our theory indicates that in the i.i.d. case, the degeneration happens just slowly enough so that the asymptotic cost of replenishment is at most of the same order (with respect to the sample size) of fitting the model once. Because the heuristic is conservative, if the rate of degeneration is faster than in the i.i.d. case, it will introduce more replenishment steps, leading to a higher asymptotic cost and rendering the method inefficient. However, if the rate of degeneration is slower, the greedy heuristic will adapt by delaying replenishment, thus maintaining efficiency. 

For dependent models, such as the geostatistical model in Section 4, we expect new observations to be on average less informative due to the partial redundancy with what was previously observed. Therefore, one can expect the decay in quality of the weights to be slower than in an i.i.d. setting. Another interesting case is of fitting a model to a data set with highly ``surprising'' observations (e.g., outliers). In this case, one would expect a significant drop in weight quality after each update that contains such an observation, and consequently degeneration would likely happen faster than expect. Therefore, we expect the effectiveness of the proposed approach to be robust to the dependence of the data generating process and of the assumed model, as long as uncertainty is properly accounted for to minimize the impact of influential observations.

\end{document}